\documentclass[iop]{emulateapj}
\usepackage{graphicx}
\usepackage{color}
\usepackage{amsmath,amssymb}
\usepackage{type1cm}
\usepackage{ulem}
\usepackage{mathrsfs}
\def\vec#1{\mbox{\boldmath $#1$}}

\begin{document}

\title{Matter mixing in aspherical core-collapse supernovae: a search for possible conditions for  
conveying $^{56}$Ni into high velocity regions}

\author{
Masaomi Ono\altaffilmark{1},  
Shigehiro Nagataki\altaffilmark{1}, 
Hirotaka Ito\altaffilmark{1}, 
Shiu-Hang Lee\altaffilmark{1},  
Jirong Mao\altaffilmark{1},\\
Masa-aki Hashimoto\altaffilmark{2}, 
and 
Alexey Tolstov\altaffilmark{1}
}
\affil{$^1$Astrophysical Big Bang Laboratory, RIKEN, Saitama 351-0198, Japan; masaomi.ono@riken.jp\\
$^2$Department of Physics, Kyushu University, Fukuoka 812-8581, Japan
}

\shorttitle{Mixing in aspherical core-collapse supernovae}
\shortauthors{Ono et al.}

\begin{abstract}
We perform two-dimensional axisymmetric hydrodynamic simulations of matter mixing in 
aspherical core-collapse supernova explosions of a 16.3 $M_{\odot}$ star with a compact hydrogen 
envelope. Observations of SN~1987A have provided evidence that $^{56}$Ni synthesized by 
explosive nucleosynthesis is mixed into fast moving matter ($\gtrsim$ 3,500 km s$^{-1}$) 
in the exploding star. In order to clarify the key conditions for reproducing such high velocity 
of $^{56}$Ni, we revisit matter mixing in aspherical core-collapse supernova explosions. 
Explosions are initiated artificially by injecting thermal and kinetic energies around the interface 
between the iron core and the silicon-rich layer. 
Perturbations of 5\% or 30\% amplitude in the radial velocities are introduced at several points in time. 
We found that no high velocity $^{56}$Ni can be obtained if we consider bipolar explosions with 
perturbations (5\% amplitude) of pre-supernova origins. 
If large perturbations (30\% amplitude) are introduced or exist due to some unknown mechanism 
in a later phase just before the shock wave reaches the hydrogen envelope, $^{56}$Ni with a velocity 
of  3,000 km s$^{-1}$ can be obtained. 
Aspherical explosions that are asymmetric across the equatorial plane with clumpy structures 
in the initial shock waves are investigated. We found that the clump sizes affect the penetration 
of $^{56}$Ni. Finally, we report that an aspherical explosion model that is asymmetric across 
the equatorial plane with multiple perturbations of pre-supernova origins can cause the penetration of $^{56}$Ni clumps into fast moving matter of 3,000 km s$^{-1}$. 
We show that both aspherical explosion with clumpy structures and perturbations of 
pre-supernova origins may be necessary to reproduce the observed high velocity of $^{56}$Ni. 
To confirm this, more robust three-dimensional simulations are required. 
\end{abstract}

\maketitle

\section{Introduction}

Morphologies of supernova explosions is a topic of hot debate. Many observations of supernovae 
and supernova remnants have indicated an aspherical nature of the supernova explosions. 
SN~1987A, a supernova occurred in the Large Magellanic Cloud on February 23rd, 
has provided many interesting features to be explained by astronomers and astrophysicists. 
Observations of SN~1987A have implied large-scale matter mixing in the supernova explosion 
from several aspects.  Early detection of hard X-ray \citep{dot87,sun87} and $\gamma$-ray lines 
from decaying $^{56}$Co \citep{mat88} have indicated that radioactive $^{56}$Ni synthesized 
by explosive nucleosynthesis is mixed into fast moving matter composed of helium and hydrogen. 
The sudden development of the fine-structure of the H$_{\alpha}$ line \citep[Bochum event :][]{han88} 
implies the existence of a high velocity ($\sim$ 4,700 km s$^{-1}$) clump of $^{56}$Ni with a mass 
of several 10$^{-3}$ $M_{\odot}$ \citep{utr95}. 
The observed line profiles of [Fe II] in SN~1987A show that the maximum velocity of $^{56}$Ni 
(or its decay products $^{56}$Co and $^{56}$Fe) reaches $\sim$ 4,000 km s$^{-1}$ and 
the position of the peak of the flux distribution as a function of Doppler velocity 
is located in the red-shifted side \citep{haa90,spy90}. 
The shape of the flux distribution is asymmetric across the peak. 
Modeling the light curve of SN~1987A using one-dimensional radiation hydrodynamics calculations 
requires the mixing of $^{56}$Ni into high velocity regions to reproduce the observed features 
of the light curve \citep{woo88,shi88,shi90,bli00,utr04}. 
\citet{shi90}, \citet{bli00}, and \citet{utr04} have 
insisted that mixing of $^{56}$Ni into high velocity regions up to 3,000 km s$^{-1}$, 4,000 km s$^{-1}$, 
and  2,500 km s$^{-1}$, respectively. Therefore, the clear consensus about the maximum velocity of 
$^{56}$Ni has not been obtained from modelings the light curve and H$_{\alpha}$ line. 
However, at least 4\% of total mass of $^{56}$Ni would have $>$ 3,000 km s$^{-1}$ \citep{haa90}. 
In addition to $^{56}$Ni, mixing of hydrogen into inner cores have been inferred and the 
minimum hydrogen velocity can be $\lesssim$ 800 km s$^{-1}$ \citep{shi90,koz98}. 
Asphericity of core-collapse supernova explosions have also been implied from other Type II-P supernovae. 
Observations of He I lines in the IR band from Type II-P supernovae indicate the mixing of 
$^{56}$Ni into the helium regions \citep[SN~1995V:][]{fas98}. 
Clumped structures of ejecta have been revealed by the observations of metal lines of other 
Type II supernovae (SN~1988A: \citealt{spy91}, SN~1993J: \citealt{spy94}). 
Recent optical observations of the inner ejecta of the supernova remnant of SN~1987A have 
revealed that the morphology of the ejecta is elliptical and the ratio of the major to minor axises 
of the ejecta is 1.8 $\pm$ 0.17 \citep{kja10}. 
The three-dimensional structure of supernova remnant Cassiopeia A demonstrates clearly 
that the ejecta is rather clumpy \citep{del10}. 

Theoretically, there is a growing awareness of multi-dimensional effects in supernova 
explosion mechanisms. In the context of the neutrino heating mechanism, convection in the 
neutrino heating layers and standing accretion shock instability (SASI) may result in 
a globally anisotropic structure inside a supernova shock wave \citep[e.g.,][]{kot06}. 
Magnetohydrodynamic (MHD) simulations of the core-collapse of massive stars 
\citep{kot04,saw05,bur07,tak09} have demonstrated magnetorotationally driven jetlike explosions. 
For more detailed descriptions of multi-dimensional effects of supernova explosions, 
see the recent reviews by \citep[e.g.,][]{kot12,jan12}. 

Rayleigh-Taylor (RT) instabilities have been thought to be a promising mechanism to facilitate 
large-scale matter mixing in supernova explosions. Other hydrodynamic instabilities, 
such as Richtmeyer-Meshkov (RM) instabilities and Kelvin-Helmholtz (KH) instabilities, 
may also contribute to the mixing in supernovae along with the RT instability. 
The condition for the RT instability for a compressible fluid is given by 
$(d P/ d r) \,(d \rho / dr) < 0$ \citep{che76}, where $P$ is the pressure, $r$ is the radius, 
and $\rho$ is the density. Stability analyses of supernova shock wave propagations in 
a pre-supernova model of SN~1987A using one-dimensional  hydrodynamics have depicted 
that the composition interfaces  between the hydrogen- and helium-rich layers (He/H) and 
that between the helium-rich layer and C+O core (C+O/He) can become unstable against 
RT instabilities \citep{ebi89,ben90}. 

We note that \citet{arn89a} commented on the possible sources of the perturbations for 
initiating the hydrodynamic instabilities. The authors considered three possibilities. 
One is thermonuclear shell flashes in the oxygen-rich layer. Second is hydrogen shell burning 
at the edge of the helium core of a pre-supernova star and which makes a jump in density  
at the composition interface of He/H. 
Note that the authors stated that the jump is not significant for RT instabilities. 
Third is the `nickel bubble', i.e. 
the heating via decays of $^{56}$Ni competing with the adiabatic cooling. 
Two-dimensional hydrodynamic simulations of a pre-collapse star have depicted 
significant fluctuations (up to 8\% in density) due to convective oxygen-shell burning 
at the edges of a burning shell \citep{baz98}. Recent two-dimensional hydrodynamic 
simulation of progenitor evolution of a 23 $M_{\odot}$ star demonstrated the growth 
of instabilities of low-order modes and a large anisotropy in each burning shell \citep{arn11}. 

Motivated by the observational evidence of matter mixing in supernovae, two or three-dimensional hydrodynamic simulations have been performed in early papers to investigate the effects of 
RT instabilities on mixing in shock wave propagations in the progenitor star of SN~1987A 
\citep{arn89a,hac90,fry91,mul91,her91,hac92}. 
All studies above have combined one-dimensional hydrodynamic simulations of supernova 
explosions with multi-dimensional simulations of late time evolutions of the shock wave propagations. 
The explosions have been implemented through some ad hoc ways, e.g. thermal bombs or piston models. 
However, such simulations have revealed that RT instabilities are insufficient to explain the 
high velocity metals. The obtained maximum velocity of $^{56}$Ni is $\sim$ 2,000 km s$^{-1}$ 
at 90 day after the explosion using a two-dimensional smoothed particle hydrodynamic (SPH) 
code \citep{her91}. \citet{her92a} referred to this gap between observations and models 
as the `nickel discrepancy'. \citet{her92a} suggested that premixing in regions of the inner 
1.5 $M_{\odot}$ above the mass cut is required to reproduce the high velocity wings 
of the [Fe II] line profiles. 

Explosive nucleosynthesis in jetlike explosions have been investigated in several papers 
\citep{nag97,nag03,nag06,fuj07,fuj08,ono09,win12,ono12}. In the context of jetlike explosions, 
matter mixing in mildly asymmetric explosions of the progenitor star of SN~1987A with 
monochromatic perturbations have been investigated \citep{yam91,nag98b,nag00}. 
\citet{yam91} concluded that an asymmetric explosion with initial perturbations of 
30\% amplitude causes strong mixing and high velocity innermost metals ($\sim 4,000$ km s$^{-1}$). 
\citet{nag98b} and \citet{nag00} have reproduced the high velocity component of $^{56}$Ni 
($\sim 3,000$ km s$^{-1}$) and line profiles of [Fe II] observed in SN~1987A by a mildly 
aspherical supernova explosion model with large monochromatic perturbations 
(amplitude of $30\%$). Additionally, \citet{nag00} suggested that the strong alpha-rich freeze-out 
in a jetlike explosion is favored to explain the amount of $^{44}$Ti in SN~1987A. 
\citet{nag98a} applied a high ratio of $^{44}$Ti/$^{56}$Ni in an asymmetric core-collapse 
explosion to Cassiopeia A. 
Note that recently, direct-escape (Hard X-ray) emission lines from the decay of $^{44}$Ti 
have been detected \citep{gre12} in the remnant of SN~1987A and the mass of $^{44}$Ti is 
estimated to be (3.1$\pm$0.8)$\times$10$^{-4}$ $M_{\odot}$. 
In \citet{yam91}, \citet{nag98b}, and \citet{nag00}, the resolutions of the simulations are 
rather low and they have not taken into account the effects of fallback of the ejecta.  
In \citet{nag98b}, large perturbations of 30\% amplitude are introduced when the shock front 
reaches the composition interface of He/H. However, as the author noted, such large 
perturbations should be introduced only in the explosion itself. 
\citet{hun03} and \citet{hun05} have investigated the effects of aspherical explosions 
on the $\gamma$-ray lines using a three-dimensional SPH code. 
The authors have shown that aspherical explosions change significantly the velocity 
distribution of $^{56}$Ni compared to that in spherical explosions, and aspherical models 
may reproduce mixing of $^{56}$Ni into the edge of hydrogen and red-shifted [Fe II] lines. 
\citet{cou09} performed two-dimensional simulations of bipolar, jetlike explosions of 
Type II supernovae using an adaptive mesh refinement (AMR) hydrodynamic code and 
commented on the observational features of jetlike explosions against those associated to 
Type II-P supernovae. Recently, \citet{ell12} studied RT mixing in a series of aspherical 
core-collapse supernova explosions using a three-dimensional SPH code and the authors 
discussed the sizes of the arising clumps. 

\citet{jog09} investigated matter mixing due to RT instabilities and fallback in spherical 
core-collapse supernova explosions of solar- and  zero-metallicity stars with a 
two-dimensional AMR code. The results depict that the growth of RT instabilities are 
significantly reduced in the zero-metallicity stars which are compact blue supergiants. 
\citet{jog10a} examined RT mixing in spherical supernova explosions of rotating zero-metallicity 
and metal-poor stars. The rotating zero-metallicity stars end their lives as red supergiants 
in contrast to non-rotating ones. Thus, more mixing and less fallback are expected in rotating 
zero-metallicity stars than that in non-rotating ones. Three-dimensional simulations of RT mixing 
in supernova explosions of rotating zero-metallicity and metal-poor stars 
indicate \citep{jog10b} that the degree of mixing at the ends of simulation time does not differ much 
from that in the two-dimension case. 

\citet{kif03} and \citet{kif06} have investigated matter mixing in neutrino-driven core-collapse 
supernova explosions aided by convection and SASI using AMR hydrodynamic codes. 
The authors have found that if the shock wave has only small-scale deviations from spherical 
symmetry (high-order modes), no high velocity $^{56}$Ni clump should be expected. 
On the other hand, a globally aspherical explosion (low-order modes, $l$ = 1, 2)  with a relatively 
high explosion energy (2 $\times$ 10$^{51}$ erg) causes strong RM instabilities at the 
composition interface of He/H and makes clumps of metals penetrate into a dense helium shell 
before the formation of a strong reverse shock. High velocity $^{56}$Ni clumps 
($\sim$ 3,300 km s$^{-1}$) are obtained by the globally aspherical explosion. 
\cite{gaw10} re-investigated the study of \citet{kif06} using a single computational domain 
and pointed out that it is difficult to achieve robust conclusions by two-dimensional 
axisymmetric hydrodynamic codes. 

\citet{ham10} performed a three-dimensional simulation of mixing in a neutrino-driven core-collapse 
supernova explosion of a compact blue star. The authors suggested that in the three-dimensional 
model, clumps of ejecta feel less drag force than that in the two-dimensional counterparts, 
and the high velocity iron group elements ($\sim$ 4,500 km s$^{-1}$) with a mass of 
$\sim$ 10$^{-3}$ $M_{\odot}$ are reproduced in the three-dimensional model, which cannot 
be obtained in two-dimension.  However, the resolution of the simulation is lower than that of 
two-dimensional high-resolution studies \citep[e.g.,][]{kif06} due to the limitation of 
computational resources, and the authors also neglected the effects of gravity, 
i.e., fallback of matter into the compact remnant. More robust calculations are required 
to conclude such dimensional effects on the high-velocity metals. 

As referenced above, there exists only a few models that obtained high velocity 
$^{56}$Ni clumps of $\gtrsim$ 3,000 km s$^{-1}$. However, even in such models, 
there are still several drawbacks in those simulations. 
The resolutions of simulations in \citet{yam91}, \citet{nag98b}, and \citet{nag00} and the three-dimensional 
simulation in \citet{ham10} are low compared with recent two-dimensional hydrodynamic simulations on matter 
mixing in supernovae \citep[e.g.,][]{kif06} and some hydrodynamical instabilities may not be captured in their simulations. 
The non-radial motion of initial explosion models used in \citet{kif06} and \citet{gaw10} tends to concentrate ejecta 
into polar regions. However, ejecta motion around polar regions are doubtful in axsymmetric two-dimensional simulation.
Therefore, the conditions for reproducing 
the observed high velocity of $^{56}$Ni are still unclear. 
In the present paper, we investigate matter mixing in a series of aspherical core-collapse supernova 
explosions of a 16.3 $M_{\odot}$ star with a compact hydrogen envelope using a two-dimensional AMR hydrodynamic code 
in order to clarify the key conditions for reproducing such high velocity of $^{56}$Ni. 
To survey a large variety of aspherical explosions, we adopt the stance that explosions are 
initiated artificially in similar ways as the earlier papers.  We revisit RT mixing in mildly 
aspherical bipolar explosions by introducing initial perturbations at several points in time. 
We also consider globally anisotropic explosions with clumpy structures by mimicking 
neutrino-driven core-collapse explosions. The purpose of this paper is to do a comprehensive 
search for the preferable conditions to explain the the observed high velocity of $^{56}$Ni. 
In \S 2, our numerical methods are described. \S 3 is devoted to explaining our models 
in the this paper. We will show our results in \S 4, and then discuss several important aspects 
based on the results in \S 5.  Finally, we conclude our study in \S6. 

\section{Numerical method, initial conditions}

The computations in this paper are preformed with the adaptive mesh refinement (AMR) 
hydrodynamic code, FLASH \citep{fry00}. We use the directionally split Eulerian version of 
the piecewise parabolic method (PPM) \citep{col84}, which provides a formally second-order 
accuracy in both space and time. 
To avoid an odd-even instability (decoupling) \citep{qui97} that can arise from shocks that 
are aligned with a grid, we adopt a hybrid Riemann solver which switches to 
an ELLE solver inside shocks. 
AMR is implemented using the PARAMESH package \citep{mac00}. 
We employ an error estimator based on \citet{loh87} adopted originally in PARAMESH package 
for the refinement criteria. For the refinement, the density, pressure, velocity, and mass fractions 
of nickel, oxygen, helium, and hydrogen are selected. 
In our computations, the two-dimensional axisymmetric spherical coordinate ($r$,\,$\theta$) 
is adopted. The initial computational domain covers the region of 
$1.36 \times 10^{8} \,{\rm cm} < r < 3 \times 10^{9} \,{\rm cm}$ and $0 <\theta < \pi$. 
The initial radius of the outer boundary corresponds to the inner part of the oxygen-rich 
layer of a pre-supernova star. The pre-supernova model used in this paper will be described below. 
The numbers of grid points of the base level (level 1) are set to be 48\,($r$) $\times$ 12\,($\theta$). 
The maximum refinement level is set to be 7. 
Therefore, the effective maximum numbers of grid points are 3072\,($r$) $\times$ 768\,($\theta$). 
The minimum effective cell sizes are approximately 10 km and 0.23 
degree in the radial and $\theta$ directions, respectively. 

To follow large physical scales from the onset of a explosion to the shock breakout, we extend gradually 
the computational domain as the forward shock propagates outward and remap the physical values 
in new domains. If the forward shock reaches close to the radial outer boundary, the radial size of the 
computational region is extended by a factor of 1.2. If the radius of the inner boundary becomes 
less than 1\% of that of the outer boundary, the radius of the inner boundary is also expanded 
by keeping to be 1\% of that of the outer boundary to prevent the time steps from becoming 
too small due to the Courant-Friedrichs-Levy (CFL) condition. In particular, the propagation of 
an acoustic wave in the $\theta$ direction in a time step is restricted severely due to the CFL condition. 
The physical values of the extended region are set to be the values of the pre-supernova model. 
The propagation of the forward shock is basically supersonic, which allows us to adopt such a prescription. 
The radius of the surface of the pre-supernova star is about 3.4 $\times$ 10$^{12}$ cm. 
Therefore, about 40 remappings are required to cover the whole star. 
Note that in previous studies \citep[e.g.,][]{kif06} similar to the present paper, the factors of 
expansions are roughly between 2 and 3. However, we found that if we adopt a factor of 2, 
the hydrodynamic values of the inner part tend to be diffusive due to remapping especially 
in the accelerating phases of shocks. Additionally, the factor of 1.2 has an advantage of 
extending time steps efficiently owing to the more frequent expansions of grids. 
To see whether such procedures introduce a significant artifact in our computations, 
we check the conservation of the total mass. We confirm that the errors due to each remapping 
are ranged between 10$^{-7}$ and 10$^{-5}$. Therefore, 40 remappings may not introduce errors 
above a factor of 10$^{-3}$ for global values at a maximum. 
Note that although the maximum refinement level is constant through a simulation, 
successive remappings enlarge gradually the effective minimum grid size as the 
computational domain is extended. 
In the remapping procedures, we use a monotonic cubic interpolation scheme \citep{ste90} 
for interpolations of physical values. The computational cost is approximately 10,000 CPU hr 
for each model in the present paper. 

At the start of the simulation, a `reflection' boundary condition is employed for the radial 
inner boundary. After the forward shock has reached the composition interface of C+O/He (corresponds to the radius of 6 $\times$ 10$^{9}$ cm), 
it is switched to a `diode' boundary condition that allows matter to flow out of the 
computational domain but inhibits matter from entering the computational domain through 
the inner boundary in order to include the effects of fallback of matter. 
If we use the `diode' boundary condition for the radial inner boundary throughout 
the whole simulation, we may overestimate the fallback of matter. As we will show later, 
explosions are initiated by injecting kinetic and thermal energies artificially around the 
inner boundary. In the case of the `diode' boundary condition, a significant part of matter 
immediately above the inner boundary falls into the central object through the inner boundary 
at the initiation of the explosion. 
Since such a situation does not match our intention, 
we adopt the `reflection' boundary condition initially. 
Although changing the timing of the switch can somewhat affect the degree of the fallback 
of the innermost matter, we fix the timing of the switch by making sure that 
the mass of $^{56}$Ni remained in the computational domain does not become too small 
compared with that for SN~1987A \citep[$\sim$ 0.07 $M_{\odot}$: e.g.,][]{shi88}. 
Note that if the `diode' boundary condition is used through the simulation, 
the mass of $^{56}$Ni is approximately 1 $\times$ 10$^{-3} M_{\odot}$ in model SP1 
(see \S 3 for the description of models). 
The maximum velocity of $^{56}$Ni is also affected by the boundary 
condition. If the `diode' boundary condition is used through the simulation, 
the maximum velocity of $^{56}$Ni becomes half in model SP1. 
However, we confirm that the timing of the switch does not  affect the obtained maximum 
velocity of $^{56}$Ni much. 
If we change the corresponding radius of the timing of the switch to 3 $\times$ 10$^{9}$ cm and 
1.2 $\times$ 10$^{10}$ cm, the obtained maximum velocities of $^{56}$Ni are same as in the case of the radius 
of 6 $\times$ 10$^{9}$ cm (1,600 km s$^{-1}$ in model SP1) within the accuracy of 100 km s$^{-1}$ (see \S 4.1 for the
definition of the maximum velocity of $^{56}$Ni ). 
We fix the other boundary conditions throughout the whole simulations. 
The `reflection' and `diode' boundary conditions are employed for the edges in $\theta$ direction 
and the radial outer boundary, respectively. 

We have included the effects of gravity in our computations as follows. Since it takes much time 
to solve correctly the Poisson equation for self-gravity,  we adopt a spherically symmetric 
approximation for gravity. Spherical density profiles are calculated by averaging the values 
in the $\theta$-direction and local gravitational potentials are estimated from enclosed masses 
at each radius. Point source gravity from the mass inside the radial inner boundary is also included. 
The total mass that passes out through the inner boundary at each time step is added to the point mass. 

Explosive nucleosynthesis is calculated using a small nuclear reaction network including 
19 nuclei (Aprox19) n, p, $^1$H, $^4$He, $^{12}$C, $^{14}$N, $^{16}$O, $^{20}$Ne, $^{24}$Mg, 
$^{28}$Si, $^{32}$S, $^{36}$Ar, $^{40}$Ca, $^{44}$Ti, $^{48}$Cr, $^{52}$Fe, $^{54}$Fe, 
and $^{56}$Ni (see \cite{wea78} for the network chain). 
The MA28 sparse matrix package \citep{duf86} and the Bader-Deuflhard method, a time 
integration scheme, \citep[e.g.,][]{bad83} are used. The feedback of nuclear energy generation 
is included in the hydrodynamic code. Among our models, the maximum temperature reached 
in the simulations is roughly 10$^{10}$ K, and in such high temperature 
($\gtrsim$ 5$ \times $10$^{9}$ K), nuclear statistical equilibrium (NSE) is established. 
Thus the time scales of nuclear burning can be much smaller than that of the hydrodynamics. 
In the paper, we do not intend to focus on the effects of the feedback of nuclear reactions. 
Therefore, we do not impose a time step limiter for the coupling of nuclear burning 
with hydrodynamics to save computational time. Hence, the obtained mass fractions of e.g., 
$^{56}$Ni may be overestimated. Additionally, since we use the small nuclear reaction network 
including only 19 nuclei, neutron-rich matter is eliminated and cannot be calculated. 
In our models, the electron fraction at the initial radial inner boundary 
($1.36 \times 10^{8} \,{\rm cm}$) is approximately 0.493. In the electron fraction of $\sim$ 0.49, 
$^{56}$Ni is the dominant product of the explosive nucleosynthesis. 
However, if the explosion is rather aspherical, more neutron-rich matter can be potentially ejected. 
If more neutron rich matter is ejected by the explosion, neutron-rich nuclei and weak interactions 
should be definitely taken into account in the nucleosynthesis calculation. 
A detailed quantitative discussion on the mass fractions, e.g., the abundance ratio of isotopes, 
is beyond the scope of the present paper and will be left for our followup studies. 
To trace the distribution of elements, the advection equations for 19 elements, 
\begin{equation}
\frac{\partial \rho X_i}{\partial t} + \nabla \cdot (\rho X_i \vec{v}) = 0,
\end{equation}
are solved in addition to the hydrodynamic equations, where $X_i$ is the mass fraction of 
the element of index $i$, $t$ is the time, and $\vec{v}$ is the velocity. 

In order to close the hydrodynamic equations, an equation of state (EOS) is required, we adopt 
the Helmholtz EOS \citep{tim00}, which includes contributions from radiation, completely 
ionized nuclei, and degenerate/relativistic electrons and positrons. Since Helmholtz EOS only 
covers the physical region of  $10^{-10} < \rho < 10^{11}$ g cm$^{-3}$ and $10^{4}< T <10^{11}$ K, 
for the region of $\rho < 10^{-9}$ g cm$^{-3}$, we adopt another EOS that includes contributions 
from radiation and ideal gas of elements as follows. 
\begin{equation}
P = f (\rho, T) \, \frac{1}{3} \, a T^4 + \frac{k_{\rm B}}{\mu \,m_{\rm H}} \, \rho T,
\label{eq:eosp}
\end{equation}
\begin{equation}
E = \frac{a T^4}{\rho} + 1.5 \, \frac{k_{\rm B}}{\mu \, m_{\rm H}} \, T,
\label{eq:eose}
\end{equation}
where $a$ is the radiation constant, $T$ is the temperature, $k_{\rm B}$ is the 
Boltzmann constant, $\mu$ is the mean molecular weight, $m_{\rm H}$ is the 
atomic mass unit, and $E$ is the specific internal energy. 
In an optically thin region, the pressure from radiation should be neglected. 
However, in our hydrodynamic code, we cannot treat separately radiation and the gases 
of nuclei in an appropriate manner. Therefore, we control the contribution of the pressure 
from radiation by a multiplicative factor $f (\rho, T)$. We take the form of $f (\rho, T)$ from \citet{jog10a}: 
\begin{equation}
f (\rho, T) = 
\begin{cases}
\ 1 & \rho \geqslant 10^{-9}\,{\rm g \,cm}^{-3} \\
     & \text{   or } T \leqslant T_{\rm neg} \\
 \ \exp \left( - \frac{T - T_{\rm neg}}{T_{\rm neg}} \right) & \rho < 10^{-9} \,{\rm g \,cm}^{-3} \\
     & \text{ and } T > T_{\rm neg} ,\\
\end{cases}
\end{equation}
where $T_{\rm neg} = ( 3 \,\rho\, k_{\rm B} / \,100 \,\mu\,m_{\rm H} \,a )^{1/3}$. 
In the hydrodynamic steps, input values of the EOS are ($\rho$,  $E$, $\mu$). 
First, $T$ is derived from Equation~(\ref{eq:eose}), then $P$ is calculated by 
Equation~(\ref{eq:eosp}). For the transition region of $10^{-8} < \rho < 10^{9}$ g cm$^{-3}$, 
we blend smoothly the Helmholtz EOS and the EOS expressed by Equations 
(\ref{eq:eosp}) and (\ref{eq:eose}). 

Energy depositions due to radioactive decays of $^{56}$Ni to $^{56}$Fe are included 
in the hydrodynamic code by the same method as described in \citet{jog09}. 
We assume that full energy depositions take place locally. The energy deposition rate 
$\dot{E}_{\rm Ni}$ due to the decay of $^{56}$Ni to $^{56}$Co is estimated as
\begin{equation}
\dot{E}_{\rm Ni} = \lambda_{\rm Ni} X_{\rm Ni} \, e^{- \lambda_{\rm Ni} \, t} q_{\rm Ni} 
\ \ \ \ \ {\rm erg \ g}^{-1} \ {\rm s}^{-1}, 
\label{eq:ni}
\end{equation}
where $\lambda_{\rm Ni}$ is the decay rate of $^{56}$Ni, $X_{\rm Ni}$ is the mass fraction of $^{56}$Ni, 
and $q_{\rm Ni}$ is the q-value of the decay of $^{56}$Ni to $^{56}$Co. 
We take the values of $\lambda_{\rm Ni}$ and $q_{\rm Ni}$ to be 1.315 $\times$ 10$^{-6}$ s$^{-1}$ 
and 2.96 $\times$ 10$^{16}$ erg g$^{-1}$, respectively. 
The energy deposition rate $\dot{E}_{\rm Co}$ due to the decay of $^{56}$Co to $^{56}$Fe is given by
\begin{equation}
\begin{split}
\dot{E}_{\rm Co} = \frac{\lambda_{\rm Ni}}{\lambda_{\rm Co} - \lambda_{\rm Ni}} \, X_{\rm Ni} 
\left( e^{-\lambda_{\rm Ni}\,t} - e^{-\lambda_{\rm Co}\,t}\right) \, \lambda_{\rm Co} \, q_{\rm Co} \\
{\rm erg \ g}^{-1} \ {\rm s}^{-1},
\end{split}
\label{eq:co}
\end{equation}
where $\lambda_{\rm Co}$ is the decay rate of $^{56}$Co and $q_{\rm Co}$ is the q-value of the decay of $^{56}$Co to $^{56}$Fe. 
The values of $\lambda_{\rm Co}$ and $q_{\rm Co}$
are taken to be 1.042 $\times$ 10$^{-7}$ s$^{-1}$ and 6.4 $\times$ 10$^{16}$ erg g$^{-1}$, respectively. 

The pre-supernova model used in the paper is a 16.3 $M_{\odot}$ star with a 6 $M_{\odot}$ 
helium core \citep{nom88} and a 10.3 $M_{\odot}$ compact hydrogen envelope. The radius of the surface 
of the hydrogen envelope is 3.4 $\times 10^{12}$ cm. SN~1987A is known to be a blue supergiant 
and our pre-supernova model is preferable to study the case of SN~1987A \citep[see e.g.,][]{shi90}. 
To follow the simulations after the shock breakout, a stellar wind component is required. 
Therefore, we attach a wind component of the density profile of $\rho \propto r^{-2}$ 
and a uniform temperature of $T = 10^4$ K. 
The inner density of the wind component is 3.0 $\times$ 10$^{-10}$ g cm$^{-3}$. 
The wind component is extended to the radius of 4.5 $\times$ 10$^{12}$ cm and simulations are 
carried out until just before shock waves reach the radius. 
The density of the wind component are 
smoothly connected to that of the stellar surface. 

To initiate the explosions, we inject kinetic and thermal energies artificially around the 
composition interface of the iron core and silicon-rich layer at the start of the simulations. 
For aspherical explosions, the initial radial velocities are set to be
\begin{equation}
v_r \propto \frac{r \,[\, 1 + \alpha \cos \,(2 \theta) \,]}{1+\alpha},
\label{eq:v_r}
\end{equation}
where $v_r$ is the radial velocity and $\alpha$ is the parameter which determines the degree 
of asymmetry as in \citet{nag00}. The ratio of the radial velocity on the polar axis to that on 
the equatorial axis is given by 
$v_{\rm pol}/ v_{\rm eq} = (1+\alpha )/ (1-\alpha)$, where $v_{\rm pol}$ ($v_{\rm eq}$) 
is the radial velocity on the polar (equatorial) axis at a radius. 
Thermal energy is also injected such that the ratio of the kinetic energy to the thermal energy 
is 1 locally. In the present paper, the total injected energies are fixed to be 2 $\times$ 10$^{51}$ erg, 
unless it is explicitly stated otherwise. 

\section{Models}

In this section, we will provide a description for our models. In order to clarify the 
preferable conditions for reproducing the observed high velocity of $^{56}$Ni, 
we investigate the effects of aspherical supernova explosions on matter mixing. 
Then, we consider some types of perturbations as follows. As mentioned in \S 1, 
there are several possible seeds of perturbations. Shell burning at the bottom of each 
composition layer is  one of the possible seeds \citep{arn11}. 
In particular, oxygen shell burning is promising for perturbations of a large amplitude 
(up to 8\% in density) \citep{baz98}. Oxygen shell flashes ($^{16}$O + $^{16}$O), 
which may occur when a shock wave reaches the inner part of the oxygen-rich layer, 
are also promising \citep{arn89a}. If perturbations are introduced due to shell burning, 
perturbations may be introduced in a supernova shock in multiple times. 
On the other hand,  the asphericity of the explosion itself is another candidate. 
As shown in recent theoretical studies of core-collapse supernova explosion mechanisms 
\citep[see][]{kot06}, convection in neutrino heating layers and SASI may cause significant 
anisotropy inside a shock wave. 

In this paper, we will explore mixing in aspherical explosions considering perturbations 
of both a pre-supernova and explosion origins. We will also revisit the best model for SN~1987A 
in \citet{nag98b} and \citet{nag00} as our baseline. 

\subsection{Aspherical explosions with perturbations of pre-supernova origins}

Motivated by previous study of mixing in aspherical supernova explosions 
\citep{nag98b,nag00,yam91}. We revisit RT mixing in mildly aspherical (bipolar jetlike) 
explosions. In this section, we consider scenarios in which perturbations are introduced 
by the anisotropy of the pre-supernova star due to e.g., shell burning. Note that we do not 
intend to specify the origin of perturbations. 

We explore the cases of $\alpha$ = 0, 1/3, and 3/5, which correspond to 
$v_{\rm pol}/ v_{\rm eq} = (1+\alpha )/ (1-\alpha)$ = 1, 2, and 4, respectively, 
Note that the case of $\alpha$ = 0 corresponds to a spherical explosion and we calculate 
it as a reference. As in early studies of RT mixing 
\citep{arn89a,hac90,fry91,mul91,her91,hac92,nag98b,nag00}, 
we introduce perturbations in the radial velocities. \citet{hac92} and \citet{fry91} concluded 
that if the initial amplitude of the perturbations is larger than 5\%, the resultant mixing 
lengths of RT fingers are only slightly affected by the resolution of the simulation, unless 
the resolution is too low. 
Therefore, we adopt an amplitude of 5\% for the perturbations. 
Since we consider here perturbations introduced by pre-supernova origins, 
we do not consider amplitude larger than 5\% in this section. 
Two types of perturbations are applied. One is the `sinusoidal' (monochromatic) perturbation 
whose form is $1 + \epsilon \sin(m \,\theta)$, where $\epsilon$ is the amplitude of the 
perturbation and $m$ is the integer parameter related to the wave length of the perturbations. 
The other is the `random' perturbation given by $1 + \epsilon \,(2\,{\rm rand} \,[m \,\theta/\pi ] - 1)$, 
where `rand' is random numbers as a function of $\theta$, which varies between 0 and 1. 
We take $m+1$ sample random numbers for perturbations at 
$\theta$ = 0, 1/$\pi$, 2/$\pi$, ..., $m$/$\pi$. For perturbations between the sample points, 
values of `rand' are interpolated from values of the adjacent sample points. 
We adopt $m =$ 20 ($m =$ 128) for the `sinusoidal' (`random') perturbations. 
Note that RT mixing in aspherical supernova explosions with `random' perturbations have not 
been explored in previous studies. 
Additionally, we perform the simulation of a spherical explosion without any imposed perturbation 
for reference. we find a growth of some perturbations in the simulation. 
The detail will be described in \S 4.1. 

\begin{figure}
\hspace{-3cm}
\begin{center}
\includegraphics[width=8cm,keepaspectratio,clip]{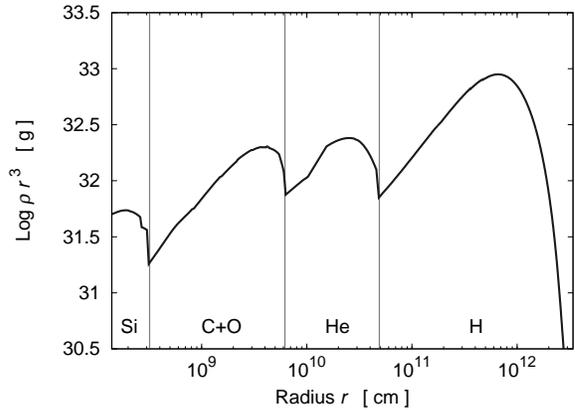}
\caption{$\rho \,r^{3}$ profile of the pre-supernova model. The composition interfaces of C+O/He 
and He/H are found at 6 $\times$ 10$^{9}$ cm and 5 $\times$ 10$^{10}$ cm, respectively. 
Regions of increasing $\rho \,r^{3}$ with increasing $r$ 
($\rho \propto$ $r^{-\omega}$ with $\omega < 3$) tend to be unstable against RT instability. }
\label{fig:rho_r3}
\end{center}
\end{figure}

Perturbations are introduced in the radial velocities inside the shock wave when it reaches 
a set of radii. For the perturbations, we employ two onset radii of 6 $\times$ 10$^{9}$ cm 
and 5 $\times$ 10$^{10}$ cm that correspond to the composition interfaces of C+O/He 
and He/H, respectively. 
Note that \citet{arn89a} considered that the jump in density at the composition interface is not significant for 
RT instability. However, it has not been clearly proved 
that fluctuations up to 5\% in e.g. density around the interface could not 
be introduced by not only observations but also multi-dimensional hydrodynamic simulations. 
Therefore, it is worth investigating the potential significance of perturbations around He/H interface. 
In similarity solutions of point explosions \citep{tay46,sed59} 
in a power-law density profile of $\rho \propto r^{-\omega}$, the radius of the shock front 
is given by $R_{\rm sh}(t) = A^{1/(5-\omega)} t^{2/(5-\omega)}$, 
where $A$ is a constant. Therefore, the velocity of the shock front is 
$v_{\rm sh}(t) \propto t^{(\omega - 3)/(5 - \omega)}$, which is rewritten as 
$v_{\rm sh}$ = $v_{\rm sh, 0}$ ($R_{\rm sh}$/$R_{\rm sh, 0}$)$^{\omega /2 - 3/2}$, 
where $R_{\rm sh, 0}$ and $v_{\rm sh, 0}$ are the radius of the shock wave and velocity 
of the shock at $t = t_{0}$, respectively. The shock wave is decelerated if $\omega < 3$, 
which produces a reverse shock, and the part of the inner region of the shock wave 
tends to be unstable against RT instabilities \citep[e.g.,][]{bet90}. 
Figure~\ref{fig:rho_r3} shows the profile of $\rho\,r^{3}$ of the pre-supernova model. 
Regions of increasing $\rho\,r^{3}$ with increasing $r$ correspond to density profiles of 
$\rho \propto r^{-\omega}$ with $\omega < 3$. The composition interfaces of C+O/He 
and He/H correspond to the radii of 6 $\times$ 10$^{9}$ cm and 5 $\times$ 10$^{10}$ cm, 
respectively. As we can see in Figure~\ref{fig:rho_r3}, a shock wave will be decelerated after 
the shock wave passes through the composition interfaces. 

In Table~\ref{table:models}, we summarize the models and the corresponding model parameters. 
The first column is the name of the model, the second is the parameter $\alpha$, 
the third is the $v_{\rm pol}/ v_{\rm eq}$ corresponding to $\alpha$, the fourth is 
the $v_{\rm up}/ v_{\rm down}$ (the definition will be described in \S 3.4), 
the fifth is the type of perturbations, the sixth is the 
amplitude of perturbations $\epsilon$, the seventh is the parameter $m$, and the eighth 
is the timing of perturbations. 
The fifth column, which is the type of perturbations, is either `random', `sinusoidal' or `clump'.  
`random' and `sinusoidal' denote that the forms of perturbations are 
$1 + \epsilon \,(2\,{\rm rand} \,[m \,\theta/\pi ] - 1)$ and $1 + \epsilon \sin(m \,\theta)$, 
respectively. `clump' will be explained in \S 3.4. The seventh column, which is the timing of 
introducing the perturbations, is either `C+O/He', `He/H', `multi', `shock' or `full'. 
`C+O/He' and `He/H' mean that the perturbations are introduced when the shock wave 
reaches the composition interfaces of C+O/He and He/H, respectively. 
`multi', `shock', and `full' will be explained in detail in \S 3.2, \S 3.4, and \S 3.5, respectively. 
The nomenclature for the names of models in the paper is as follows. 
The first character indicates whether the explosion is spherical (S) or aspherical (A), i.e., 
$\alpha =$ 0 or not.  The second character is either `P', `S', `M' or `T'. 
`P' and `S' mean `Pre-supernova' and `Shock' denoting the origins of the perturbations. 
`M' means `Multiple' whose perturbations are introduced in multiple times. `T' means `Test'. 
Models with a second characters of `S', `M', and `T' are described in later sections. 
If there are more than two models that have the same first two characters, a number is 
added to the name to distinguish the models. 
The models related to this particular section are SP1, SP2, and AP1 to AP8. 

\begin{deluxetable*}{lrrrrrrr}
\tabletypesize{\footnotesize}
\tablenum{1}
\tablecolumns{8}
\tablecaption{Models and parameters.}
\label{table:models}
\tablehead
{
\multicolumn{1}{l}{Model} &
\multicolumn{1}{r}{$\alpha$} & 
\multicolumn{1}{r}{$v_{\rm eq}/v_{\rm pol}$} &
\multicolumn{1}{r}{$v_{\rm up}/v_{\rm down}$} &
\multicolumn{1}{r}{Type of perturb.\tablenotemark{a}} &
\multicolumn{1}{r}{$\epsilon$} &
\multicolumn{1}{r}{$m$} &
\multicolumn{1}{r}{Timing of perturb.\tablenotemark{b}}
}
\startdata
SP1 & 0 & 1 & 1 & random & 5\% & 128 & C+O/He \\
SP2 & 0 & 1 & 1 & random & 5\% & 128 & He/H \\
SM & 0 & 1 & 1 & random & 5\% & 128 & multi \\
AP1 & 1/3 & 2 & 1 & random & 5\% & 128 & C+O/He \\
AP2 & 3/5 & 4 & 1 & random & 5\% & 128 & C+O/He \\
AP3& 1/3 & 2 & 1 & sinusoidal & 5\% & 20 & C+O/He \\
AP4& 3/5 & 4 & 1 & sinusoidal & 5\% & 20 & C+O/He \\
AP5 & 1/3 & 2 & 1 & random & 5\% & 128 & He/H \\
AP6 & 3/5 & 4 & 1 & random & 5\% & 128 & He/H \\
AP7 & 1/3 & 2 & 1 & sinusoidal & 5\% & 20 & He/H \\
AP8 & 3/5 & 4 & 1 & sinusoidal & 5\% & 20 & He/H \\
%
%
AT1\tablenotemark{c} & 1/3 & 2 & 1 & sinusoidal & 30\% & 20 & He/H \\
AT2\tablenotemark{c} & 1/3 & 2 & 1 & sinusoidal & 30\% & 20 & He/H \\
AS1 & 1/3 & 2 & 1 & sinusoidal & 30\% & 20 & shock \\
AS2 & 1/3 & 2 & 2 & clump & 30\% & 3 & shock \\
AS3 & 1/3 & 2 & 2 & clump & 30\% & 5 & shock \\
AS4 & 1/3 & 2 & 2 & clump & 30\% & 7 & shock \\
AS5 & 1/3 & 2 & 2 & clump & 30\% & 9 & shock \\
AS6 & 1/3 & 2 & 2 & clump & 30\% & 11 & shock \\
AS7 & 1/3 & 2 & 2 & clump & 30\% & 13 & shock \\
AS8 & 1/3 & 2 & 2 & clump & 30\% & 15 & shock \\
AM1 & 3/5 & 4 & 1 & random & 5\% & 128 & multi \\
AM2\tablenotemark{d,e} & 1/3 & 2 & 2 & clump/random & 30\,/\,5\% & 15\,/\,128 & full \\
AM3\tablenotemark{e} & 1/3 & 2 & 2 & random & 5\% & 128 & multi
\enddata
\scriptsize{
\tablenotetext{a}{Types of perturbations. `random', `sinusoidal', and `clump' denote shapes 
of perturbations, $1 + (2\,\epsilon \,{\rm rand} \,[m \,\theta/\pi ] - 1)$, $1+\epsilon \sin(m\theta)$, 
and $1 + \sum^{4}_{n=1} \frac{\epsilon}{2^{(n-1)}} \sin(m \, n \, \theta )$ 
(Equation~(\ref{eq:clump})), respectively.}
\tablenotetext{b}{Timings that perturbations are introduced. `C+O/He', `He/H', and `multi' denote 
that perturbations are introduced when shock waves reach at the composition interfaces of 
C+O/He, He/H, and both of C+O/He and He/H, respectively. `shock' denotes that perturbations 
are introduced in the initial radial velocities. `full' indicates perturbations are fully introduced 
(see the note $^{\rm d}$).} 
\tablenotetext{c}{Models AT1 and AT2 are test models of which setups of simulations are similar 
to that of model A1 in \citet{nag00}. For model AT1, gravity is turned off, the inner boundary 
condition is `reflection', and energy of 1 $\times$ 10$^{51}$ erg is initially injected. 
Model AT2 has same model parameters but the treatments of gravity, inner boundary condition, 
and injected energy are same as other models in this paper.}
\tablenotetext{d}{Perturbations are imposed fully multiply, i.e., `clump' perturbations of 
30\% amplitude are introduced in initial radial velocities, and `random' perturbations of 
5\% amplitude are introduced when the shock wave reaches the composition interfaces of
C+O/He and He/H.}
\tablenotetext{e}{Energy of 2.5 $\times$ 10$^{51}$ erg is initially injected for the initiation 
of the explosion.}
}
\end{deluxetable*}

\subsection{Aspherical explosions with multiply introduced perturbations of pre-supernova origins}

In this section, we will explain models in which perturbations are introduced in multiple times. 
If the perturbations are introduced due to shell burning in the pre-collapse star, those could 
be multiply introduced. However, in the previous studies of RT mixing in supernovae, such 
situations have not been investigated. Therefore, we simply mimic perturbations multiply 
introduced in the pre-supernova star by introducing the perturbations in the radial 
velocities at different two times when the shock wave reaches the composition 
interfaces of C+O/He and He/H, respectively. Namely, the first perturbations 
are introduced when the shock wave reaches the composition interface of C+O/He and the second 
perturbations are introduced when the shock wave reaches the composition interface of He/H. 
We investigate models of both spherical and mildly 
aspherical explosions SM and AM1, respectively. The second character of the names of models 
in this section is `M', which means `Multiple' as explained above. In the two models, `random' perturbations are employed. 
In table~\ref{table:models}, the eighth column is represented by `multi' for the models in this section. 

\subsection{Revisiting the best model in Nagataki et al.}

\citet{nag98b} and \citet{nag00} have investigated matter mixing in aspherical explosions 
using a pre-supernova mode for SN~1987A, and a mildly aspherical model of 
$v_{\rm pol}/v_{\rm eq} = $ 2 with sinusoidal perturbations of a large amplitude (30\%) 
(model A1 in Nagataki et al.) have reproduced the high velocity of $^{56}$Ni 
(up to $\sim$ 3,000 km s$^{-1}$). The pre-supernova model used in Nagataki et al. is 
the same as that in the present paper. Besides, the way of initiating the explosions is also 
basically same. However, the resolution of their simulations are rather low compared to 
that of recent studies of matter mixing in supernova explosions \citep[e.g.,][]{kif06} and 
the authors have not taken into account gravity, i.e., effects of fallback. 
Therefore, we revisit the best model in Nagataki et al. including the effects of gravity. 
We test two models AT1 and AT2, where the second character of the names 
`T' means `Test' as mentioned before. Model AT1 is the model whose setup of the 
simulation is basically the same as that of model A1 in Nagataki et al. 
In model AT1, effects of gravity is turned off, the total injected energy is set to be 10$^{51}$ erg, 
the boundary condition of the radial inner edge is the `reflection' boundary condition, 
$\alpha =$ 1/3 ($v_{\rm pol}/v_{\rm eq} = $ 2), $\epsilon =$ 30\%, and the form of 
perturbations is `sinusoidal' with $m =$ 20. Model AT2 is the counterpart of model AT1 
whose model parameters are also the same as those of AT1 except that the
effects of gravity are turned on and the boundary condition of the radial inner edge 
is switched to the `diode' boundary condition at the later phase as in the other models 
in the present paper. In model AT2, the total injected energy is set to be 2 $\times$ 10$^{51}$ erg 
because we have included gravitational potentials in this model. Note that the resultant 
explosion energy will be smaller than that of model AT1, if we inject the same 10$^{51}$ erg 
as in model AT1. In both models, perturbations are introduced when the shock wave 
reaches the composition interface of He/H as in Nagataki et al. However, as the authors 
mentioned in their paper, such large perturbations with $\epsilon =$ 30\% should be 
introduced in the supernova explosions itself. Therefore, we investigate the model AS1 
whose model parameters and the setup of the simulation 
are the same as those of AT2 except for the timing of introducing the perturbations. 
In model AS1, the perturbations are introduced in the initial radial velocities as in the 
models described in the next section. 

\subsection{Aspherical explosions with clumpy structures}

As mentioned in \S 1, theoretically, multi-dimensional effects are essential for a 
successful core-collapse supernova explosion. Recent multi-dimensional radiation 
hydrodynamic simulations of core-collapse supernova explosions have revealed that 
in the context of neutrino heating mechanisms, convection and SASI cause large anisotropy 
inside the standing shock and low-order unstable modes ($l$ = 1, 2) can grow 
dominantly \citep[e.g.,][]{mar09,suw10,nor10,tak12}. Some models of neutrino-driven 
explosions aided by SASI have demonstrated that explosions may become stronger in 
either the north or south direction than those in the other directions across the 
equatorial plane \citep[e.g.,][]{mar09,suw10}. Such asymmetry in explosions have thought 
to be the one of origins of neutron star kicks and proper motions of young pulsars 
\citep{sch06,won10}. For example, we can see a globally anisotropic supernova shock 
wave whose morphology looks very clumpy (see e.g., Figure~1 in \citet{ham10}). 
As mentioned in \S 1, \citet{kif06} and \citet{gaw10} have successfully reproduced high 
velocity clumps of $^{56}$Ni in some models with neutrino-driven explosions. 
The authors have explained that the globally anisotropic explosion and the relatively 
large explosion energy (2 $\times$ 10$^{51}$ erg) result in high velocity clumps of 
metals and strong RM instabilities at the composition interface of He/H. 
Such high velocity clumps can penetrate the dense helium core before the formation 
of a strong reverse shock. Strong RM instabilities at the interface of He/H cause a global 
anisotropy of the inner ejecta at late phases. However, their successful models remain small 
in number and the explosion energies involved are relatively large. Therefore, the conditions 
for reproducing the observed high velocity of $^{56}$Ni are still not fully understood. 

We explore matter mixing in such globally anisotropic explosions parametrically by mimicking 
the morphology of the explosion. We can see radially averaged physical values as a 
function of $\theta$ for an anisotropic explosion e.g., in Figure~11 in \citet{gaw10}. 
The distribution of radial velocity is relatively smooth but the distributions of density and 
velocity exhibit smaller-scale clumpy structures. 

\begin{figure}
\hspace{-1.5cm}
\begin{center}
\includegraphics[width=8cm,keepaspectratio,clip]{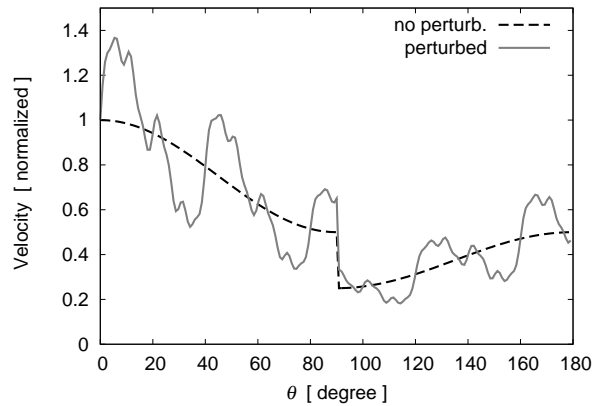}
\caption{Distribution of initial radial velocity at a radius inside the shock wave as a 
function of $\theta$ for model AS5. Radial velocities with perturbations given by 
Equation (\ref{eq:clump}) (solid line) and those with no perturbation (dashed line) 
are shown. Values of the velocities are arbitrarily normalized. }
\label{fig:vel_clump}
\end{center}
\end{figure}

We mimic such globally anisotropic explosions as follows. 
First, we consider mildly aspherical explosion with $v_{\rm eq}/v_{\rm pol}$ = 2 ($\alpha$  = 1/3). 
Second, perturbations of a large amplitude (30\%) with several smaller-scales are introduced 
in the initial radial velocities as 
\begin{equation}
1 + \sum^{4}_{n=1} \frac{\epsilon}{2^{(n-1)}} \sin(m \, n \, \theta ),
\label{eq:clump}
\end{equation}
where $\epsilon$ is the amplitude and $m$ is the integer parameter. 
We simply adopt the superposition of sinusoidal functions with different wave lengths and 
assume that the larger (smaller) the wavelength of the perturbations, the larger (smaller) 
the amplitude is. Third, we impose asymmetry across the equatorial plane by changing the 
normalizations of $v_{\rm r}$ across the equatorial plane as $v_{\rm up}/v_{\rm down} =$ 2, 
where $v_{\rm up}$ and $v_{\rm down}$ are the initial radial velocities at a radius inside 
the shock before imposing above perturbations (i.e., Equation (\ref{eq:clump})) at 
$\theta =$ 0$^{\circ}$ and $\theta =$ 180$^{\circ}$, respectively. 
The values of $v_{\rm up}/v_{\rm down}$ are shown in the fourth column of 
Table \ref{table:models}. We also test models having different base clump sizes 
(models AS2 to AS8) by changing the parameter $m$ ($m =$ 3 -- 15). 
Figure~\ref{fig:vel_clump} shows the distribution of the initial radial velocities 
at a radius inside the shock as a function of $\theta$ for model AS5. 
The second character of the names of models `S' means `Shock', which means 
that perturbations are imposed in the initial radial velocities. In Table \ref{table:models}, 
the eighth column is represented by `shock' for the models described in this section. 

\subsection{Aspherical explosions with clumpy structures and multiply introduced 
perturbations of pre-supernova origins}

Finally, we consider aspherical explosions with clumpy structures and multiply introduced 
perturbations of pre-supernova origins, i.e. multiple perturbations in a complete sense, 
which can be thought of as the combination of \S 3.2 and \S 3.4. For the perturbations 
introduced in the initial radial velocities, we adopt the perturbations given by 
Equation~(\ref{eq:clump}) ($\epsilon =$ 30\% and $m =$ 15). 
For the perturbations of pre-supernova origins, `random' perturbations 
($\epsilon =$ 5\% and $m =$ 128) are employed. We consider a globally aspherical explosion 
given by $v_{\rm pol}/v_{\rm eq} =$ 2 ($\alpha =$ 1/3) and $v_{\rm up}/v_{\rm down} =$ 2 
as in models in \S 3.4. We refer to the model as AM2. The model parameters are 
listed in Table~\ref{table:models}. The eighth column, the timing of the perturbations, 
is denoted by `full'. To see the impact of initial clumpy structures on 
the mixing, we add the model AM3 that have the same model parameters 
but with no perturbation in the initial radial velocities as a reference. Note that 
in the models in this sections AM2 and AM3, an energy of 2.5 $\times$ 10$^{51}$ erg 
is injected to initiate the explosions. 

\section{Results}

\subsection{Spherical explosions with perturbations of pre-supernova origins}

\begin{figure*}[htbp]
\begin{tabular}{ccc}
\hspace{-0.5cm}
\begin{minipage}{0.33\hsize}
\begin{center}
\includegraphics[width=5.5cm,keepaspectratio,clip]{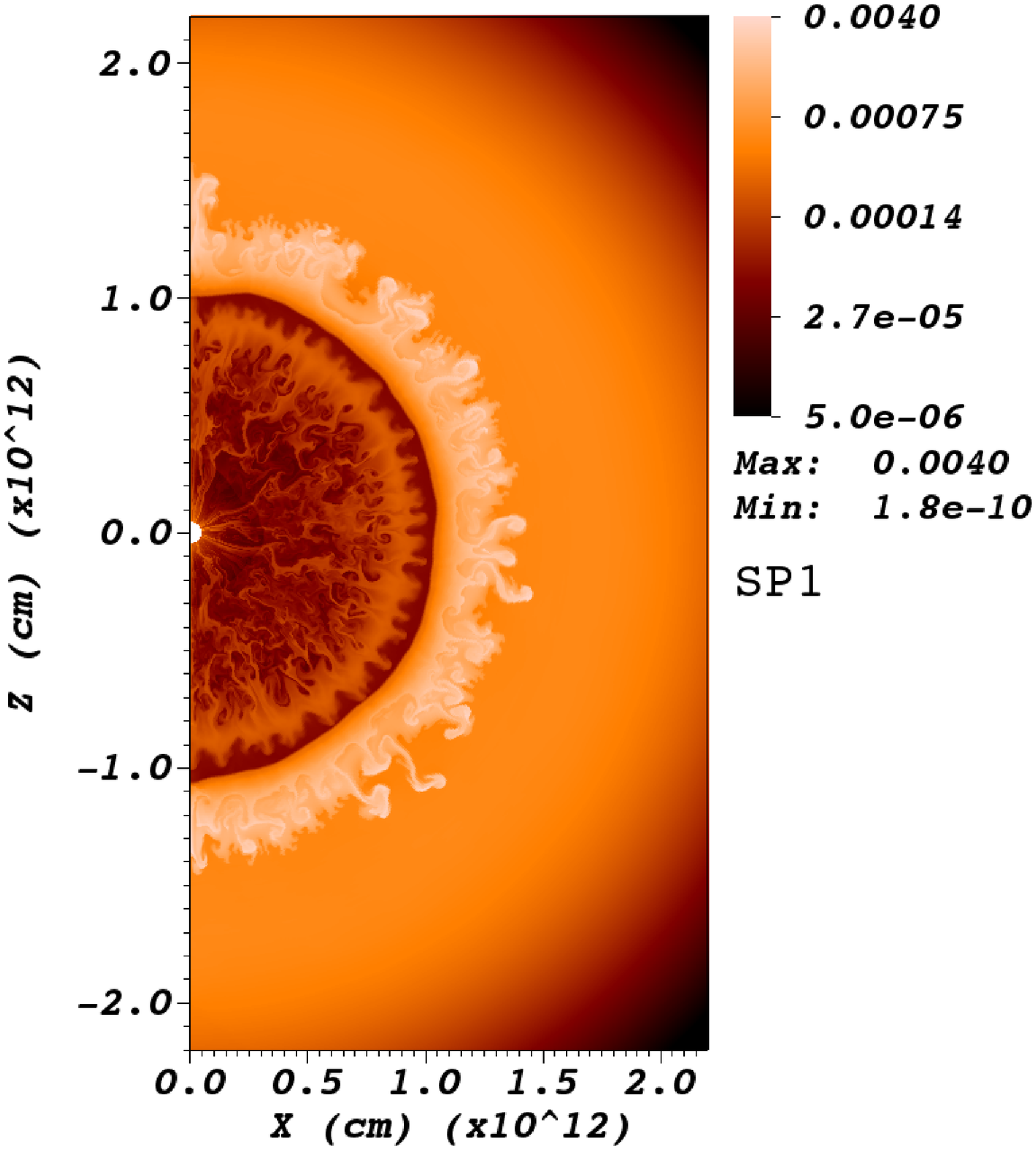}
\end{center}
\end{minipage}
\begin{minipage}{0.33\hsize}
\begin{center}
\includegraphics[width=5.5cm,keepaspectratio,clip]{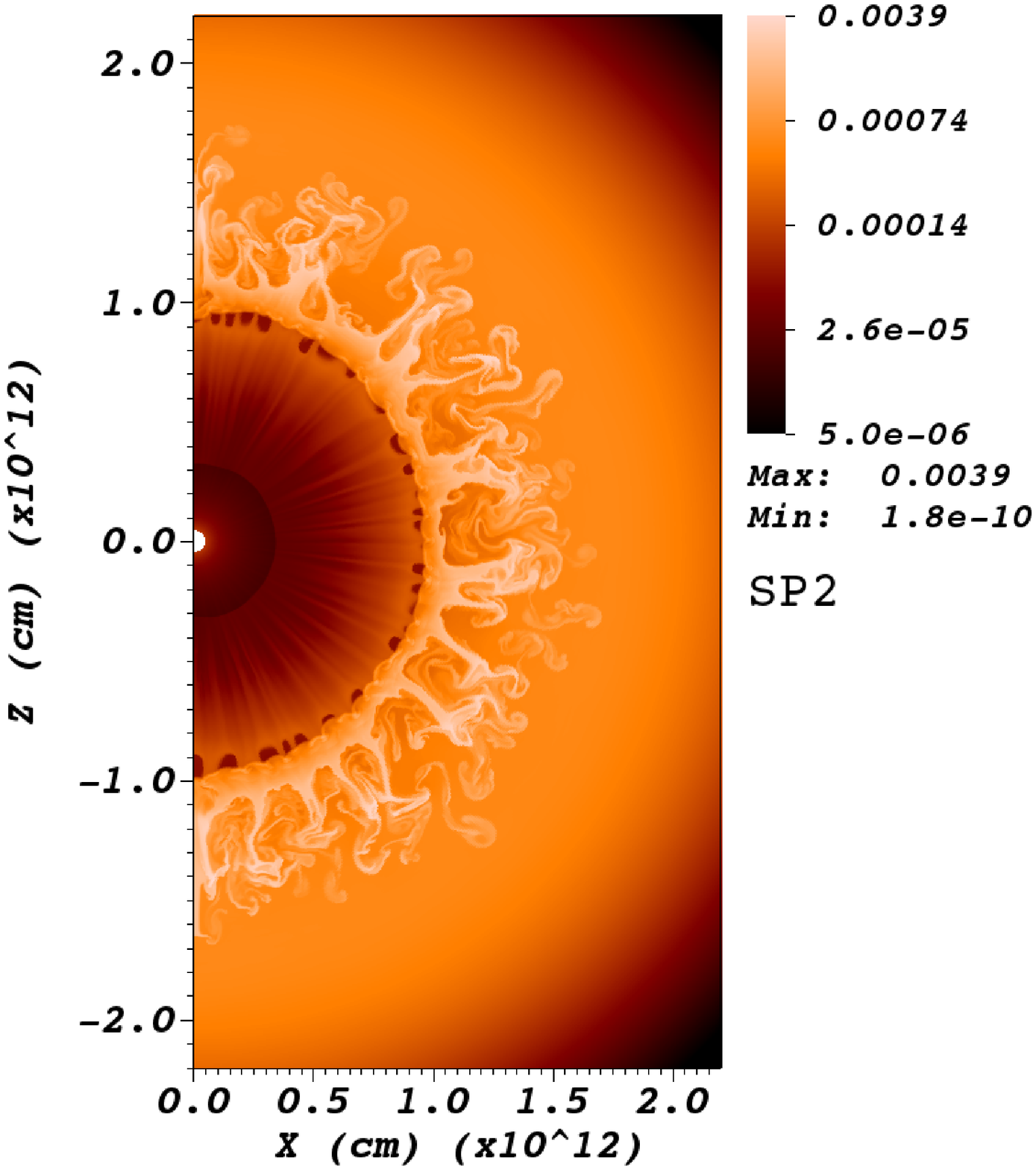}
\end{center}
\end{minipage}
\begin{minipage}{0.33\hsize}
\begin{center}
\includegraphics[width=5.5cm,keepaspectratio,clip]{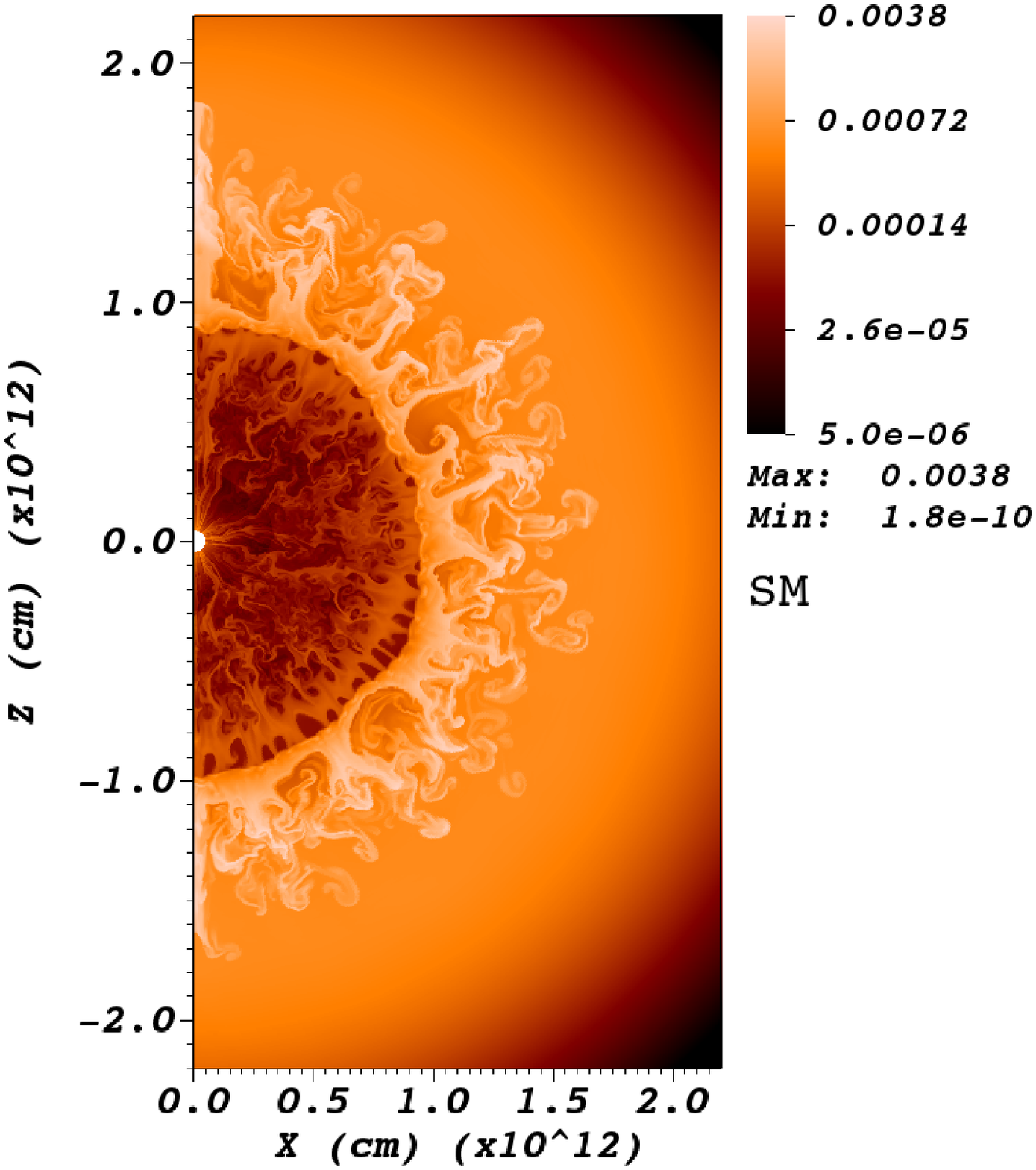}
\end{center}
\end{minipage}
\end{tabular}
\caption{Density distributions in the $X$--$Z$ plane at the ends of simulation time for 
models SP1 (left), SP2 (middle), and SM (right), which correspond to the time 
of 5986 s, 6006 s, and 5958 s, respectively. The unit of the values in the color bars 
is g cm$^{-3}$ and the values in the color bars are logarithmically scaled.}
\label{fig:dens_SPM}
\end{figure*}

First, we will show the results from models of spherical explosions with perturbations 
of pre-supernova origins, i.e., models SP1, SP2, and SM. The density distributions in 
the $X$--$Z$ plane ($X = r\sin \theta$ and $Z = r\cos \theta$) at the ends of simulation 
time for models SP1, SP2, and SM are shown in Figure~\ref{fig:dens_SPM}. 
We stop the calculation when the forward shock reaches close to the radius 
of 4.5 $\times$ 10$^{12}$ cm after the shock breakout. Models SP1 and SP2 are those 
in which random perturbations are introduced when the shock waves reach 
the composition interfaces of C+O/He and He/H, respectively. We can see the prominent 
RT fingers in both models above the radius of 1 $\times $10$^{12}$ cm. However, the lengths 
of the RT fingers are different between the two models. The lengths of RT fingers 
(hereafter the mixing lengths) in model SP1 is approximately 0.3 $\times$ 10$^{12}$ cm. 
On the contrary, the mixing length of model SP2 is roughly 0.6 $\times$ 10$^{12}$ cm. 
We find that in model SP1, perturbations grow around the composition interface 
of C+O/He due to RT instabilities but the fluctuations do not grow much after the 
shock wave has reached the composition interface of He/H. In model SP2, perturbations 
grow significantly around the composition interface of He/H. 
The morphology of RT fingers are also different between the two models. 
In model SP2, RT fingers are clearly distinguished. 
In model SP1, we can see prominent complex structures in the inner regions 
compared to model SP2. From above, the growth of RT instabilities around the 
composition interface of He/H is larger than that around the interface of C+O/He 
in our models. On the other hand, mixing of the inner regions is larger in 
model SP1 than that in model SP2. 

\begin{figure}
\hspace{-3cm}
\begin{center}
\includegraphics[width=8cm,keepaspectratio,clip]{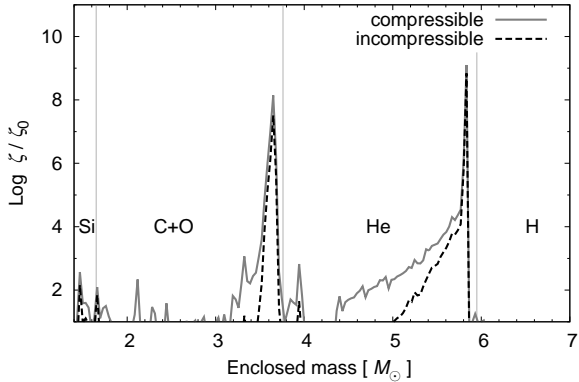}
\caption{Growth factors $\zeta/ \zeta_0$ of an initial seed perturbation as a function of 
enclosed mass at the time of 5925 s. The total injected energy is 2 $\times$ 10$^{51}$ erg 
as in other models. Two cases of growth factors are depicted. One is estimated by the 
growth rate for the incompressible fluid (solid line) and the other is estimated by that 
for the compressible fluid (dashed line). Overall, the growth factor for the compressible 
fluid is greater than that for the incompressible fluid. Growth factors are outstanding 
around the composition interfaces of C+O/He and He/H. Growth factor around the 
interface of He/H is about one order-of-magnitude greater than that around 
the interface of C+O/He.}
\label{fig:growth}
\end{center}
\end{figure}

To find the cause of the differences seen between models SP1 and SP2, 
we perform a one-dimensional simulation of a spherical explosion with no perturbation. 
The total injected energy (2 $\times$ 10$^{51}$ erg) and setups are the same as in 
other spherical models but now with no perturbation introduced. 
We estimate the growth factors of an initial seed perturbation using two growth 
rates as follows. One is the growth rate $\sigma$ for the incompressible fluid given by
\begin{equation}
\sigma = \sqrt{-\frac{P}{\rho} \mathscr{P} \,\mathscr{R}},
\label{eq:growth_incomp}
\end{equation}
where $\mathscr{P} = \partial \ln P / \partial \,r$ and $\mathscr{R} 
= \partial \ln \rho / \partial \,r$. 
The other is the growth rate for the compressible fluid given by
\begin{equation}
\sigma = \frac{c_s}{\gamma} \sqrt{{\mathscr P}^2 - \gamma \mathscr{P} \,\mathscr{R}},
\label{eq:growth_comp}
\end{equation}
where $c_s$ is the sound speed and $\gamma$ is the adiabatic index.
The growth factor of an initial seed perturbation $\zeta/ \zeta_0$ is given by
\begin{equation}
\frac{\zeta}{\zeta_0} = \exp\,\left(\int_0^{t} {\rm Re}\,[\,\sigma \,]\, \mathrm{d} t^{\prime} \right),
\end{equation}
where $\zeta_0$ is the amplitude of the initial perturbation and $\zeta$ is the amplitude 
at the time of $t$ (see e.g., \citet{mul91}). The growth factors just after the shock breakout 
are shown in Figure~\ref{fig:growth}. Overall, the growth factor for the compressible fluid 
is greater than that for the incompressible fluid. The growth factors are prominent around 
the composition interfaces of C+O/He and He/H. The growth factor around the interface 
of He/H is about one order-of-magnitude larger than that around the interface of C+O/He, 
which indicates that the growth of RT instabilities around the interface of He/H may be larger 
than that around the interface of C+O/He. We find that in model SP1, after the shock wave 
has passed through the interface, RT instabilities grow only around the interface of C+O/He 
and the forward shock propagates by roughly keeping a spherical symmetry. Therefore, 
in model SP1, when the shock wave reaches the interface of He/H, regions around the 
interface of He/H remain almost unperturbed and RT instabilities around 
the interface of He/H cannot grow well. While in model SP2, after the shock wave reaches 
the interface of He/H, RT instabilities start to grow. From the growth factors estimated above, 
the growth of RT instabilities around the interface of He/H may be larger than that around 
the interface of C+O/He, which is consistent with the results that the mixing lengths 
in model SP2 are larger than those in model SP1. 

In model SM, perturbations are introduced at different two times when the shock wave reaches 
the composition interfaces of C+O/H and He/H, respectively. Model SM has the features of both 
SP1 and SP2 (the right panel of Figure~\ref{fig:dens_SPM}), i.e., the strong mixing 
of the inner regions and the prominent extension of RT fingers. The mixing length 
of model SM is nearly comparable to that of SP2 although more complex structures 
of RT fingers are observed. The structures of the inner regions are similar to that 
in model SP1. Note that somewhat more extended RT fingers are found around 
the polar region ($\theta \sim$ 0$^{\circ}$) compared with those in other directions in model SP1 and SM, which 
may be responsible for discretization errors around the polar axis 
but the deviation from the basic spherical symmetry is not large. 

\begin{figure*}[htbp]
\begin{tabular}{cc}
\hspace{-0.5cm}
\begin{minipage}{0.5\hsize}
\begin{center}
\includegraphics[width=6cm,keepaspectratio,clip]{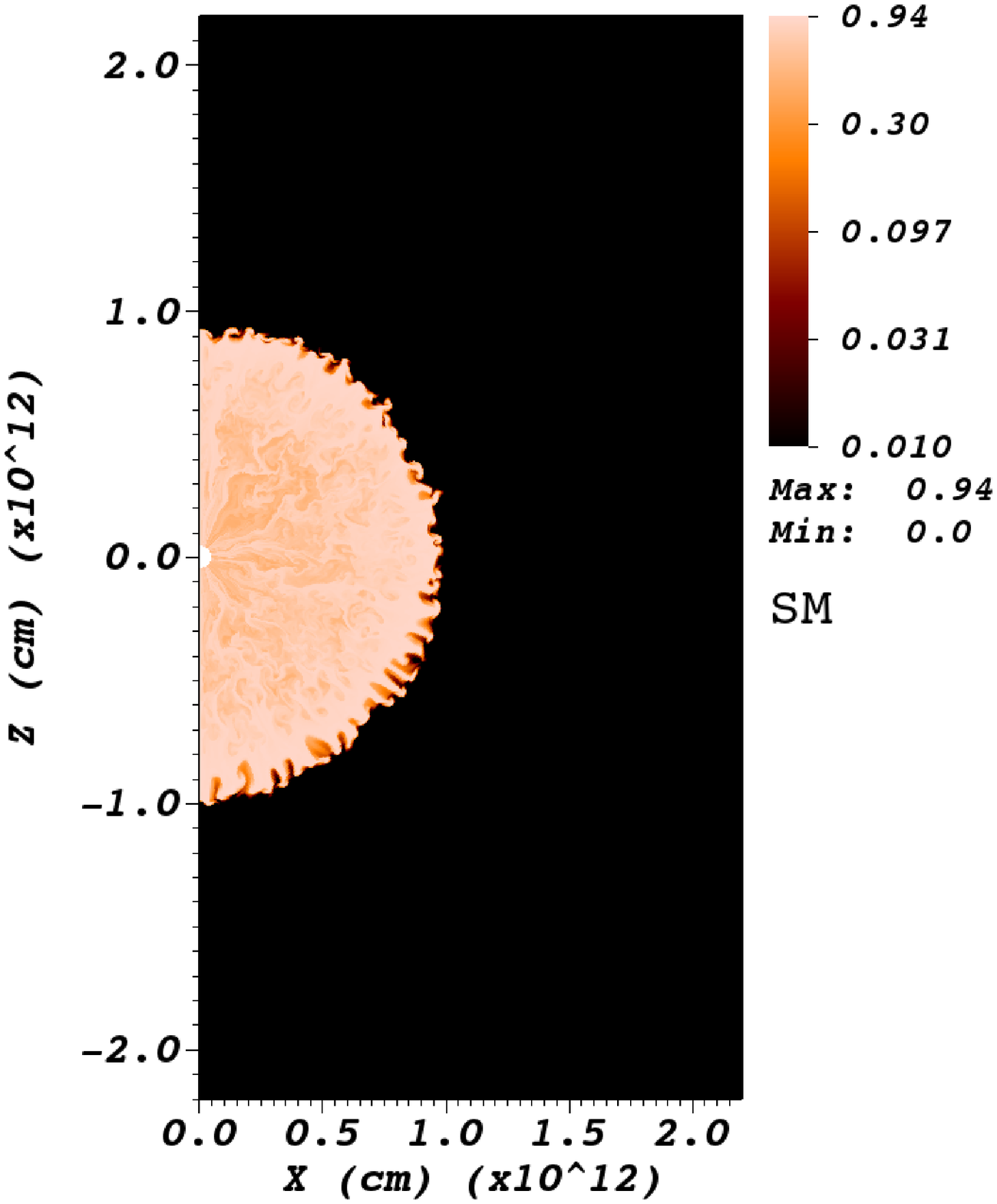}
\end{center}
\end{minipage}
\begin{minipage}{0.5\hsize}
\begin{center}
\includegraphics[width=6cm,keepaspectratio,clip]{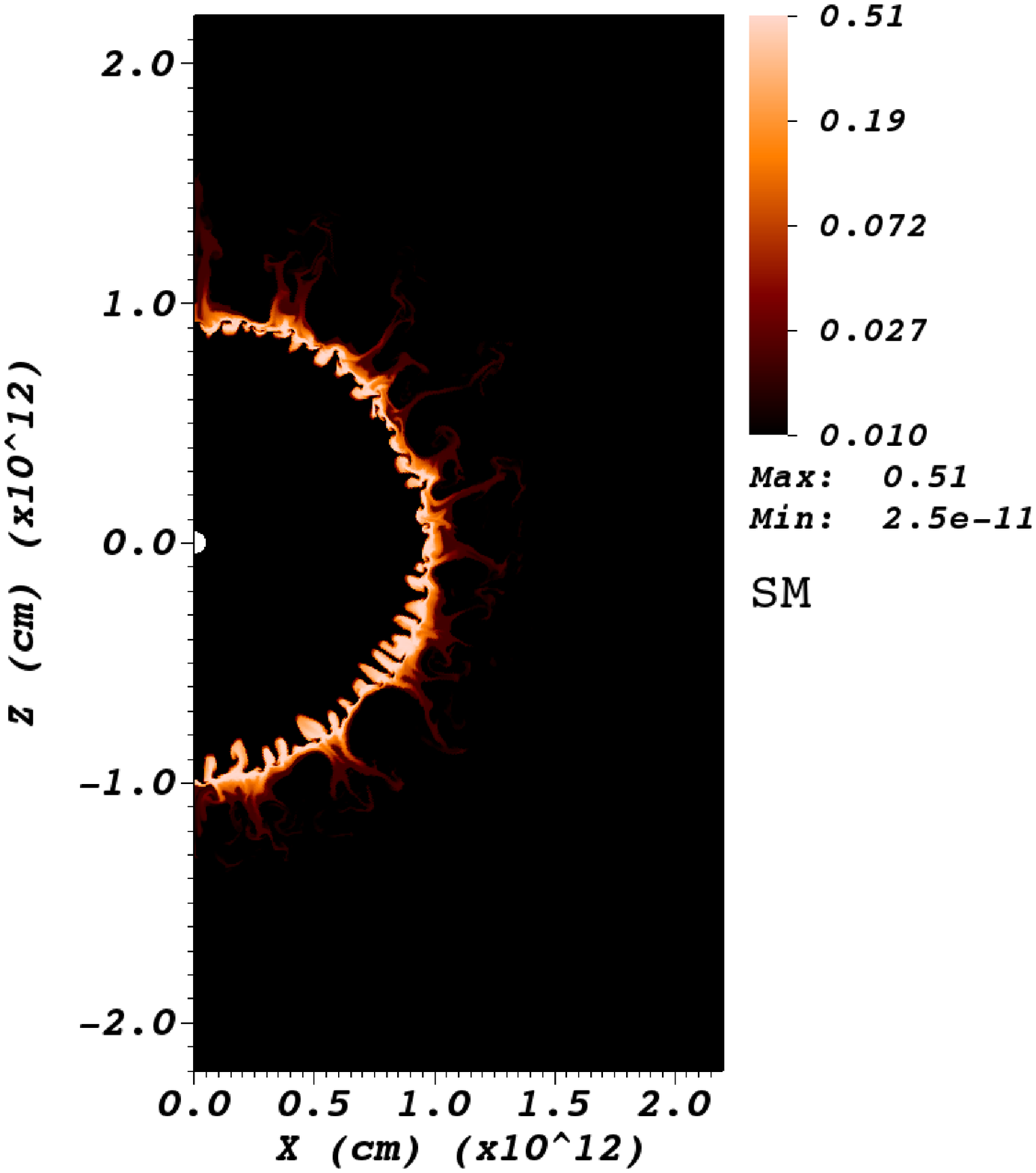}
\end{center}
\end{minipage}
\\
\hspace{-0.5cm}
\begin{minipage}{0.5\hsize}
\begin{center}
\includegraphics[width=6cm,keepaspectratio,clip]{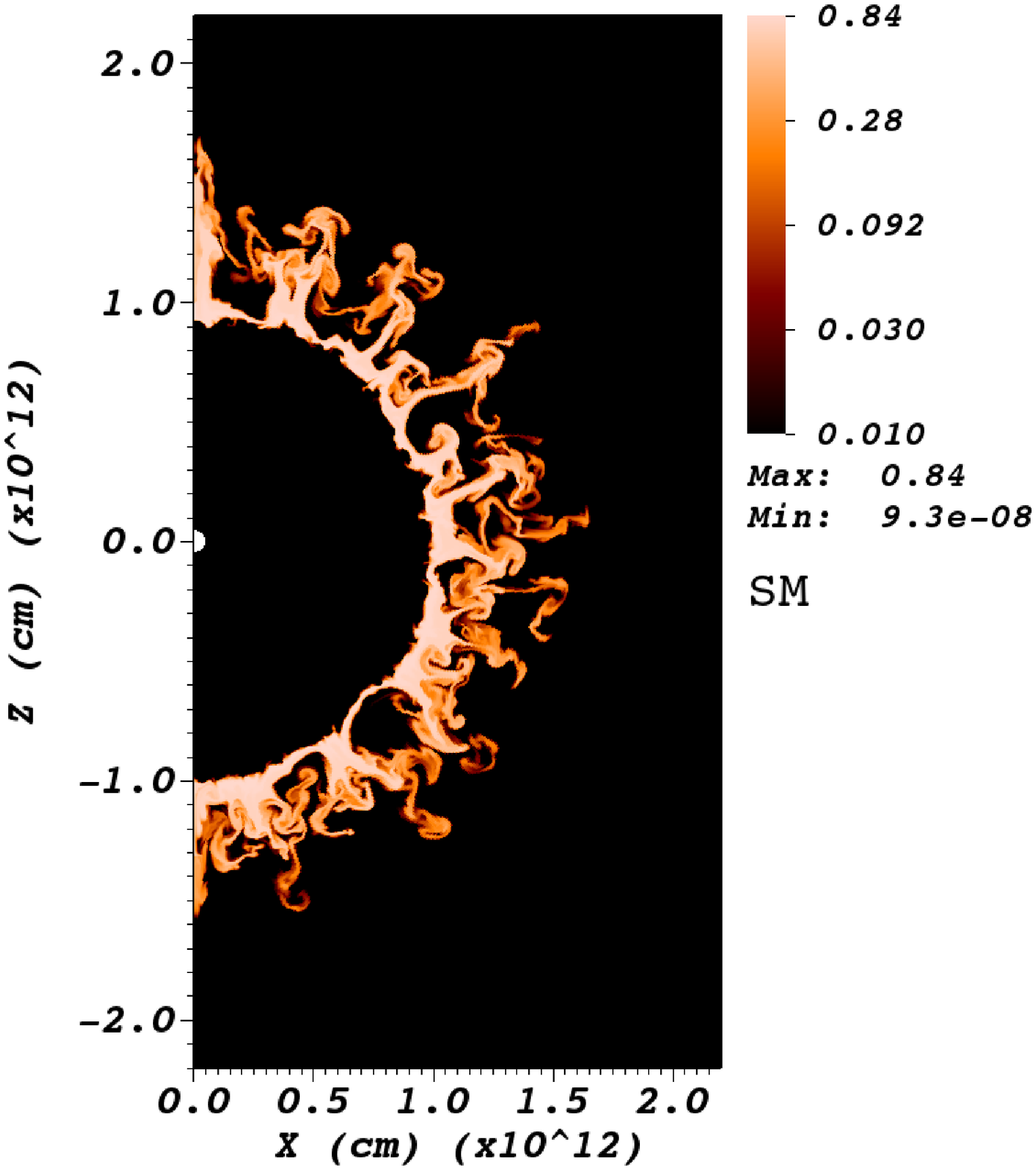}
\end{center}
\end{minipage}
\begin{minipage}{0.5\hsize}
\begin{center}
\includegraphics[width=6cm,keepaspectratio,clip]{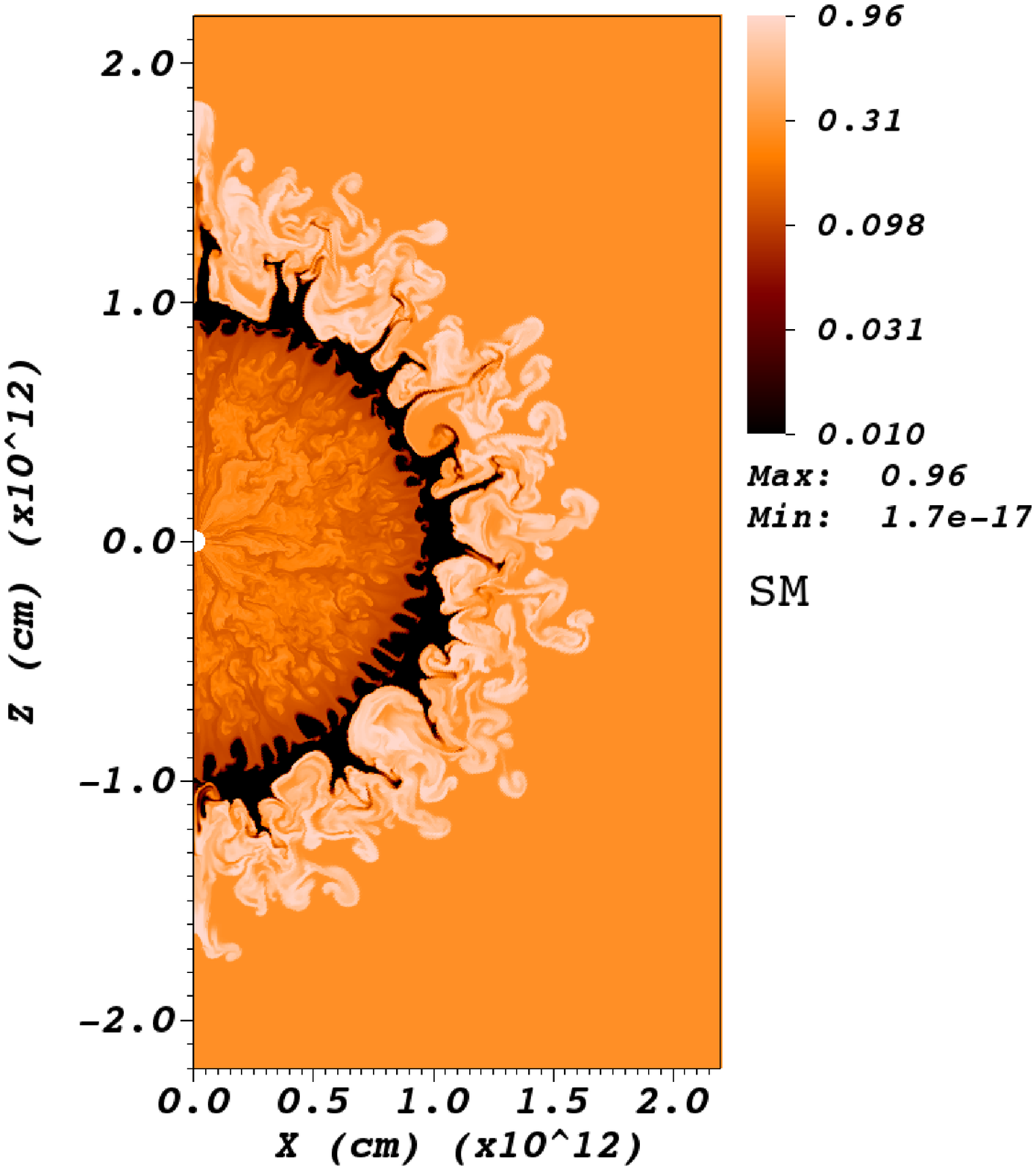}
\end{center}
\end{minipage}
\end{tabular}
\caption{Distributions of mass fractions of elements $^{56}$Ni (top left), $^{28}$Si (top right), 
$^{16}$O (bottom left), and $^{4}$He (bottom right) at the end of simulation time 
in the $X$--$Z$ plane for model SM, which corresponds to the time of 5958 s. 
Values in the color bars are logarithmically scaled and the minimum value 
in the color bars is set to be 1 $\times$ 10$^{-2}$.}
\label{fig:element_SM}
\end{figure*}

The distributions of mass fractions for the elements $^{56}$Ni, $^{28}$Si, $^{16}$O, 
and $^{4}$He at the end of simulation time for model SM are shown 
in Figure~\ref{fig:element_SM}. $^{56}$Ni is concentrated inside the dense helium 
shell around the radius of 1 $\times$ 10$^{12}$ cm. $^{28}$Si encompasses the 
inner $^{56}$Ni and a small fraction of $^{28}$Si is conveyed outward along the 
RT fingers. $^{16}$O is prominent at the bottom of the helium shell and inside 
the RT fingers. $^{4}$He is found to be the most abundant around the RT fingers. 
$^4$He are also seen inside the helium shell, which is responsible for the explosive 
nucleosynthesis. 

\begin{figure*}[htbp]
\includegraphics[width=18cm,keepaspectratio,clip]{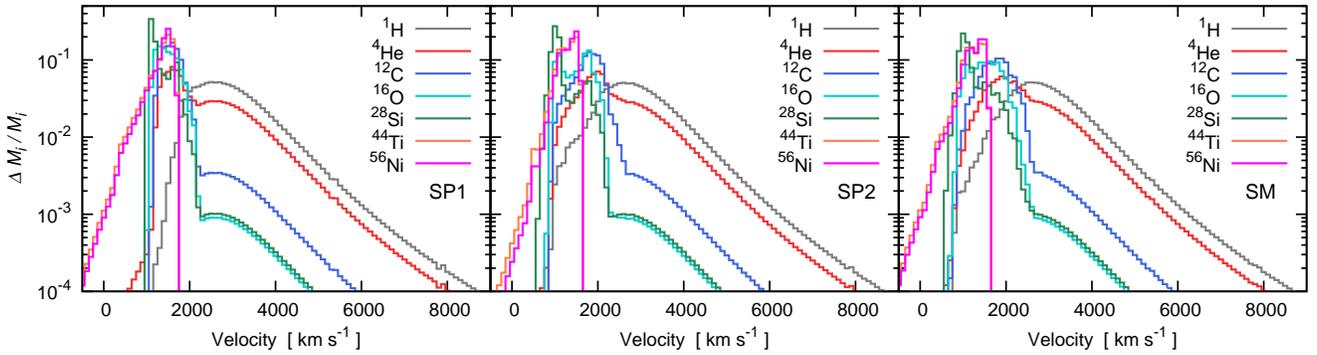}
\caption{Mass distributions of elements, $^1$H, $^{4}$He, $^{12
}$C, $^{16}$O, $^{28}$Si, $^{44}$Ti, and $^{56}$Ni, as a function of radial velocity 
at the ends of simulation time for models SP1 (left), SP2 (middle), and SM (right), 
which correspond to the time of 5986 s, 6006 s, and 5958 s, respectively. 
$\Delta M_i$ is the mass of the element with index $i$ in the velocity range of 
$v \sim v+\Delta v$. $M_i$ is the total mass of the element with index $i$. 
For the binning of radial velocity, $\Delta v =$ 100 km s$^{-1}$ is adopted.} 
\label{fig:vel_SPM}
\end{figure*}

The mass distributions of elements $^1$H, $^{4}$He, $^{12}$C, $^{16}$O, $^{28}$Si, 
$^{44}$Ti, and $^{56}$Ni as a function of radial velocity at the ends of simulation time 
for models SP1, SP2, and SM are shown in Figure~\ref{fig:vel_SPM}. In model SP2, 
we can see slight enhancements of the high velocity component of $^{12}$C, $^{16}$O, 
and $^{28}$Si around 2,000 km s$^{-1}$ and a low velocity tail of $^1$H compared 
with those of model SP1. On the other hand, in model SP1, enhancements 
of low-velocity tails of the inner most metals $^{56}$Ni and $^{44}$Ti are seen. 
RT instabilities grown around the composition interface of He/H mix up elements of
$^1$H, $^{4}$He, $^{12}$C, $^{16}$O, and $^{28}$Si more efficiently than that 
around the interfaces of C+O/He. While, RT instabilities developed around 
the interface of C+O/He convey the innermost metals farther outward than 
that around the interface of He/H. Same as Figure~\ref{fig:dens_SPM}, model SM 
has the features of both SP1 and SP2, i.e., enhancements of high velocity components 
of $^{12}$C, $^{16}$O, and $^{28}$Si and a low-velocity tail of $^1$H compared 
to SP1, as well as enhancements of low-velocity tails of $^{56}$Ni and $^{44}$Ti. 
In all three models, the distributions of $^{44}$Ti are quite similar to those of $^{56}$Ni. 
The obtained maximum radial velocity of $^{56}$Ni is approximately 1,600 km s$^{-1}$ 
and the minimum radial velocity of $^{1}$H is 800 km s$^{-1}$ among the three models, 
where we define the maximum (minimum) radial velocity as that 
among the bins with $\Delta M_i/M_i > 1 \times$ 10$^{-3}$. 

For reference, 
we also perform a simulation of a spherical explosion without any imposed perturbation. 
The setup and the initial conditions are same as in models SP1, SP2, and SM but for no imposed perturbation.
We recognize a growth of some perturbations in this reference model. At the end of the simulation time, 
radial folds above the reverse shock and slight rippled structures around the forward 
shock in density are seen. We find with a touch of surprise that the maximum velocity of $^{56}$Ni (1,700 km s$^{-1}$) 
is larger than those of any other spherical explosion models in this section, i.e., SP1, SP2, and SM. 
However, the growth of RT instability 
around the composition interface of He/H are rather small and the mixing of $^1$H into 
inner cores is negligible. The obtained minimum velocity of $^1$H is 1,700 km s$^{-1}$ and which is 
the largest among spherical explosion models.  
The perturbations may be introduced by grids and/or remappings and the wavelengths of the perturbations 
could be smaller than those of the imposed perturbations in the models in the paper. 
Since the growth of  the perturbation with a smaller wavelength is faster than that of the 
perturbation with a larger wavelength, the introduced perturbations can grow even in a small 
dynamical time scale in a relatively early phase. 

In Table~\ref{table:results}, we summarize the results of our models. 
The first column is the explosion energy, $E_{\rm exp}$, at the end of simulation time, 
the second column is the obtained minimum radial velocity of hydrogen 
$v_{r, {\rm min}}$\,($^1$H)  and the third column is the obtained maximum 
radial velocity of $^{56}$Ni $v_{r, {\rm max}}$\,($^{56}$Ni). The explosion energy 
$E_{\rm exp}$ is estimated as
\begin{equation}
E_{\rm exp} =2\pi \int_{r_1}^{r_2} \int_{0}^{\pi} \left( \frac{1}{2}\,\rho \,\vec{v}^2 
+ \rho\, E +\rho \,\Phi \right) r^2 \sin \theta \, \mathrm{d} r \,\mathrm{d} \theta,
\label{eq:exp}
\end{equation}
where $r_1$ ($r_2$) is the radius of the inner (outer) edge of the computational domain, 
$\Phi$ is the gravitational potential and the integrand is summed up only 
when it is positive. In models SP1, SP2, and SM, the obtained explosion energies are 
approximately 1.4 $\times$ 10$^{51}$ erg at the ends of simulation time. 
The maximum velocities of $^{56}$Ni are approximately 1,500 km s$^{-1}$, 
which is much smaller than the observed values of SN~1987A ($\sim$ 4,000 km s$^{-1}$) 
as mentioned above. In models SP2 and SM, the minimum velocity of $^1$H 
is 800 km s$^{-1}$, which is consistent with the theoretically inferred values \citep{shi90,koz98}. 
Therefore, inward mixing of hydrogen may be caused by the RT instability 
around not the interface of C+O/He but the interface of He/H. 

\begin{deluxetable}{lcrr}
\tabletypesize{\footnotesize}
\tablewidth{0pt}
\tablenum{2}
\tablecolumns{4}
\tablecaption{Results of models.}
\label{table:results}
\tablehead
{
\multicolumn{1}{l}{Model} & 
\multicolumn{1}{c}{$E_{\rm exp}$\tablenotemark{a}} &
\multicolumn{1}{c}{$v_{r, {\rm min}}$\tablenotemark{b}($^{1}$H)} &
\multicolumn{1}{c}{$v_{r, {\rm max}}$\tablenotemark{c}($^{56}$Ni)} 
\vspace{0.2cm} 
\\
& 
\multicolumn{1}{c}{(erg)} & 
\multicolumn{1}{c}{(km s$^{-1}$)} & 
\multicolumn{1}{c}{(km s$^{-1}$)}
}	
\startdata

SP1 & \hspace{0.15cm}1.43 (51)\tablenotemark{d} & 1,400 & 1,600 \\
SP2 & 1.43 (51) & ~~900 & 1,500 \\
SM &  1.44 (51) & ~~800 & 1,500 \\
AP1 & 1.48 (51) & 1,300 & 1,600 \\
AP2 & 1.50 (51) & 1,100 & 1,600 \\
AP3 & 1.47 (51) & 1,300 & 1,600 \\
AP4 & 1.50 (51) & 1,000 & 1,600 \\
AP5 & 1.48 (51) & ~~900 & 1,500 \\
AP6 & 1.51 (51) & ~~900 & 1,500 \\
AP7 & 1.47 (51) & ~~800 & 1,300 \\
AP8 & 1.50 (51) & ~~800 & 1,200 \\
AM1 & 1.51 (51) & ~~700 & 1,700 \\
AT1 & --\tablenotemark{e} & ~~500 & 3,300 \\
AT2 & 1.51 (51) & ~~600 & 3,100 \\
AS1 & 1.54 (51) & 1,500 & 1,900 \\
AS2 & 1.28 (51) & ~~900 & 1,900 \\
AS3 & 1.50 (51) & 1,200 & 2,200 \\
AS4 & 1.51 (51) & 1,300 & 2,100 \\
AS5 & 1.51 (51) & 1,300 & 1,800 \\
AS6 & 1.51 (51) & 1,200 & 1,900 \\
AS7 & 1.52 (51) & 1,200 & 1,800 \\
AS8 & 1.51 (51) & 1,200 & 1,900 \\
AM2 & 2.03 (51) & 1,100 & 3,000 \\
AM3 & 1.99 (51) & 1,100 & 2,100
\enddata
\scriptsize{
\tablenotetext{a}{Explosion energy estimated by Equation~(\ref{eq:exp}) 
at the end of simulation time.}
%
\tablenotetext{b}{Minimum velocity of $^{1}$H with $\Delta M$\,($^{1}$H)\,/\,$M$\,($^{1}$H) 
$>$ 1 $\times$ 10$^{-3}$ at the end of simulation time.}
\tablenotetext{c}{Maximum velocity of $^{56}$Ni with $\Delta M$\,($^{56}$Ni)\,/\,$M$\,($^{56}$Ni) 
$>$ 1 $\times$ 10$^{-3}$ at the end of simulation time.}
\tablenotetext{d}{The values in parentheses denote the powers of ten.}
\tablenotetext{e}{The explosion energy for model AT1 cannot be estimated by 
Equation (\ref{eq:exp}) because model AT1 does not include effects of gravity. Hence, 
for model AT1, we do not discuss the value.}
}
\end{deluxetable}

\subsection{Aspherical explosions with perturbations of pre-supernova origins}

\begin{figure*}[htbp]
\begin{tabular}{cc}
\hspace{-0.5cm}
\begin{minipage}{0.5\hsize}
\begin{center}
\includegraphics[width=6cm,keepaspectratio,clip]{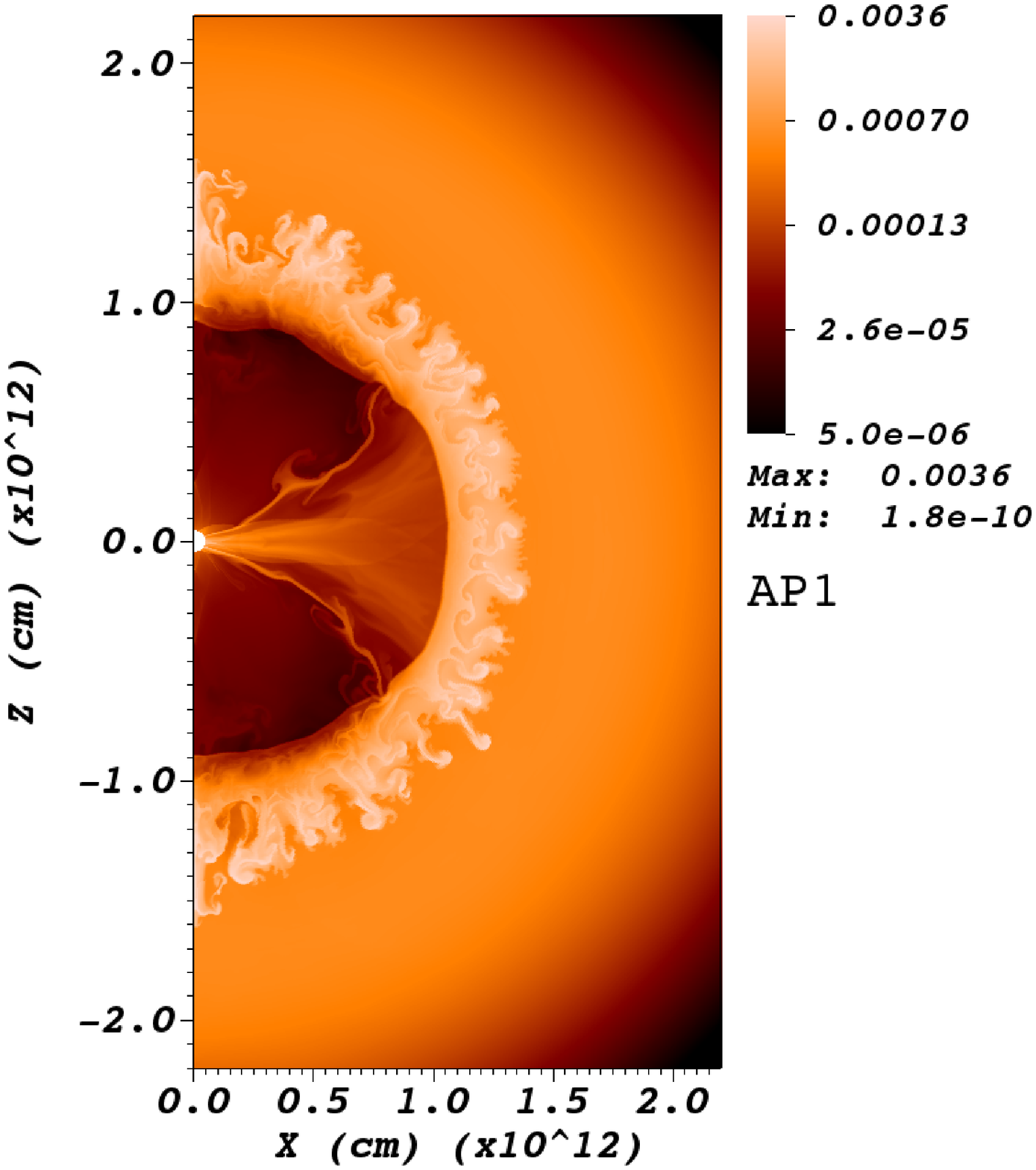}
\end{center}
\end{minipage}
\begin{minipage}{0.5\hsize}
\begin{center}
\includegraphics[width=6cm,keepaspectratio,clip]{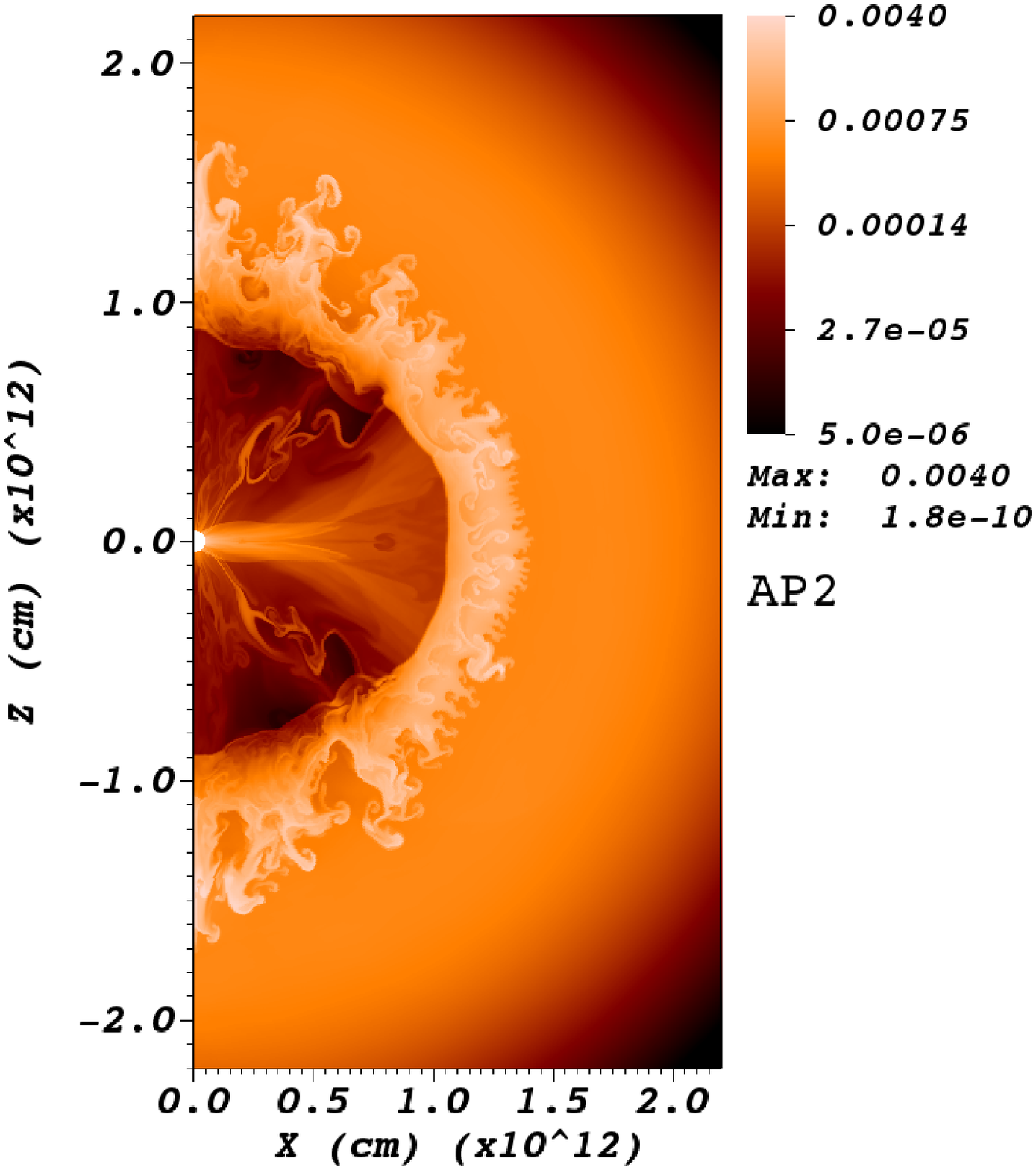}
\end{center}
\end{minipage}
\\
\hspace{-0.5cm}
\begin{minipage}{0.5\hsize}
\begin{center}
\includegraphics[width=6cm,keepaspectratio,clip]{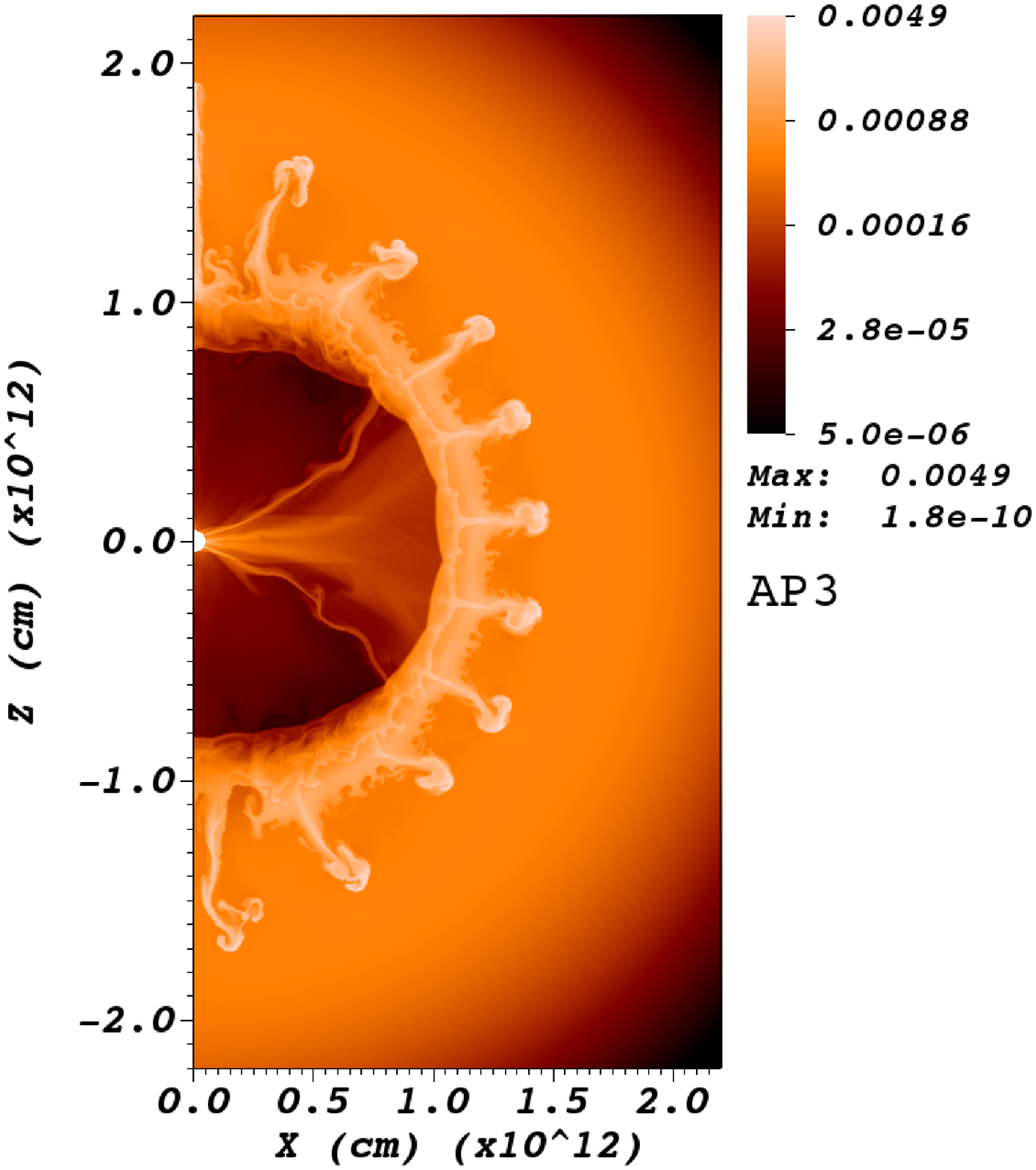}
\end{center}
\end{minipage}
\begin{minipage}{0.5\hsize}
\begin{center}
\includegraphics[width=6cm,keepaspectratio,clip]{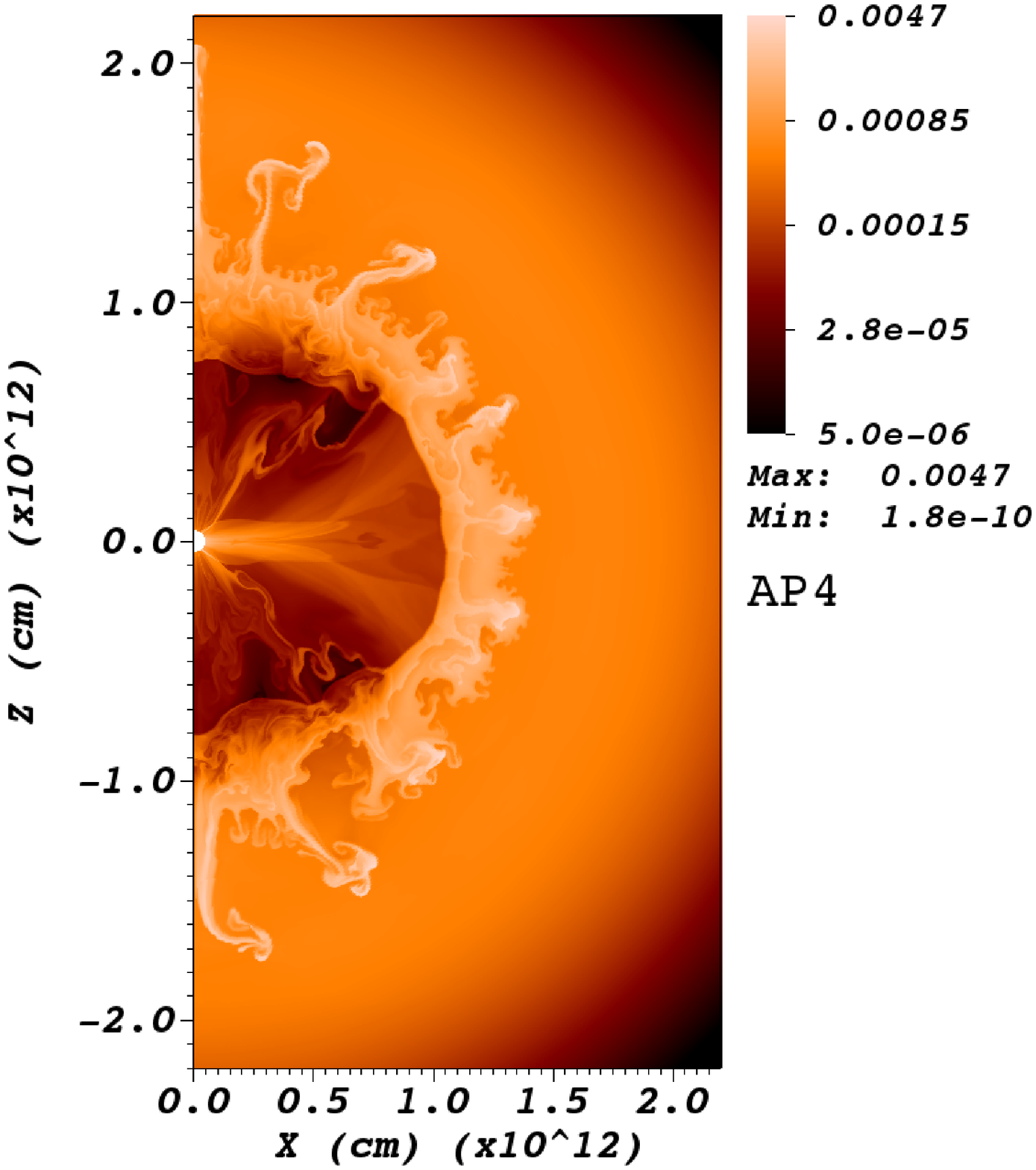}
\end{center}
\end{minipage}
\end{tabular}
\caption{Same as Figure~\ref{fig:dens_SPM} but for models AP1 (top left), AP2 (top right), AP3 (bottom left), and AP4 (bottom right)
and the time of 5851 s, 5773 s, 5861 s, and 5781 s, respectively.}
\label{fig:dens_AP-1}
\end{figure*}

In this section, we present our results for models of aspherical explosions with perturbations 
of pre-collapse origins, i.e., models AP1 to AP8 and AM1. The density distributions
for models AP1, AP2, AP3, and AP4 are shown in Figure~\ref{fig:dens_AP-1}. 
In models AP1 to AP4, perturbations are introduced when the shock waves reach 
the interface of C+O/He. In models AP1 and AP2, `random' perturbations are introduced 
but the degree of asphericity ($v_{\rm pol}$/$v_{\rm eq}$) are different. More extended 
RT fingers produced by model AP2 are seen around the polar regions than 
those produced by model AP1. In models AP3 and AP4, the situation is similar to that 
in models AP1 and AP2 but the perturbations are sinusoidally introduced. 
The mixing lengths in models AP3 and AP4 are comparable with those 
in models AP1 and AP2, respectively. 
Compared to RT fingers produced by models AP1 and AP3, those produced 
by more aspherical explosion models AP2 and AP4 have smaller-scale. 
In model AP3 and AP4, prominent protrusions along the polar axis are given. 
Compared with the spherical explosion cases, in aspherical models, AP1 to AP2, 
the shapes of the dense shells around the radius of 1 $\times$ 10$^{12}$ cm deviate 
slightly from the spherical symmetry and the corresponding positions are shifted inward 
in the regions closer to the polar axis. 

\begin{figure*}[htbp]
\begin{tabular}{cc}
\hspace{-0.5cm}
\begin{minipage}{0.5\hsize}
\begin{center}
\includegraphics[width=6cm,keepaspectratio,clip]{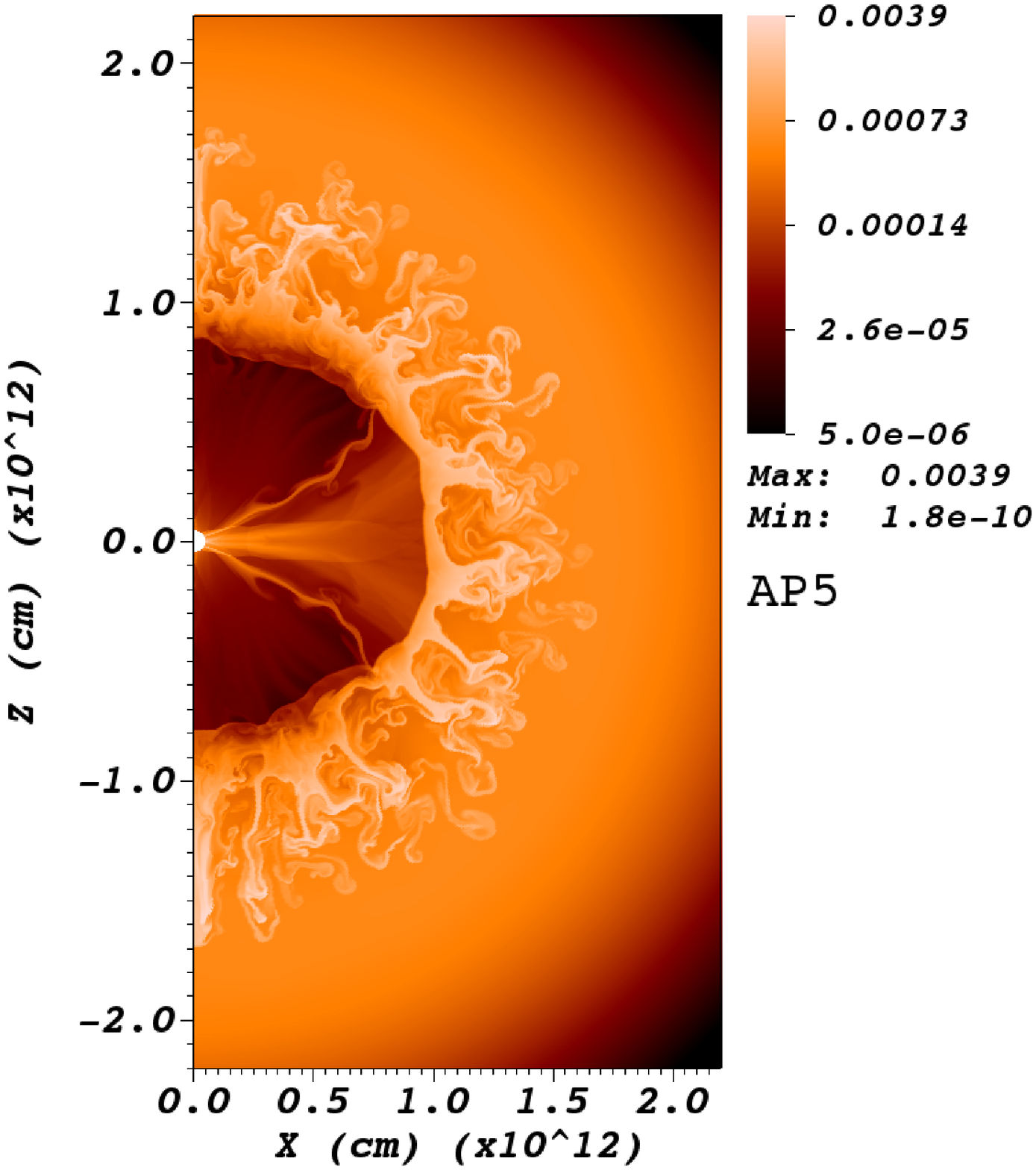}
\end{center}
\end{minipage}
\begin{minipage}{0.5\hsize}
\begin{center}
\includegraphics[width=6cm,keepaspectratio,clip]{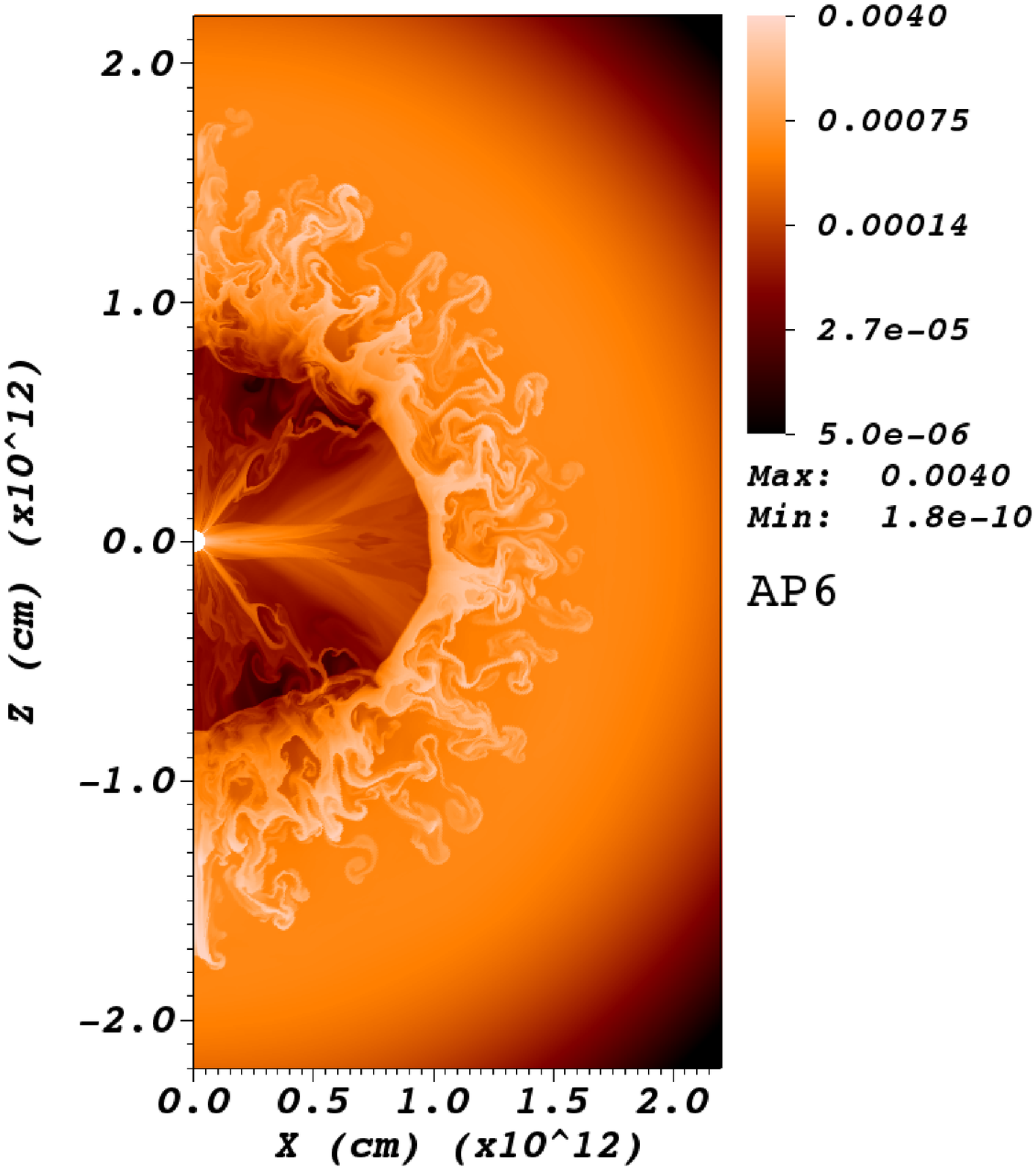}
\end{center}
\end{minipage}
\\
\hspace{-0.5cm}
\begin{minipage}{0.5\hsize}
\begin{center}
\includegraphics[width=6cm,keepaspectratio,clip]{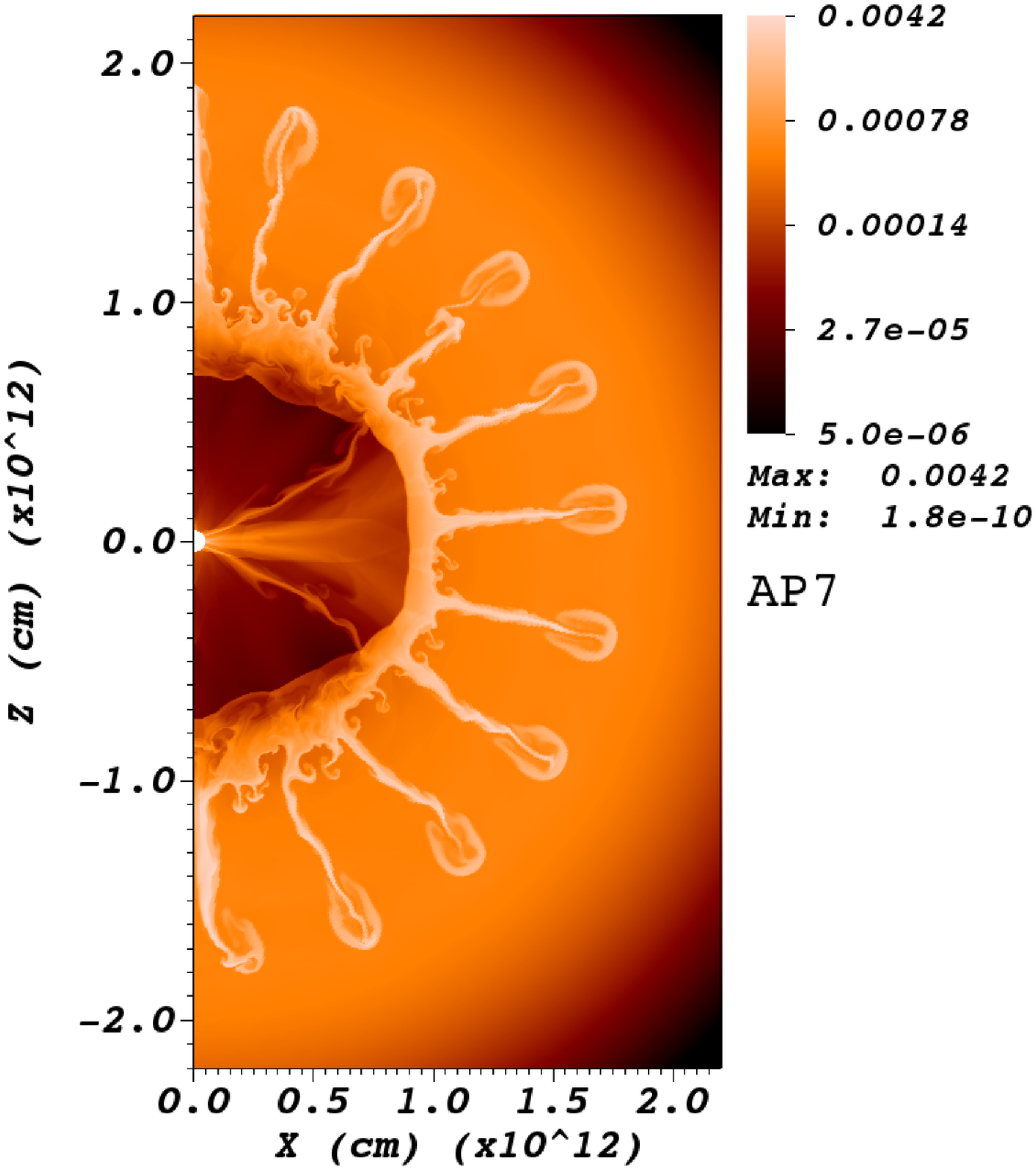}
\end{center}
\end{minipage}
\begin{minipage}{0.5\hsize}
\begin{center}
\includegraphics[width=6cm,keepaspectratio,clip]{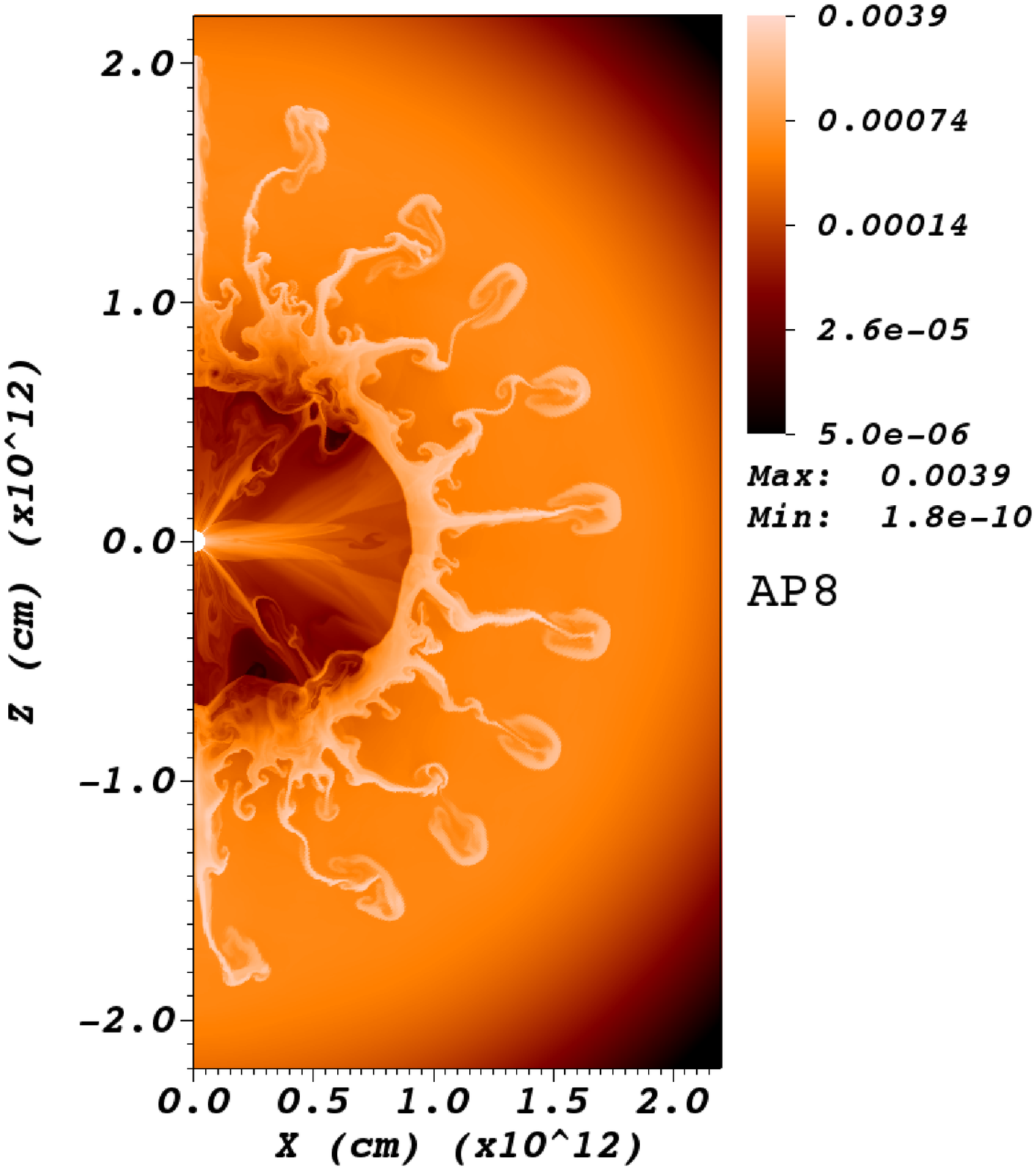}
\end{center}
\end{minipage}
\end{tabular}
\caption{Same as Figure~\ref{fig:dens_SPM} but for models AP5 (top left), AP6 (top right), 
AP7 (bottom left), and AP48 (bottom right) and the time of 5859 s, 5781 s, 5881 s, and 5793 s, 
respectively.}
\label{fig:dens_AP-2}
\end{figure*}

The density distributions for models AP5, AP6, AP7, and AP8 are shown 
in Figure~\ref{fig:dens_AP-2}. In models AP5 to AP8, perturbations are introduced 
when the shock waves reach the interface of He/H. In overall,  the mixing lengths 
in models AP5 to AP8 are apparently enhanced compared to those in models AP1 to AP4. 
We further recognize enhanced inward mixing in regions close to the polar axis 
in models AP5 to AP8 compared with those in models AP1 to AP4 from the positions 
of the inner edges of the dense shells. The mixing lengths derived from `sinusoidal' 
perturbation models AP7 and AP8 are enlarged compared with those in the counterparts 
of `random' perturbation models AP5 and AP6, respectively. The RT fingers in model AP8 
have stronger wobbling than those in model AP7, because model AP8 has clearer aspherical 
feature than model AP7. 

\begin{figure*}[htbp]
\begin{center}
\includegraphics[width=13cm,keepaspectratio,clip]{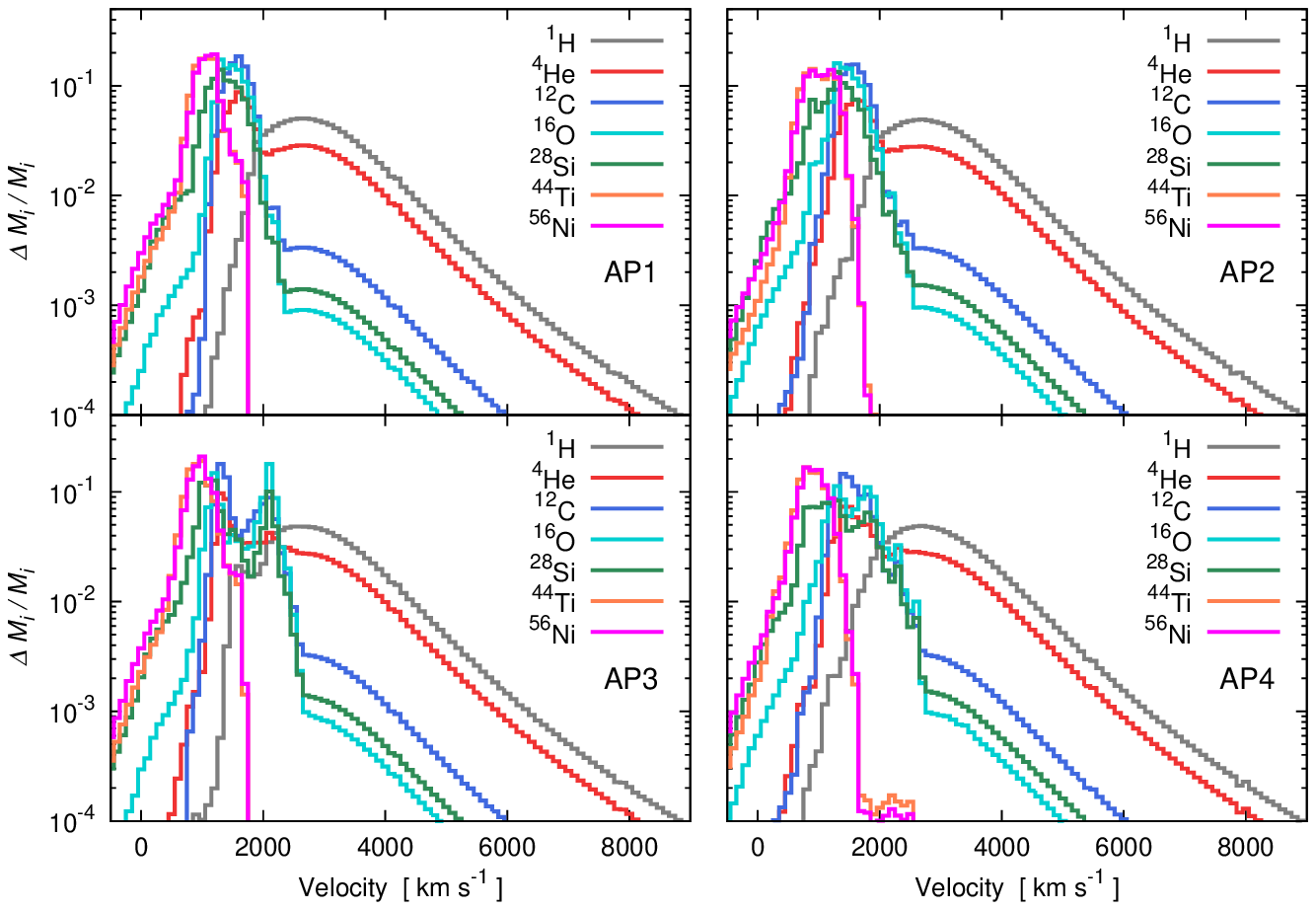}
\caption{Same as Figure~\ref{fig:vel_SPM} but for models AP1 (top left), AP2 (top right), 
AP3 (bottom left), and AP4 (bottom right) and the time of 5851 s, 5773 s, 5861 s, and 5781 s, 
respectively.}
\label{fig:vel_AP-1}
\end{center}
\end{figure*}

The mass distributions of elements as a function of radial velocity at the ends of 
simulation time for models AP1, AP2, AP3, and AP4 are shown in Figure~\ref{fig:vel_AP-1}. 
The high velocity tails of $^{56}$Ni, $^{28}$Si, $^{12}$C, and $^{16}$O in model AP2 
are slightly enhanced compared with those in model AP1, because AP2 has clearer aspherical 
feature than model AP1. As summarized in Table~\ref{table:results}, the obtained 
maximum velocity of $^{56}$Ni are approximately 1,600 km s$^{-1}$ in models AP1 and AP2. 
The low-velocity tail of hydrogen is slightly more prominent in model AP1 compared 
to that in model AP2. In models AP3 and AP4, the obtained maximum velocity of $^{56}$Ni 
are comparable to those in models AP1 and AP2. However, the high velocity components 
of $^{28}$Si, $^{12}$C, and $^{16}$O in sinusoidal perturbation models AP3 and AP4 
are enhanced compared with those in the random perturbation models AP1 and AP2. 
The minimum velocities of $^1$H range between 1,000 and 1,300 km s$^{-1}$ 
among models AP1 to AP4. The inward mixing in models AP2 and AP4 is more prominent 
than that in models AP1 and AP3. The minimum velocities of $^1$H in models AP2 and AP4 
are smaller than those in models AP1 and AP3. The reason is because models AP2 and AP4 
have clearer aspherical feature than models AP1 and AP3. 

\begin{figure*}[htbp]
\begin{center}
\includegraphics[width=13cm,keepaspectratio,clip]{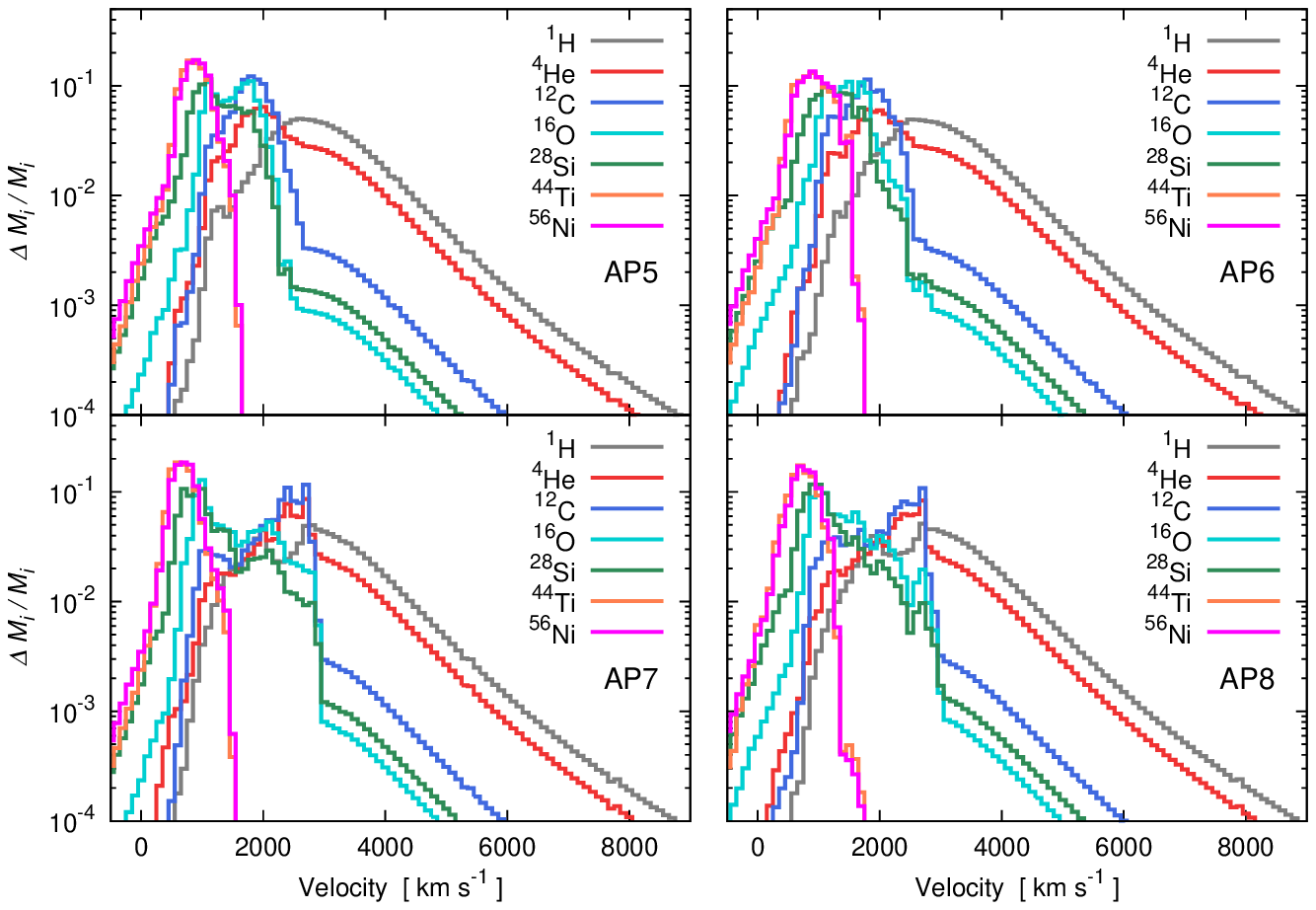}
\caption{Same as Figure~\ref{fig:vel_SPM} but for models AP5 (top left), AP6 (top right), 
AP7 (bottom left), and AP8 (bottom right) and the time of 5859 s, 5781 s, 5881 s, and 5793 s, 
respectively.}
\label{fig:vel_AP-2}
\end{center}
\end{figure*}

The mass distributions of elements, $^1$H, $^4$He, $^{12}$C, $^{16}$O, $^{28}$Si, $^{44}$Ti, 
and $^{56}$Ni, as a function of radial velocity at the ends of simulation time for 
models AP5, AP6, AP7, and AP8 are shown in Figure~\ref{fig:vel_AP-2}. 
In models AP5 to AP8, perturbations are introduced when shock waves reach 
the interface of He/H. Overall, high velocity tails of $^{28}$Si, $^{12}$C, and $^{16}$O 
in models AP5 to AP8 are enlarged compared with those in models AP1 to AP4, 
because models AP5 to AP8 have prominent RT instabilities around the composition 
interface of He/H. However, the maximum velocities of inner most metals such as 
$^{56}$Ni and $^{44}$Ti are reduced in models AP5 to AP8 compared with those 
in models AP1 to AP4. Obtained maximum velocities of $^{56}$Ni range 
between 1,200 and 1,500 km s$^{-1}$ among models AP5 to AP8. 
On the other hand, the minimum velocities of $^1$H are smaller than those 
in models AP1 to AP4 and range between 800 and 900 km s$^{-1}$. 
From above results, mixing of innermost metals, $^{56}$Ni and $^{44}$Ti 
is prominent in models that perturbations are introduced in an early phase. 
On the contrary, mixing of the other elements is prominent in models where 
perturbations are introduced in a later phase. Overall, the mixing is slightly enhanced 
in models with strong aspherical feature compared with models with weaker 
aspherical feature. In all aspherical explosion modes AP1 to AP8, 
the obtained maximum velocities of $^{56}$Ni do not reach the observed high values of SN~1987A.  

\begin{figure*}[t]
\begin{tabular}{cc}
\hspace{-0.5cm}
\begin{minipage}{0.5\hsize}
\begin{center}
\includegraphics[width=6cm,keepaspectratio,clip]{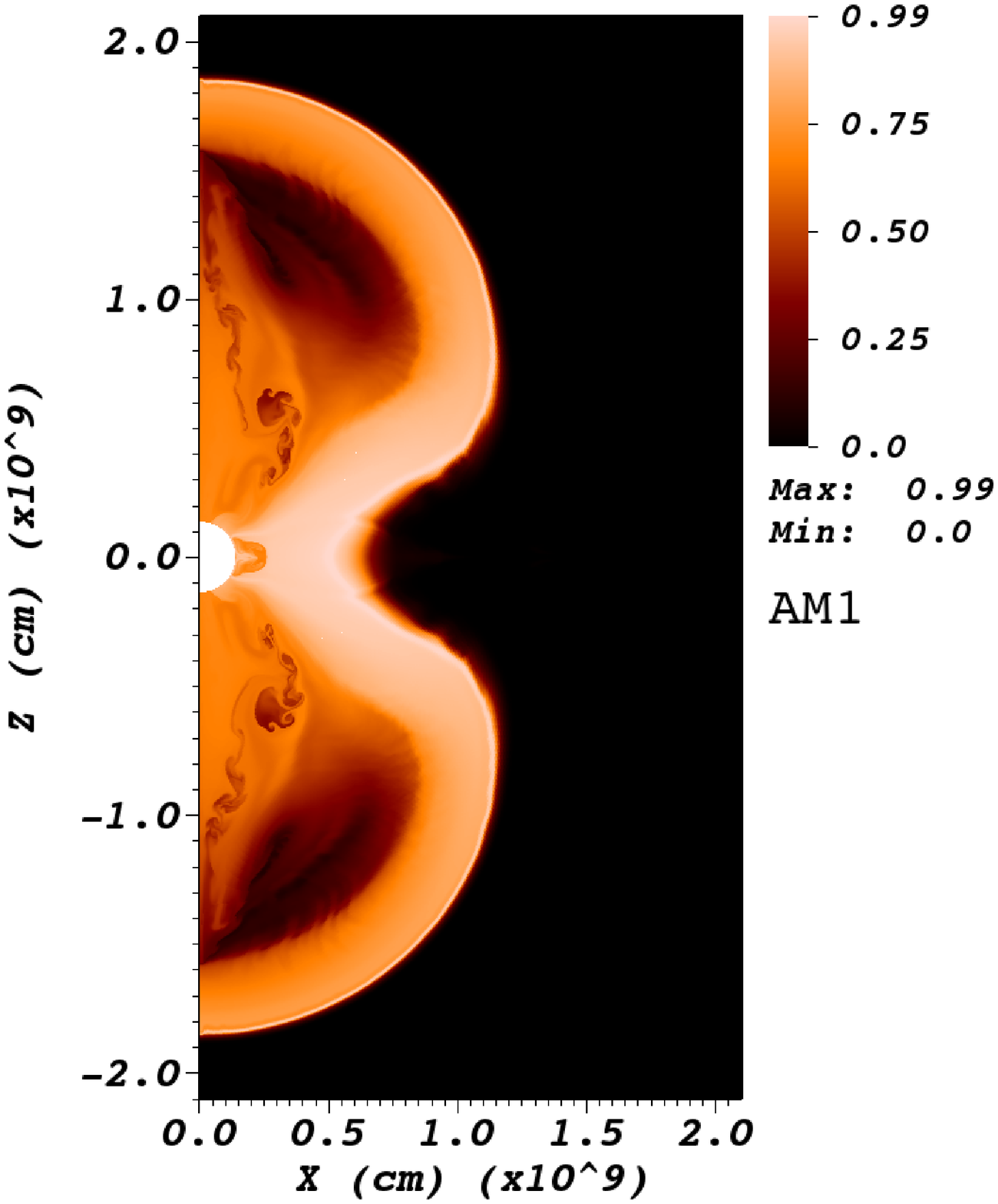}
\end{center}
\end{minipage}
\begin{minipage}{0.5\hsize}
\begin{center}
\includegraphics[width=6cm,keepaspectratio,clip]{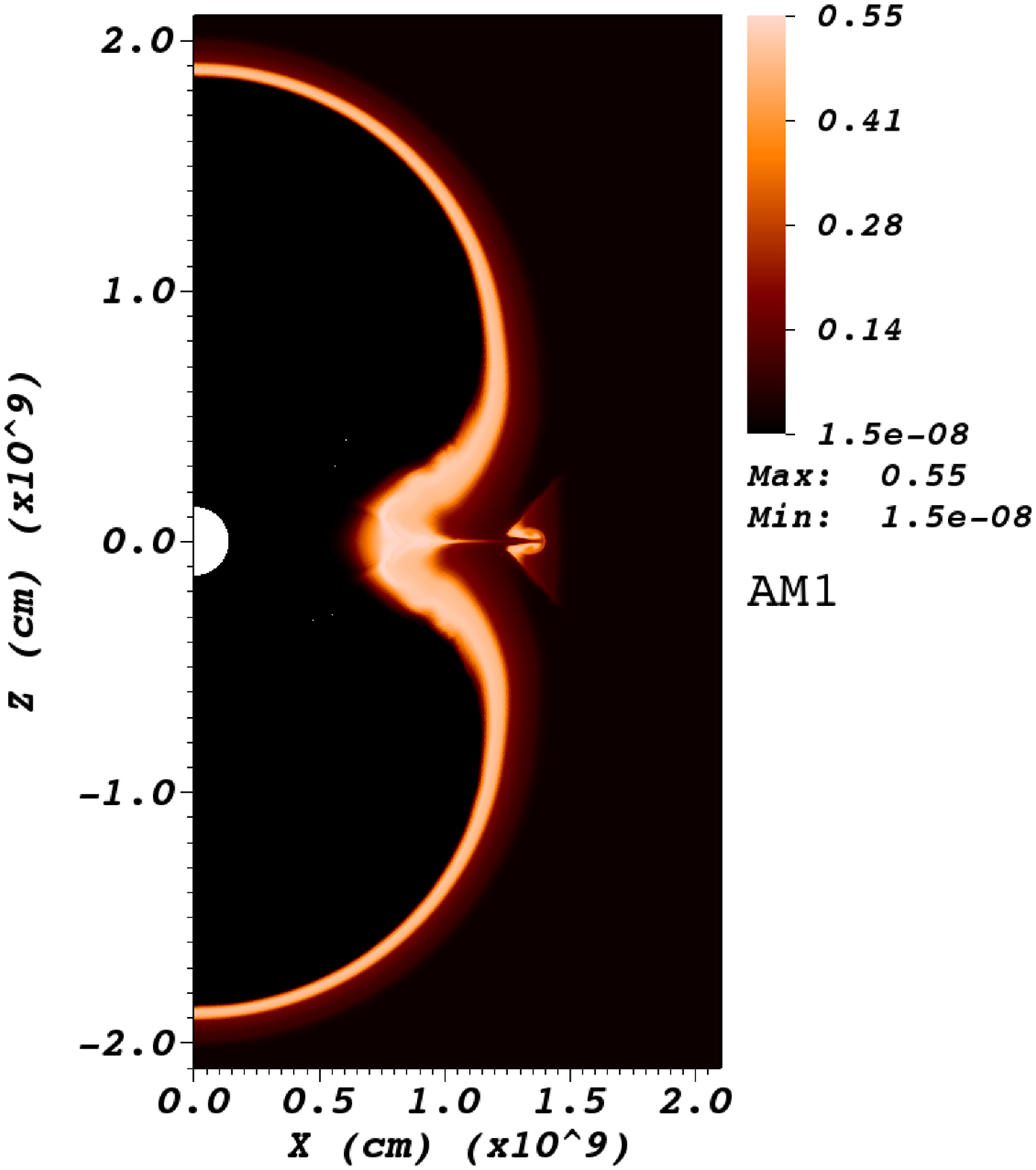}
\end{center}
\end{minipage}
\\
\hspace{-0.5cm}
\begin{minipage}{0.5\hsize}
\begin{center}
\includegraphics[width=6cm,keepaspectratio,clip]{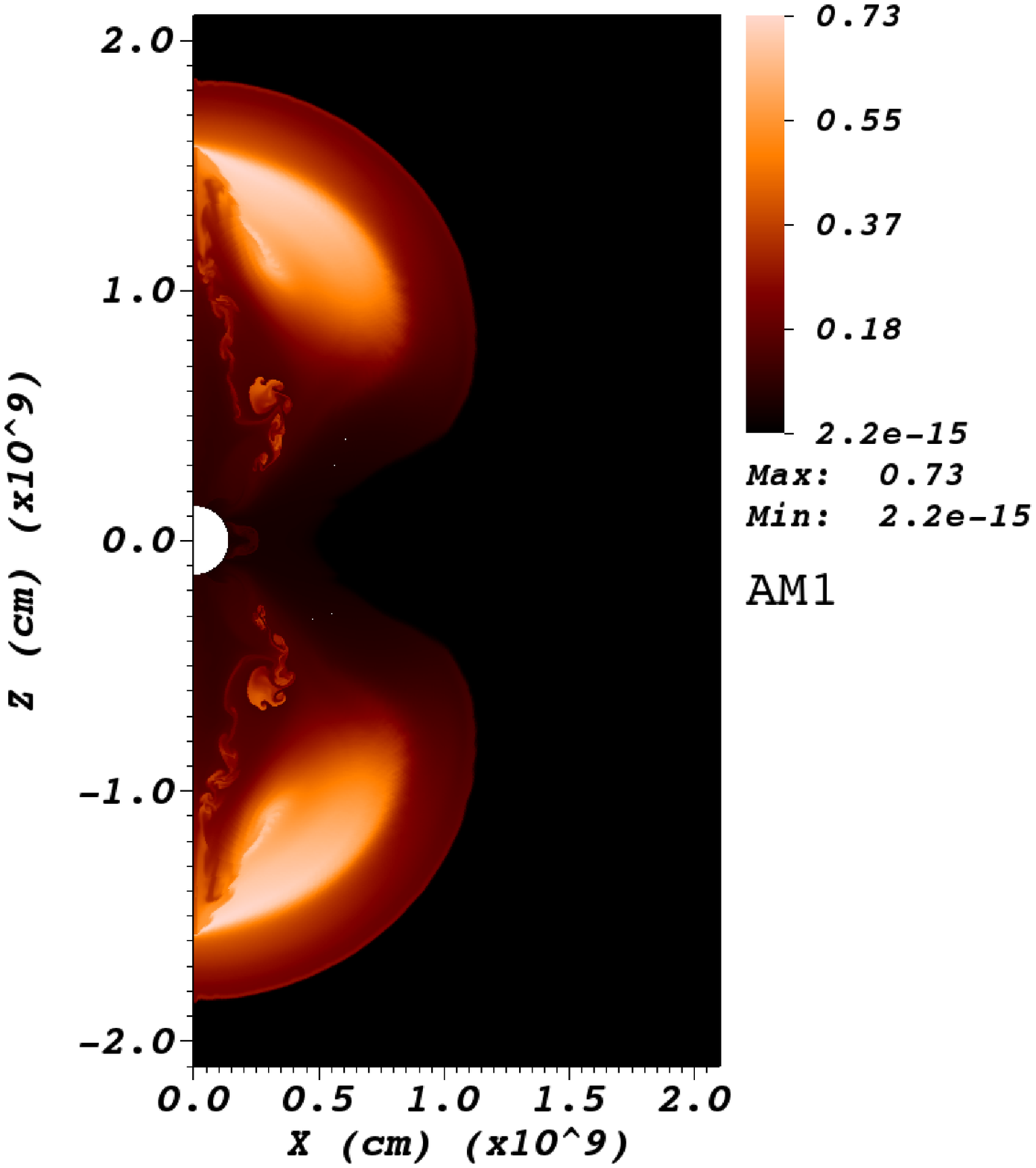}
\end{center}
\end{minipage}
\begin{minipage}{0.5\hsize}
\begin{center}
\includegraphics[width=6cm,keepaspectratio,clip]{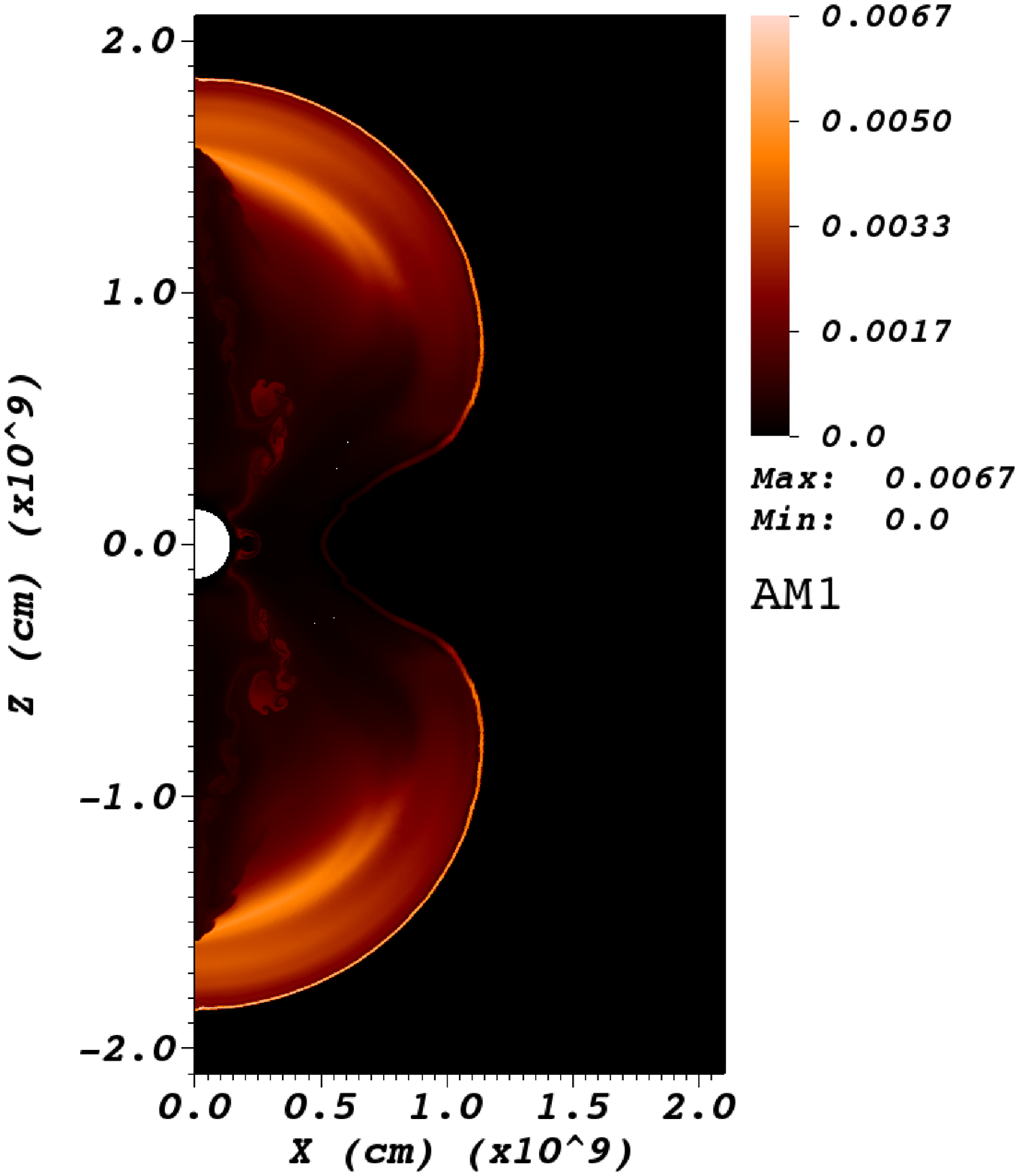}
\end{center}
\end{minipage}
\end{tabular}
\caption{Distributions of mass fractions of elements, $^{56}$Ni (top left), $^{28}$Si (top right), 
$^{4}$He (bottom left), and $^{44}$Ti (bottom right) at the time of 0.96 s 
in the $X$--$Z$ plane for model AM1. The values in color bars are linearly scaled.}
\label{fig:nucl_AM1}
\end{figure*}

Next, we show the results of the aspherical explosion model AM1 in which perturbations 
are multiply introduced. First, we explain briefly the explosive nucleosynthesis 
by taking model AM1 as an example. The distributions of mass fractions of elements, 
$^{56}$Ni, $^{28}$Si, $^4$He, and $^{44}$Ti, are shown as the results at the evolutionary 
time of 0.96 s for model AM1 in Figure~\ref{fig:nucl_AM1}. The values in color bars are 
linearly scaled. $^{56}$Ni is synthesized prominently in the edge of a gourd-like structure 
and inner regions close to the polar axis (the top left panel). 
In the thin edge of the gourd-like structure, $^{28}$Si remains unburned partly due to 
the incomplete silicon burning. 
Inside the gourd-like structure, some fraction of $^{4}$He also remains unburned. 
The regions that $^{4}$He remains unburned correspond to relatively low density regions 
inside the shock. In a low density regime, the explosive silicon burning ends up with 
so-called the alpha-rich freeze-out. $^{44}$Ti is prominent in regions that $^{4}$He 
remains unburned due to the alpha-rich freeze-out. This is consistent with 
the results of \citet{nag00}. 

\begin{figure*}[htbp]
\begin{tabular}{cc}
\hspace{-0.5cm}
\begin{minipage}{0.5\hsize}
\begin{center}
\includegraphics[width=6cm,keepaspectratio,clip]{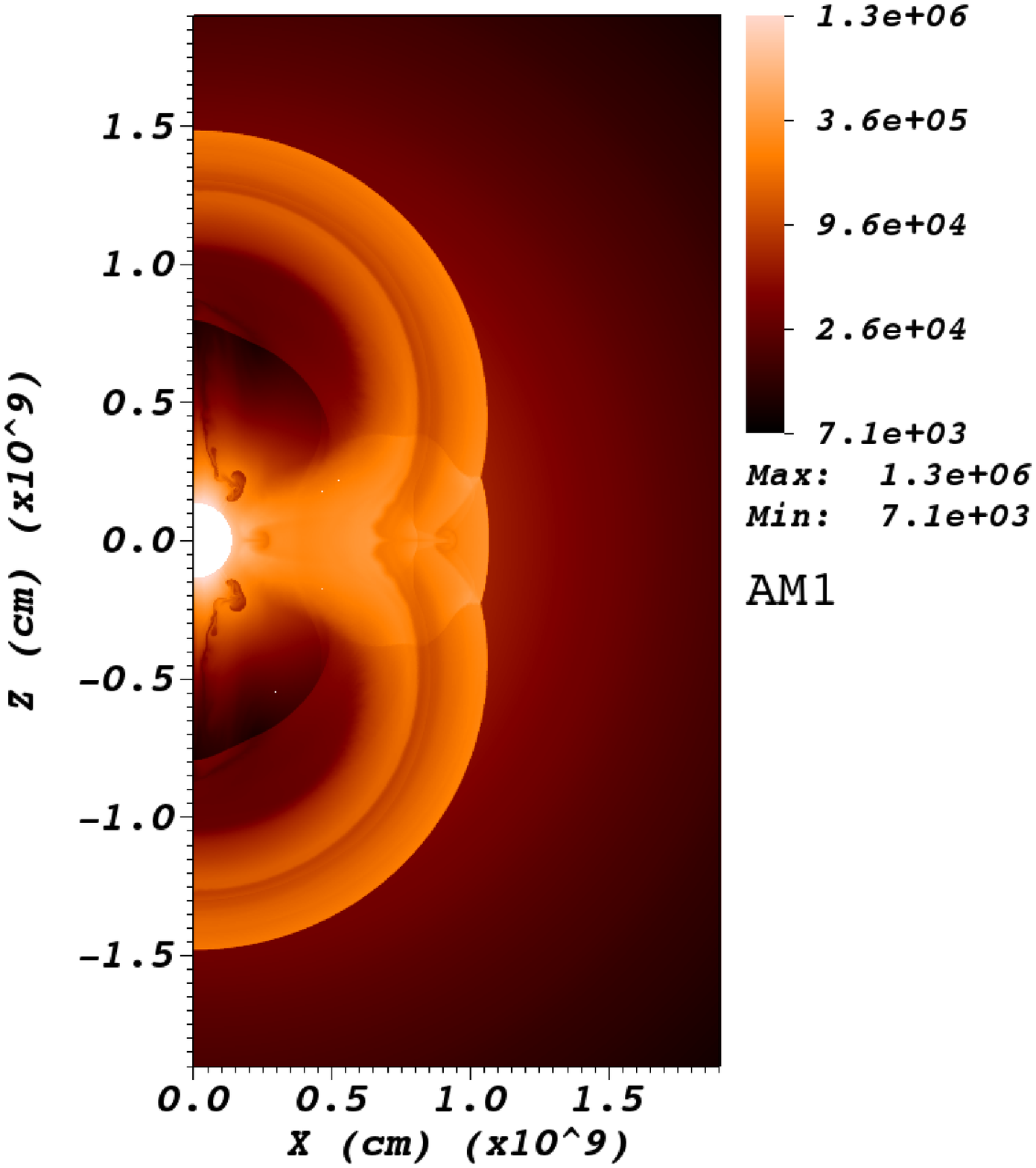}
\end{center}
\end{minipage}
\begin{minipage}{0.5\hsize}
\begin{center}
\includegraphics[width=6cm,keepaspectratio,clip]{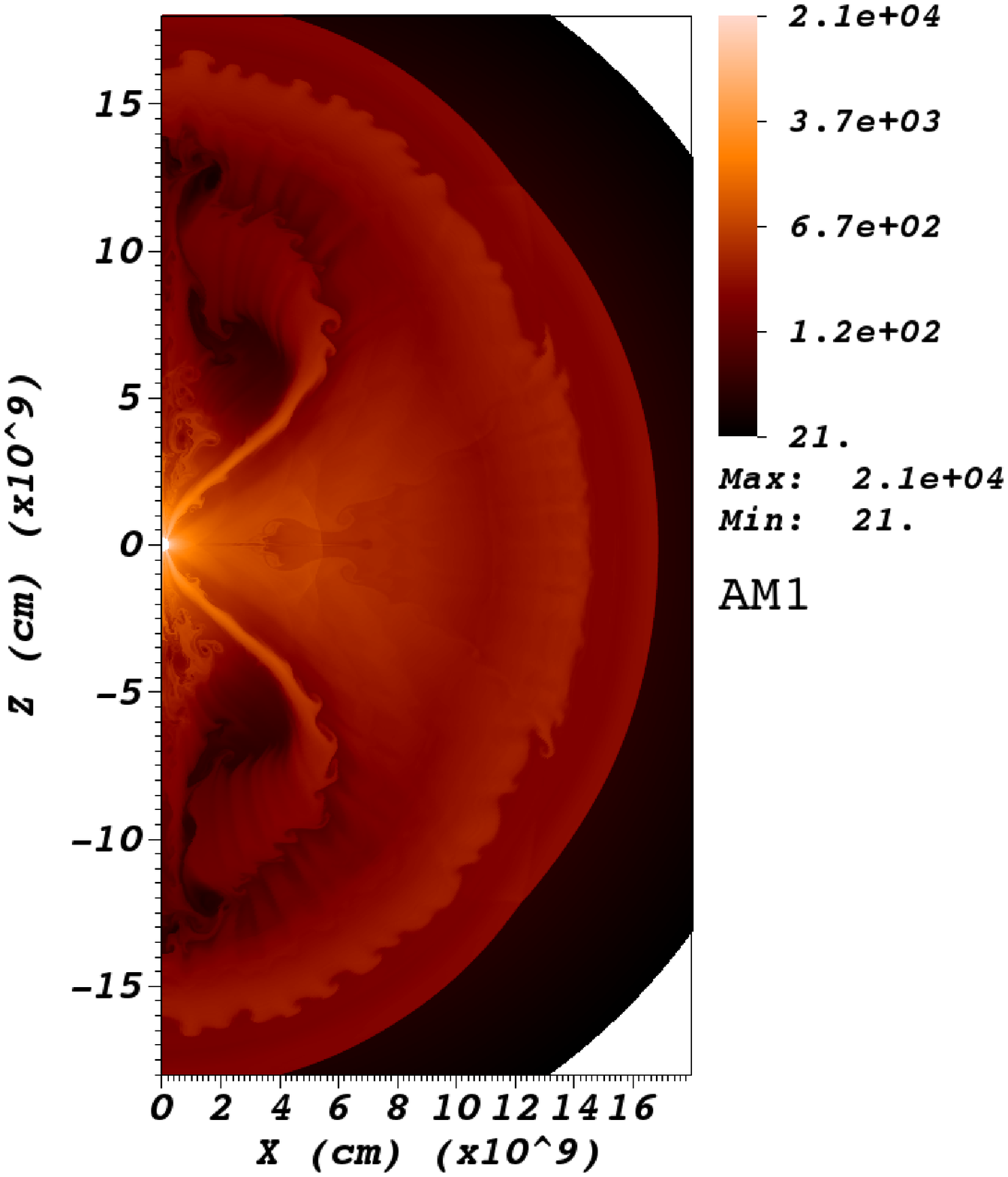}
\end{center}
\end{minipage}
\\
\hspace{-0.5cm}
\begin{minipage}{0.5\hsize}
\begin{center}
\includegraphics[width=6cm,keepaspectratio,clip]{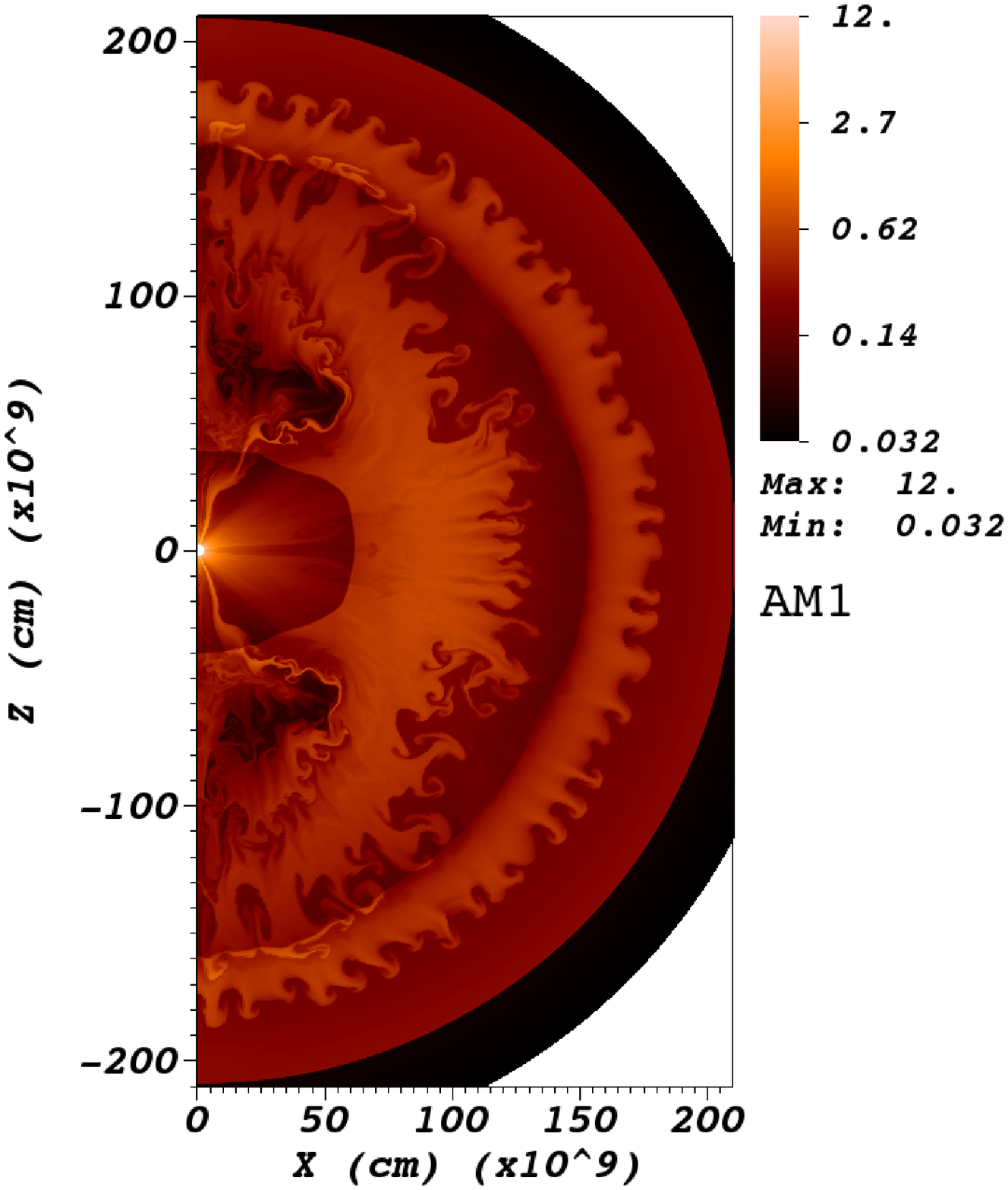}
\end{center}
\end{minipage}
\begin{minipage}{0.5\hsize}
\begin{center}
\includegraphics[width=6cm,keepaspectratio,clip]{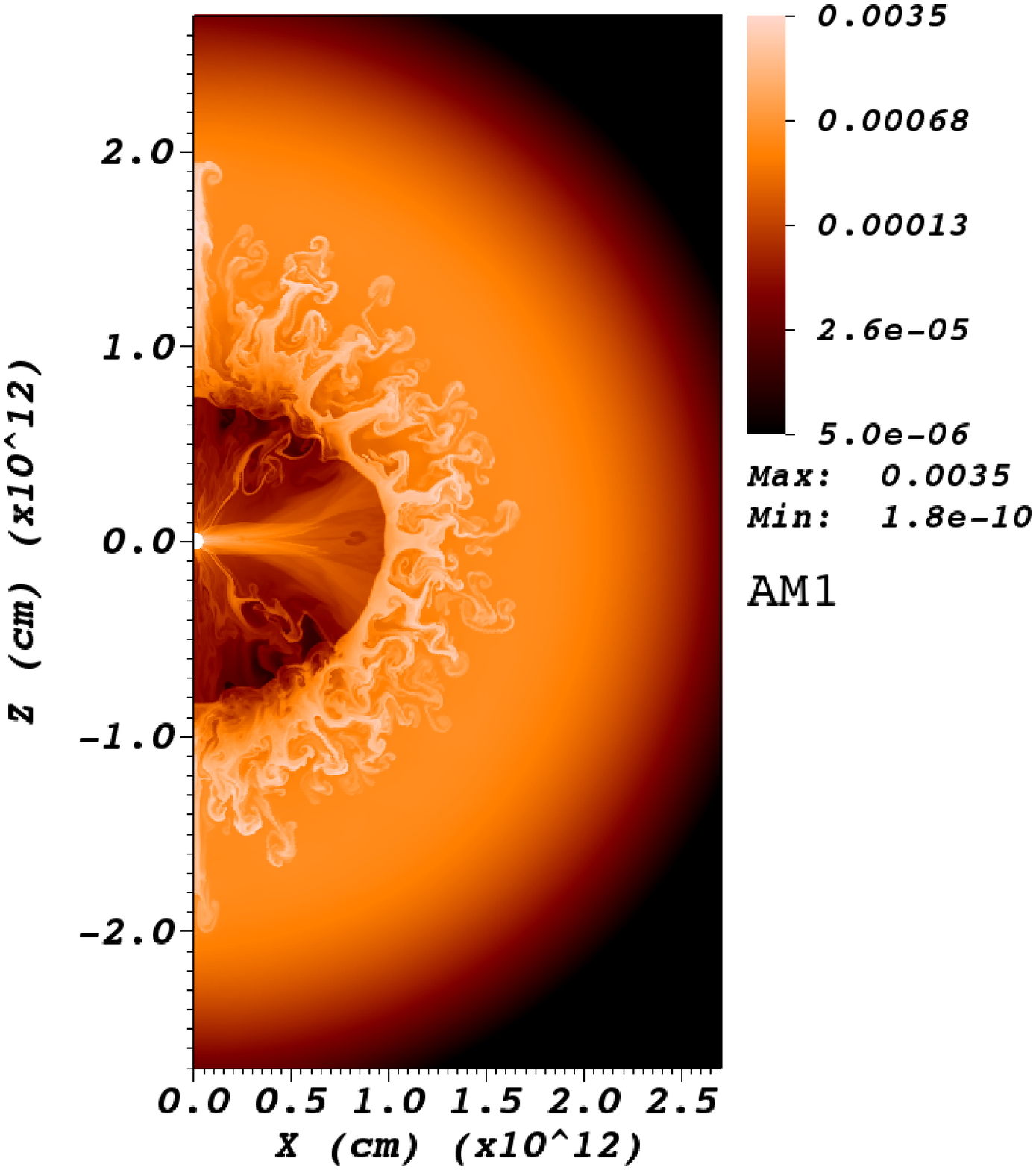}
\end{center}
\end{minipage}
\end{tabular}
\caption{Snap shots of distributions of density at the time of 
0.53 s (top left), 16.6 s (top right), 288 s (bottom left), and 5752 s (bottom right) for model AM1. 
The unit of values in color bars is g cm$^{-3}$. The values in color bars are logarithmically scaled.}
\label{fig:dens_AM1}
\end{figure*}

We show the time evolution of density distribution for model AM1. The snap shots 
of density distributions for model AM1 at the time of 0.53 s, 
16.6 s, 288 s, and 5752 s are shown in Figure~\ref{fig:dens_AM1}. 
Note that the white color regions are outside of the computational domain. 
A gourd-shaped shock is generated by the bipolar explosion as shown 
in the top left panel just after the initiation of the explosion. 
The snap shot just after the introduction of the first perturbations is shown 
in the top right panel. We recognize that the gourd-shaped shock becomes narrower 
in equatorial regions due to the fallback of matter. The snap shot after the 
introduction of the second perturbations is shown in the bottom left panel. 
Finally, the snap shot at the end of simulation time is shown in the bottom right panel. 
The appearance of the density distribution is similar to that in model AP6 
(see the top left panel in Figure~\ref{fig:dens_AP-2}). However, more prominent inward 
and outward mixing is seen around the polar regions. The mixing length around 
the polar region is approximately 1 $\times$ 10$^{12}$ cm. We can more extended RT fingers along the polar axis, which reach the radius of 2 $\times$ 10$^{12}$ cm. 

\begin{figure*}[htbp]
\begin{tabular}{cc}
\hspace{-0.5cm}
\begin{minipage}{0.5\hsize}
\begin{center}
\includegraphics[width=6cm,keepaspectratio,clip]{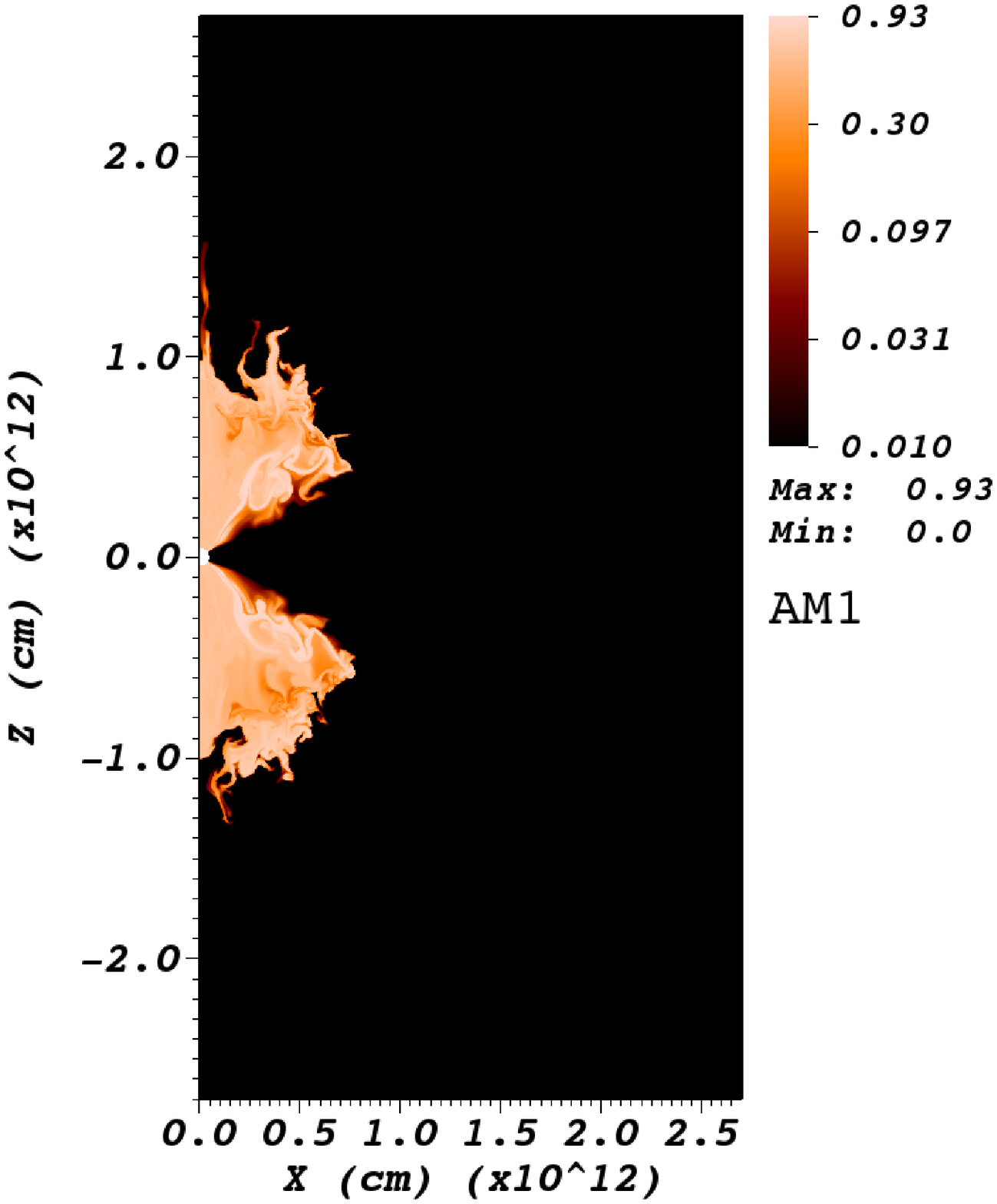}
\end{center}
\end{minipage}
\begin{minipage}{0.5\hsize}
\begin{center}
\includegraphics[width=6cm,keepaspectratio,clip]{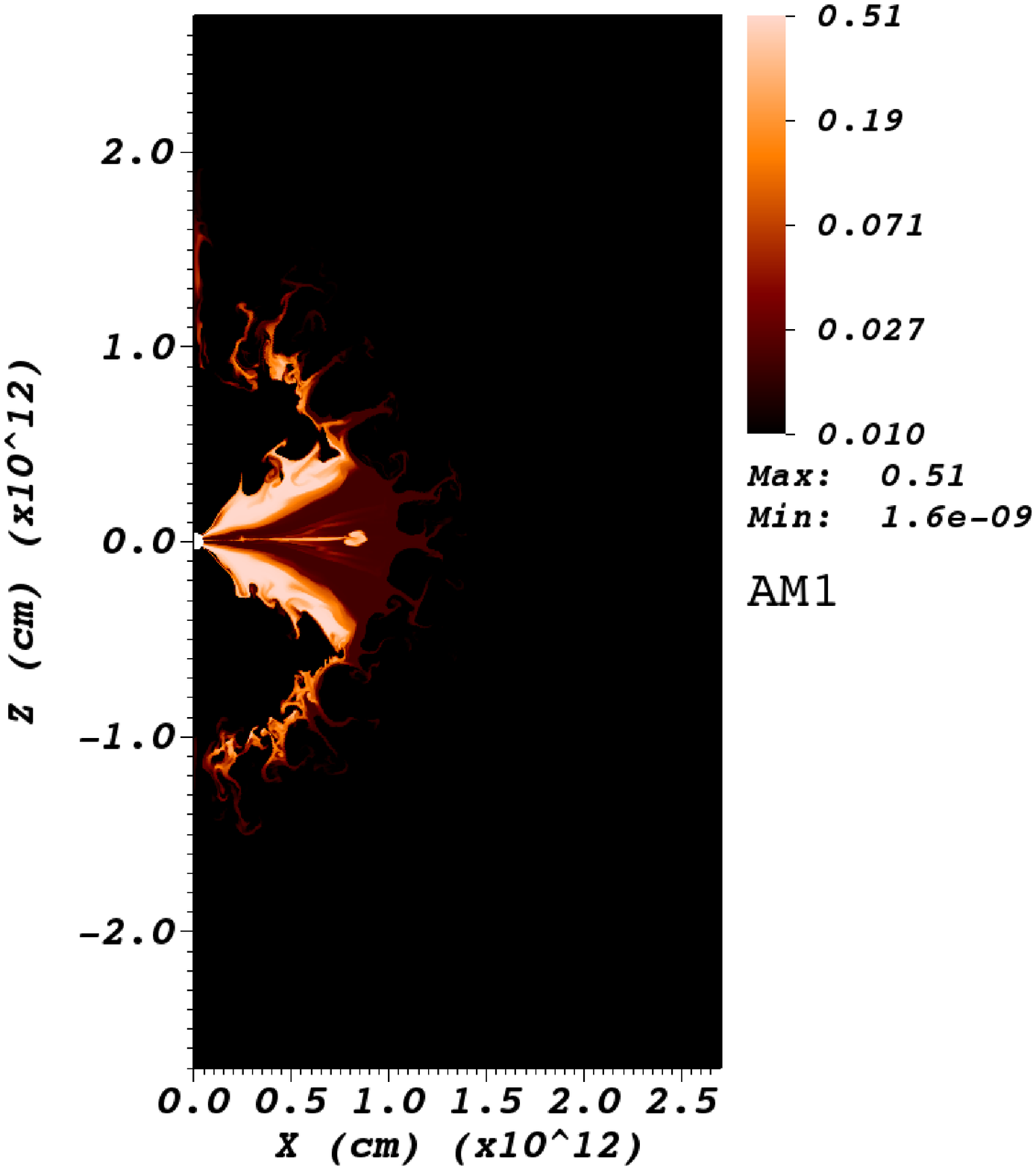}
\end{center}
\end{minipage}
\\
\hspace{-0.5cm}
\begin{minipage}{0.5\hsize}
\begin{center}
\includegraphics[width=6cm,keepaspectratio,clip]{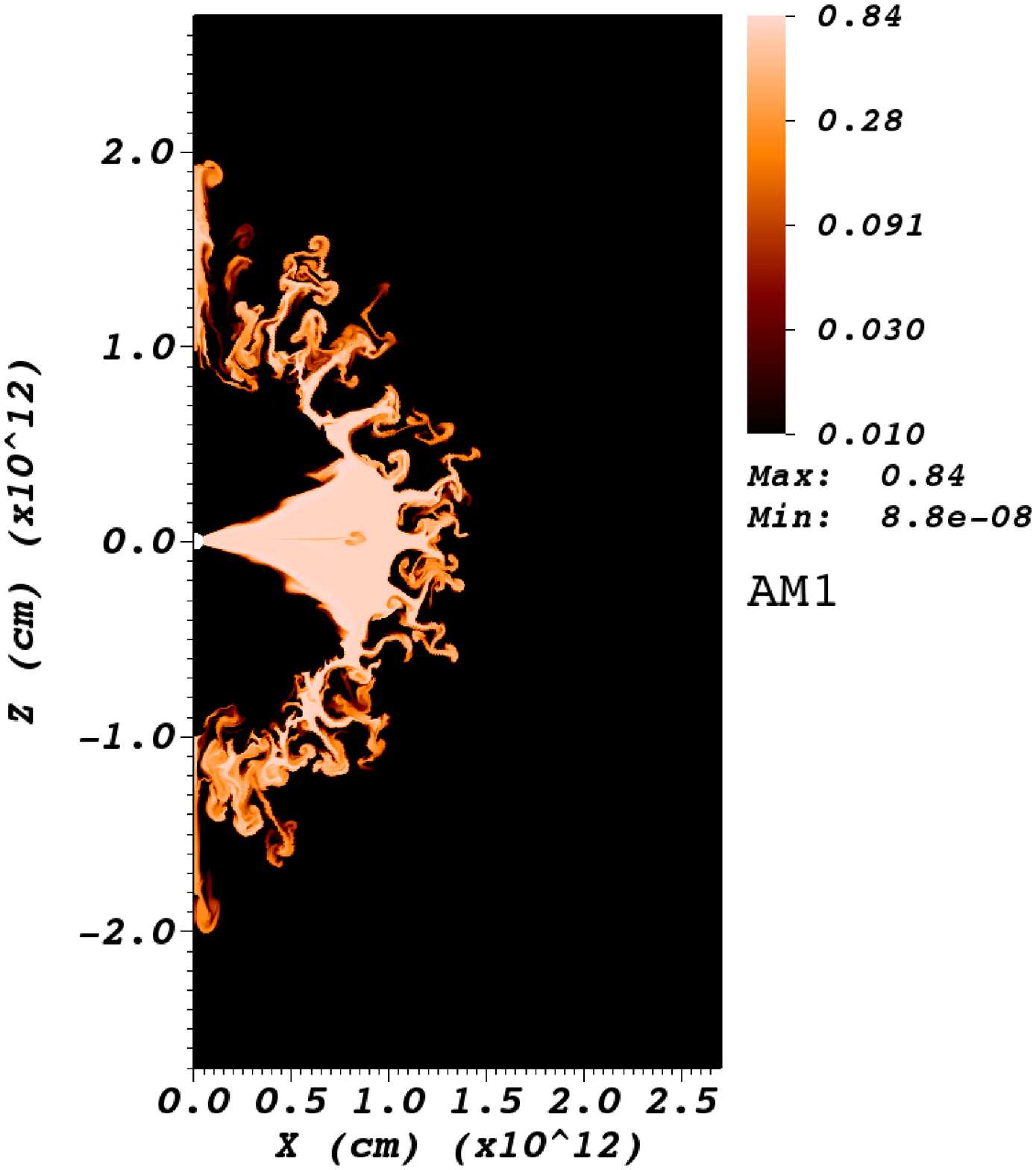}
\end{center}
\end{minipage}
\begin{minipage}{0.5\hsize}
\begin{center}
\includegraphics[width=6cm,keepaspectratio,clip]{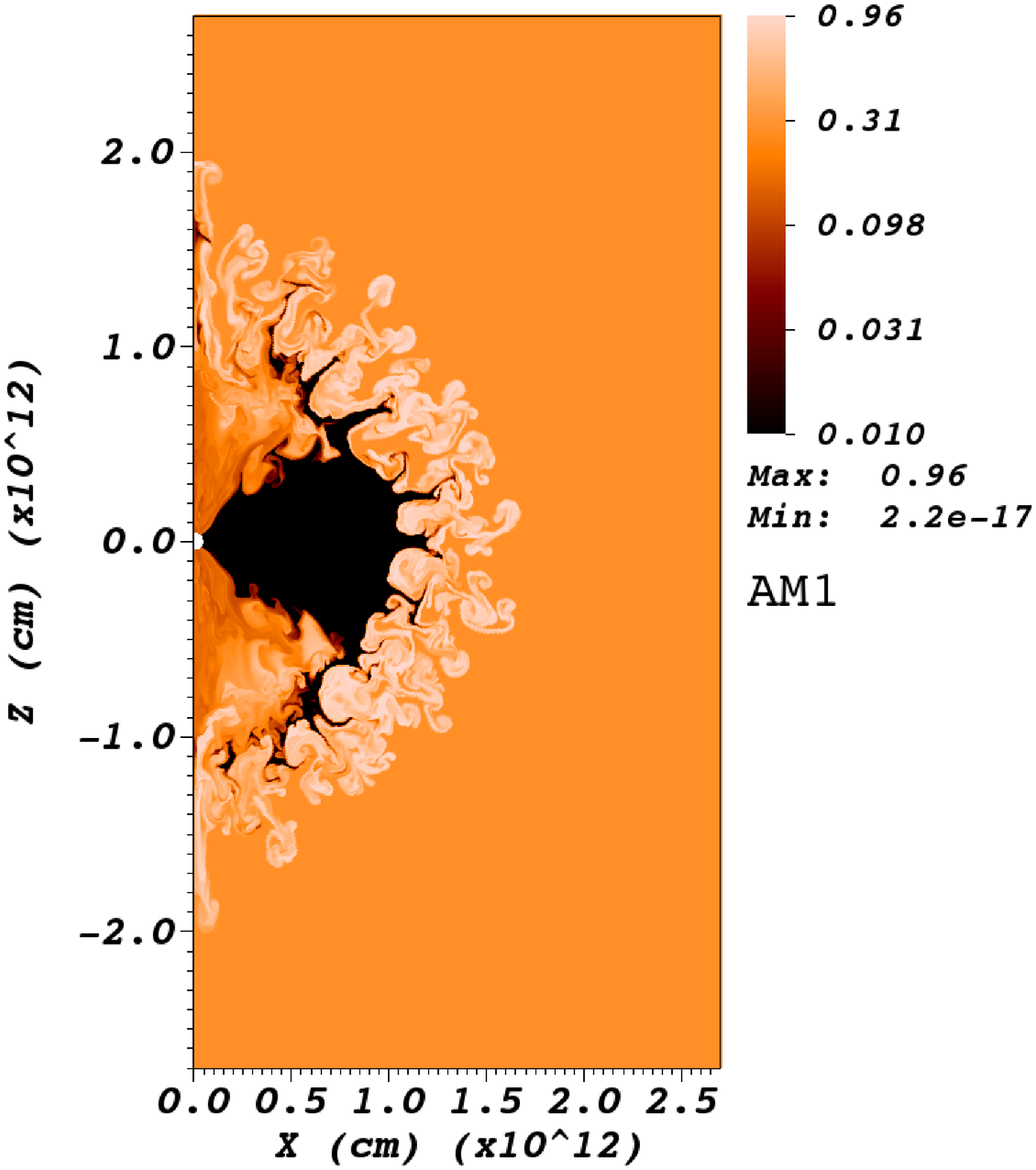}
\end{center}
\end{minipage}
\end{tabular}
\caption{Same as Figure~\ref{fig:element_SM} but for elements, $^{56}$Ni, $^{28}$Si, $^{16}$O, and $^{4}$He 
and the time of 5752 s.}
\label{fig:element_AM1}
\end{figure*}

The distributions of mass fractions of 
elements, $^{56}$Ni, $^{28}$Si, $^{16}$O, and $^4$He, for model AM1 at the 
end of simulation time are shown in Figure~\ref{fig:element_AM1}. 
Unlike the results in spherical explosion models shown in Figure~\ref{fig:element_SM}, 
here $^{56}$Ni is distributed in the wedge-shaped regions around the polar axis. 
Slight protrusions of $^{56}$Ni along the RT fingers are also seen. 
However, $^{56}$Ni is basically concentrated inside the dense helium shell. 
$^{28}$Si encompasses $^{56}$Ni. Some fractions of $^{28}$Si are conveyed outward 
along the RT fingers. $^{16}$O is prominent in a wedged-shaped region along the 
equatorial plane and inside the RT fingers. $^4$He is distributed around the 
RT fingers and the inner wedge-shaped regions along the polar axis. 
$^4$He in inner regions are synthesized by the explosive nucleosynthesis, 
as same as the process in the spherical explosion models. 

\begin{figure}
\hspace{-2cm}
\begin{center}
\includegraphics[width=6cm,keepaspectratio,clip]{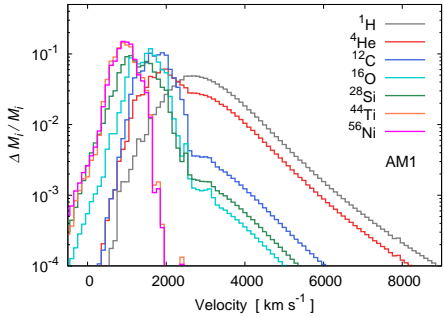}
\caption{Same as Figure~\ref{fig:vel_SPM} but for model AM1 at the time of 5752 s.}
\label{fig:vel_AM1}
\end{center}
\end{figure}

In Figure~\ref{fig:vel_AM1}, we show the mass distributions of elements, 
$^1$H, $^{4}$He, $^{12}$C, $^{16}$O, $^{28}$Si, $^{44}$Ti, and $^{56}$Ni, 
as a function of radial velocity for model AM1 at the end of simulation time. 
As expected from the previous discussion, 
the distributions have features seen in both models of AP2 and AP6. 
The high velocity tails of $^{28}$Si, $^{12}$C, and $^{16}$O in model AM1 are enhanced 
compared with those in model AP2. The innermost metals, $^{56}$Ni and $^{44}$Ti, 
in model AM1 are conveyed in higher velocity regions compared with the situation 
in model AP6. The maximum velocity of $^{56}$Ni reaches 1,700 km s$^{-1}$, which 
is the largest value among all the models mentioned above. 

\begin{figure}
\hspace{-2cm}
\begin{center}
\includegraphics[width=6.5cm,keepaspectratio,clip]{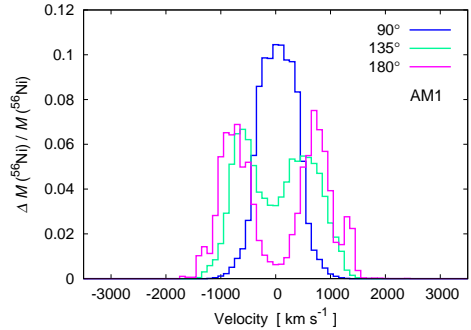}
\caption{
Mass distributions of $^{56}$Ni as a function of line of sight velocity at the end of 
simulation time (5752 s) for model AM1. The mass distributions of three observer 
angles $\theta_{\rm ob}$ = 90$^{\circ}$, 135$^{\circ}$, and 180$^{\circ}$ are shown. 
$\Delta M$\,($^{56}$Ni) is the mass of $^{56}$Ni in the velocity range of 
$v \sim v+\Delta v$. $M$\,($^{56}$Ni) is the total mass of $^{56}$Ni. For binning 
of line of sight velocity, $\Delta v =$ 100 km s$^{-1}$ is adopted.}
\label{fig:line_AM1}
\end{center}
\end{figure}

The mass distributions of $^{56}$Ni as a function of line of sight velocity 
at the end of simulation time for model AM1 are shown in Figure~\ref{fig:line_AM1}. 
The mass distributions of three observer angles $\theta_{\rm ob}$ = 90$^{\circ}$, 
135$^{\circ}$, and 180$^{\circ}$ are given. Note that the vertical values are linearly 
scaled and the shapes of the mass distributions approximately 
correspond to the observable line profiles of [Fe II]. Appearances of mass distributions 
are rather different by the observer angles as expected. 
If $\theta_{\rm ob} =$ 90$^{\circ}$, the distribution is well symmetric across 
the null velocity point. and the distribution concentrates around the point. 
If the observer angle is 180$^{\circ}$ and the bipolar explosion is seen head on, 
the distribution prominently split into the red-shifted and blue-sifted sides. 
The peaks locate around the line of sight velocities of $\pm$ 1,000 km s$^{-1}$. 
In the case of $\theta_{\rm ob} =$ 135$^{\circ}$, the split distribution is relatively 
moderate but we can recognize distinct double peaks. 
Note that other aspherical explosion models AP1 to AP8 have basically same features. 
Even if the head-on explosion is seen by an observer, the tails are extended only up to 
values of $\pm$ 1,500 km s$^{-1}$. As stated in \S 1, the observed line profile of [Fe II] 
in SN~1987A is asymmetric across the peak of the flux distribution \citep{haa90}. 
Therefore, the morphology of the explosion of SN~1987A may not be a simple 
bipolar explosion symmetric across the equatorial plane. 

\subsection{Results of revisiting the best model in Nagataki et al.}

\begin{figure*}[htbp]
\begin{tabular}{cc}
\hspace{-0.5cm}
\begin{minipage}{0.5\hsize}
\begin{center}
\includegraphics[width=6cm,keepaspectratio,clip]{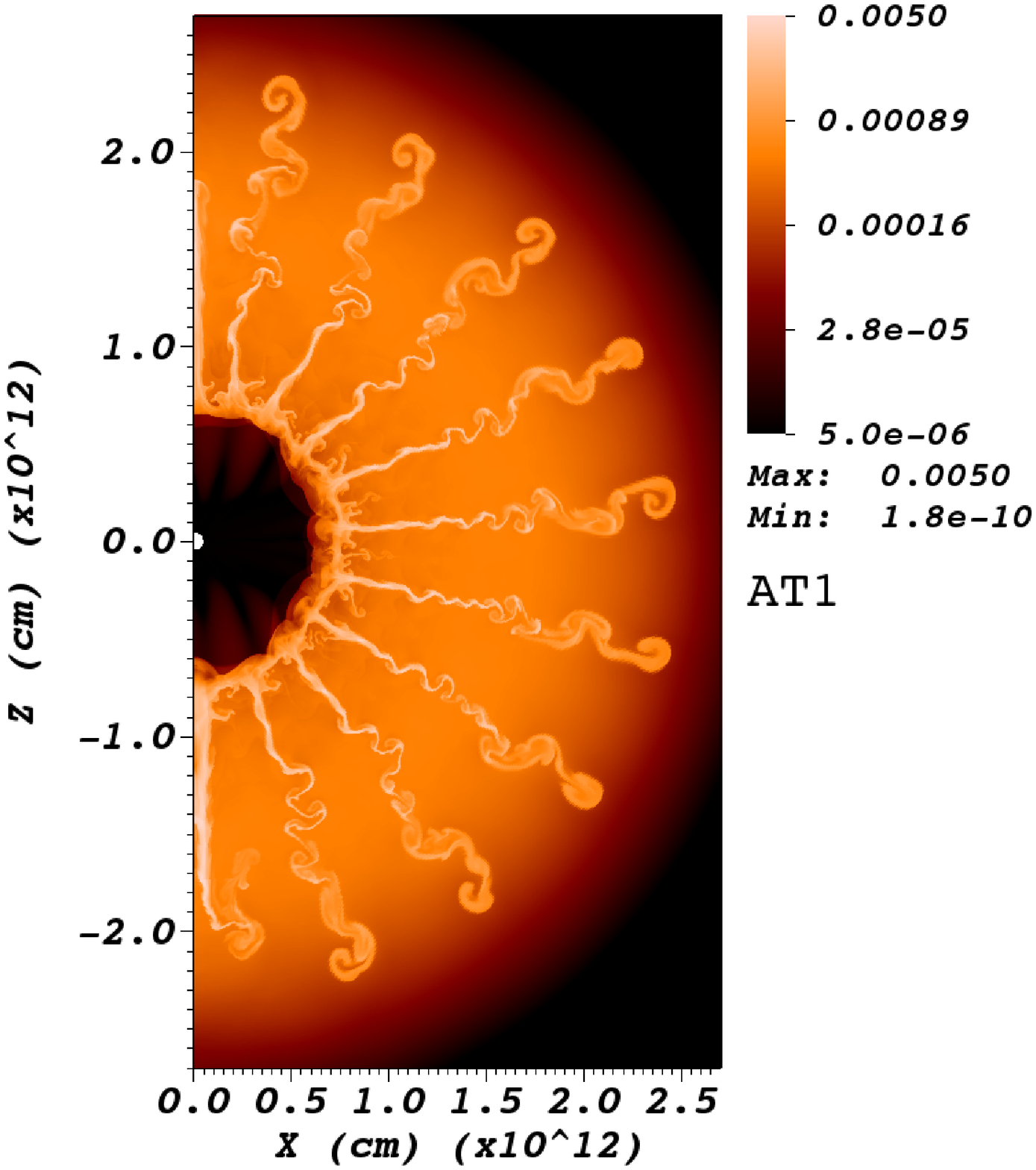}
\end{center}
\end{minipage}
\begin{minipage}{0.5\hsize}
\begin{center}
\includegraphics[width=6cm,keepaspectratio,clip]{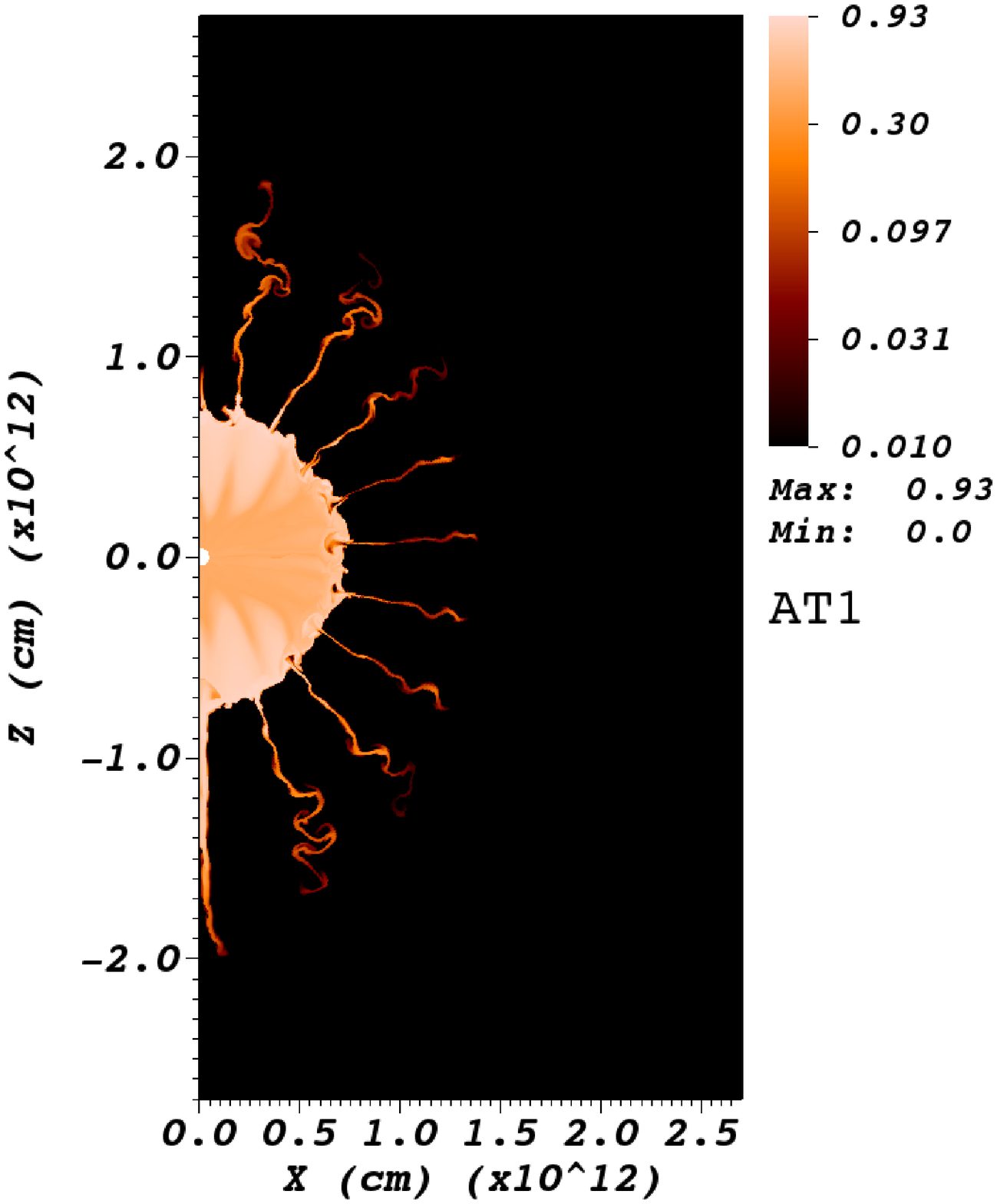}
\end{center}
\end{minipage}
\\
\hspace{-1cm}
\begin{minipage}{0.5\hsize}
\begin{center}
\includegraphics[width=6cm,keepaspectratio,clip]{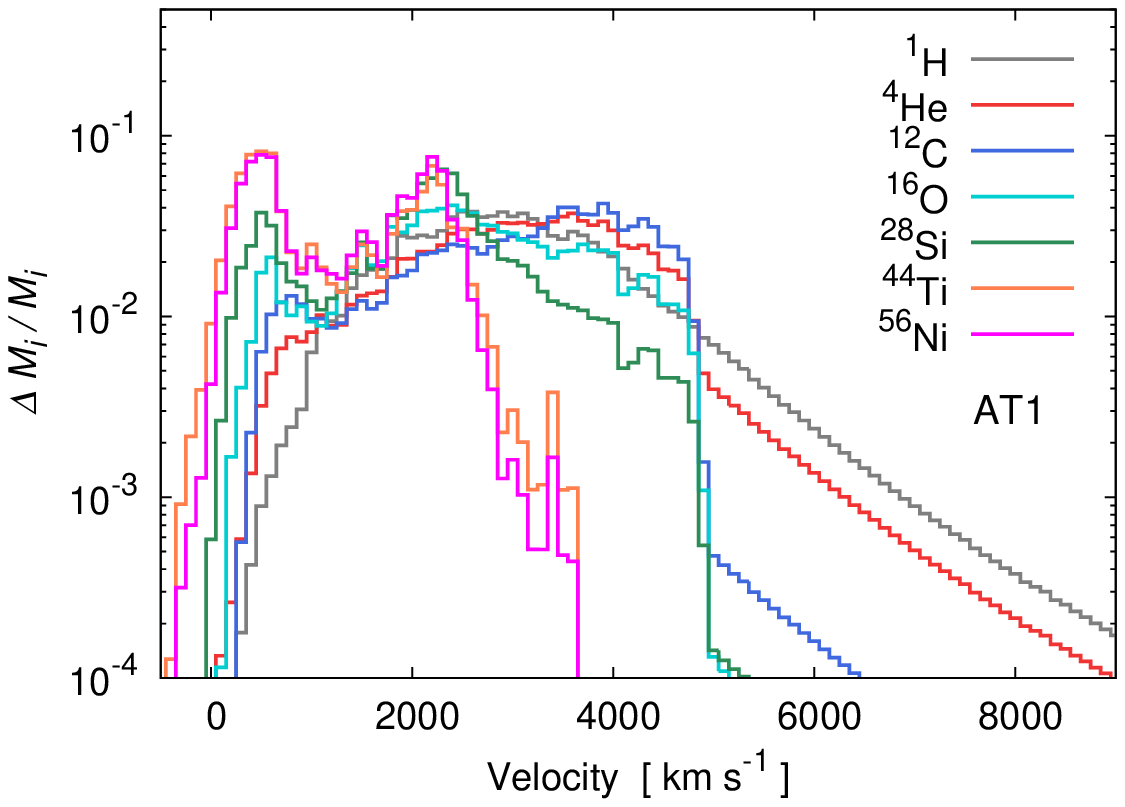}
\end{center}
\end{minipage}
\begin{minipage}{0.5\hsize}
\begin{center}
\includegraphics[width=6cm,keepaspectratio,clip]{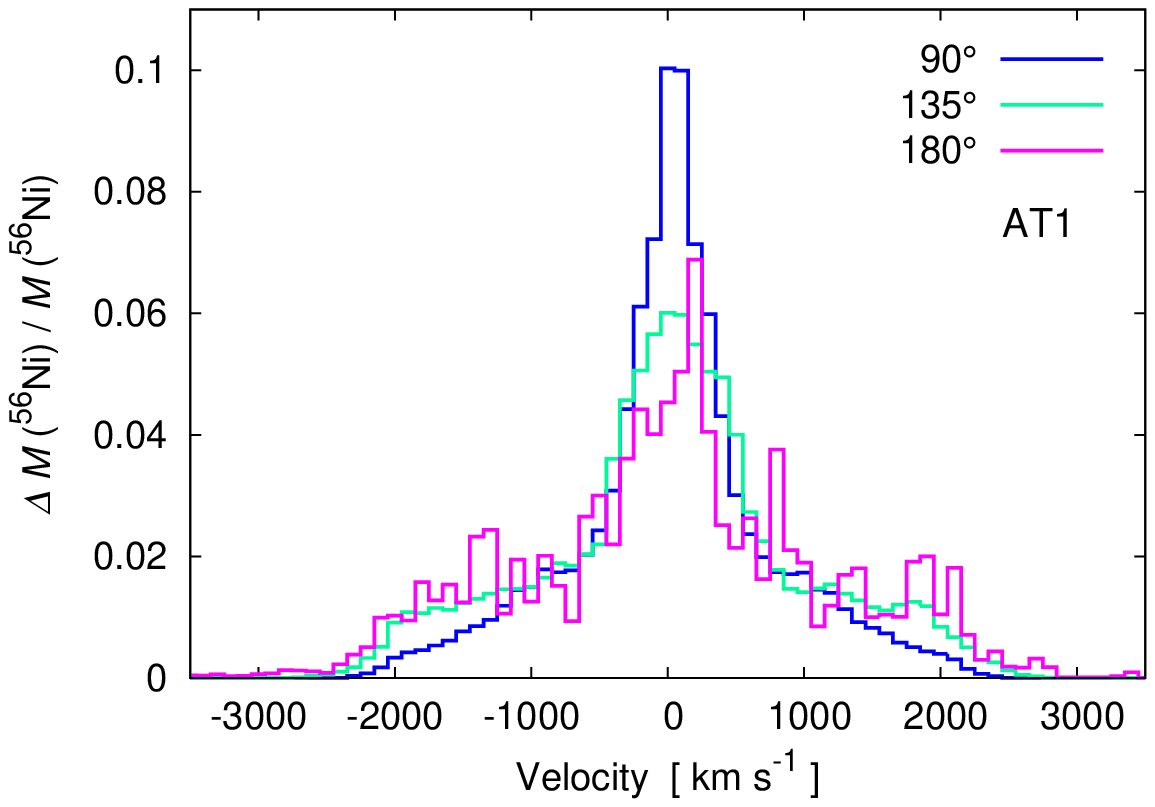}
\end{center}
\end{minipage}
\end{tabular}
\caption{Results of model AT1 at the time of 5060 s, the density distribution (top left), 
the distribution of the mass fraction of $^{56}$Ni (top right), the mass distributions 
of elements as a function of radial velocity (bottom left), and the mass distributions 
of $^{56}$Ni as a function of the line of sight velocity (bottom right).}
\label{fig:AT1}
\end{figure*}

This section is devoted to present the results of model AT1, AT2, and AS1. 
The summary of the results of model AT1 is shown in Figure~\ref{fig:AT1}. 
We can see marked RT fingers in the density distribution shown in the top left panel. 
$^{56}$Ni is distributed inside both the dense shell around the radius 
of 0.7 $\times$ 10$^{12}$ cm and inner regions of RT fingers. The number of RT fingers 
are consistent with the wave lengths of imposed perturbations, i.e., the parameter $m$. 
Strong mixing of metals $^{56}$Ni, $^{44}$Ti, $^{28}$Si, $^{16}$O, and $^{12}$C is seen 
in the bottom left panel of Figure~\ref{fig:AT1}. The obtained maximum velocity 
of $^{56}$Ni with $\Delta M$\,($^{56}$Ni)/$M$\,($^{56}$Ni) $>$ 1 $\times$ 10$^{-3}$ 
is 3,300 km s$^{-1}$ (see Table~\ref{table:results}), which is roughly consistent with 
that in \citet{nag00}. A small fraction of $^{56}$Ni with 
$\Delta M$\,($^{56}$Ni)/$M$\,($^{56}$Ni) $>$  1 $\times$ 10$^{-4}$ reaches velocity 
of 3,500 km s$^{-1}$. Strong inward mixing of $^{1}$H is also seen. The obtained 
minimum velocity of $^{1}$H is 500 km s$^{-1}$. The mass distributions of $^{56}$Ni 
as a function of line of sight velocity are depicted in the bottom right panel of 
Figure~\ref{fig:AT1}. For all observer angles, the tails of mass distributions are 
extended around $\pm$ 3,000 km s$^{-1}$. Sharp decays of the distributions across 
$\pm$ 1,000 km s$^{-1}$ are seen. These are somewhat different from the observed 
smooth flux distributions of [Fe II] in SN~1987A. The sharp decays of the distributions 
is also somewhat different from the distribution seen in model A1 in Nagataki et al. 
(see e.g., Fig.~14 in \citet{nag00}) wherein the smoother decay than those 
of model AT1 are seen. The differences may be attributed to the different hydrodynamic 
code used and the different resolutions of simulations. 

\begin{figure*}[htbp]
\begin{tabular}{cc}
\hspace{-0.5cm}
\begin{minipage}{0.5\hsize}
\begin{center}
\includegraphics[width=6cm,keepaspectratio,clip]{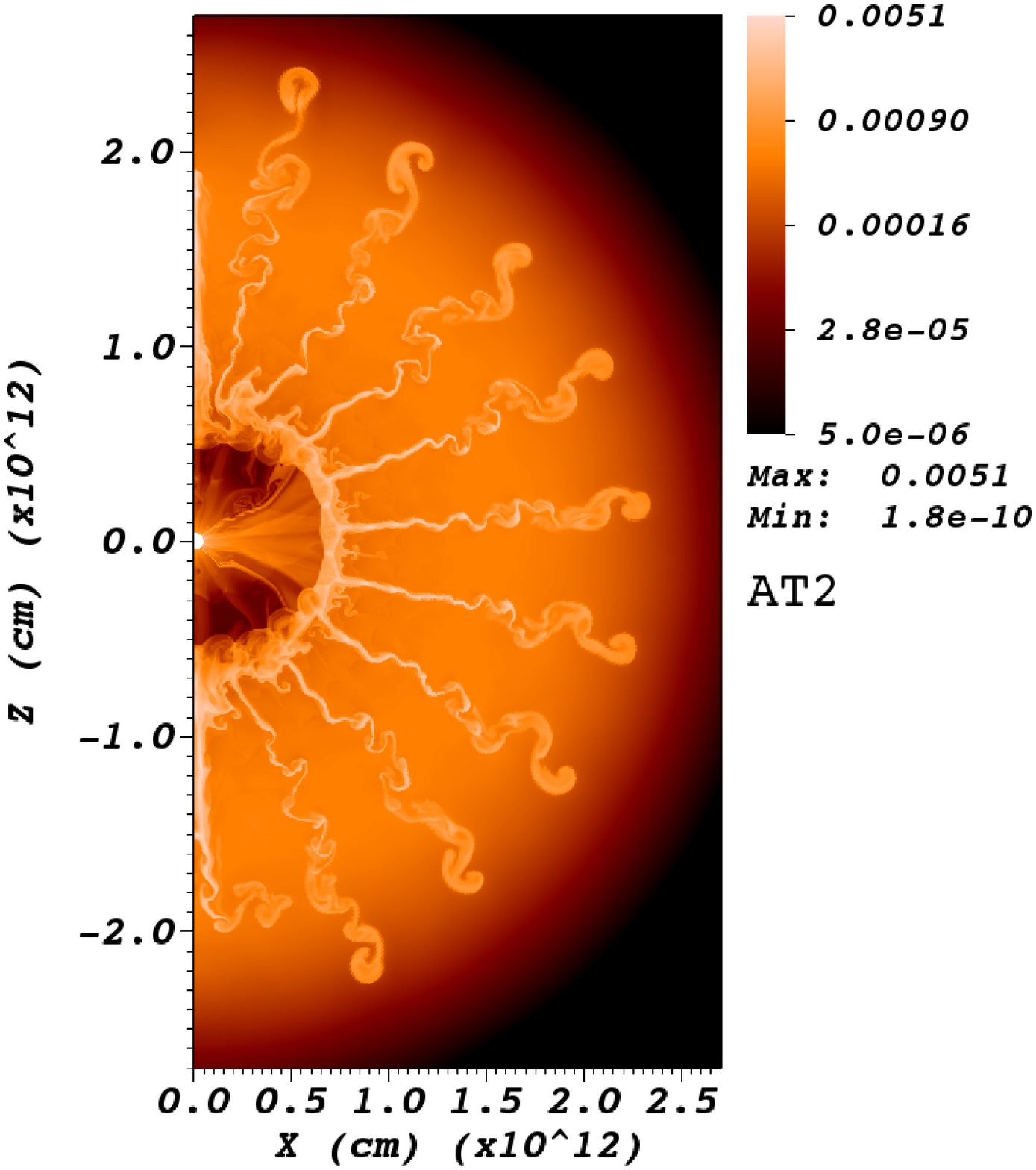}
\end{center}
\end{minipage}
\begin{minipage}{0.5\hsize}
\begin{center}
\includegraphics[width=6cm,keepaspectratio,clip]{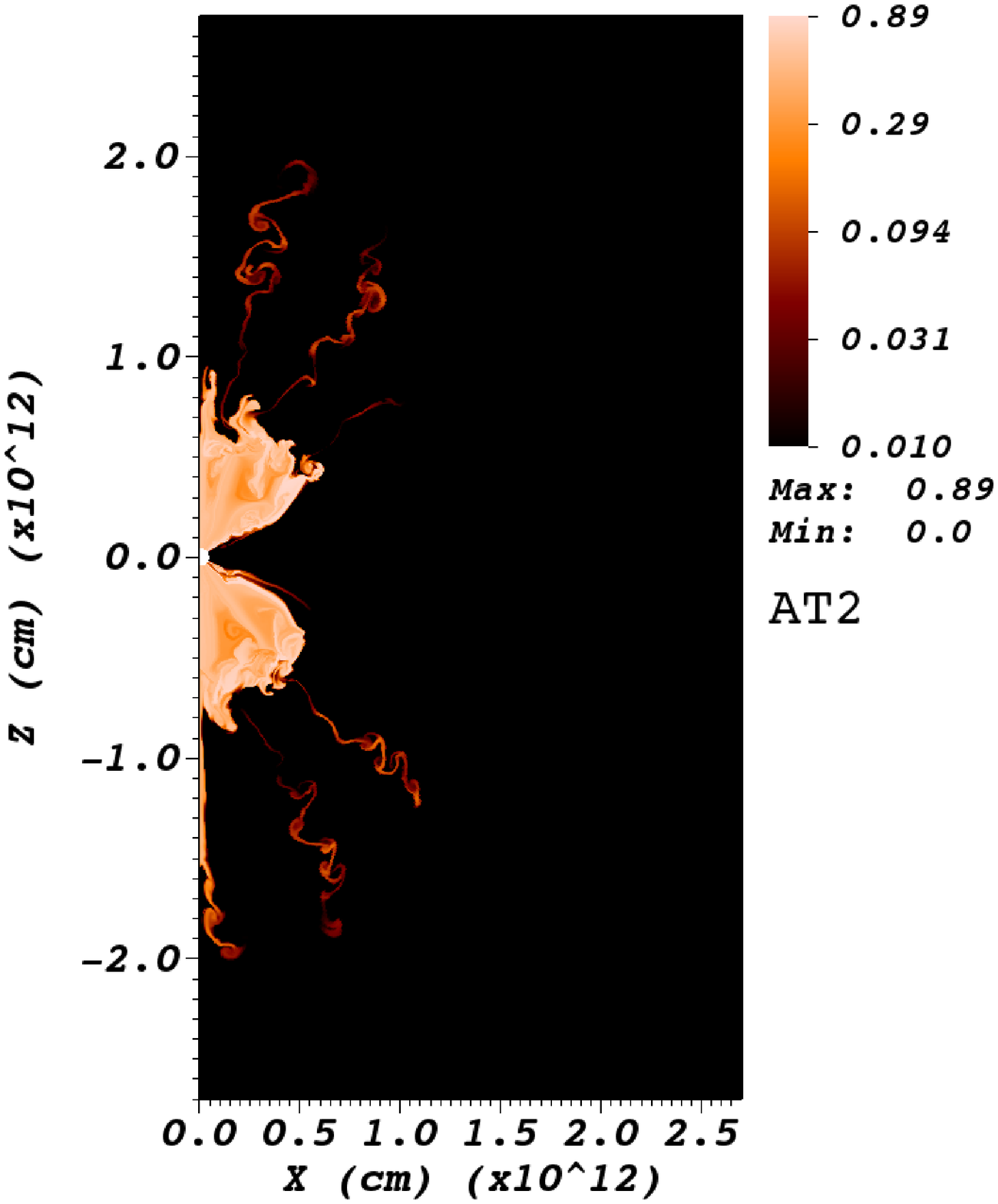}
\end{center}
\end{minipage}
\\
\hspace{-1cm}
\begin{minipage}{0.5\hsize}
\begin{center}
\includegraphics[width=6cm,keepaspectratio,clip]{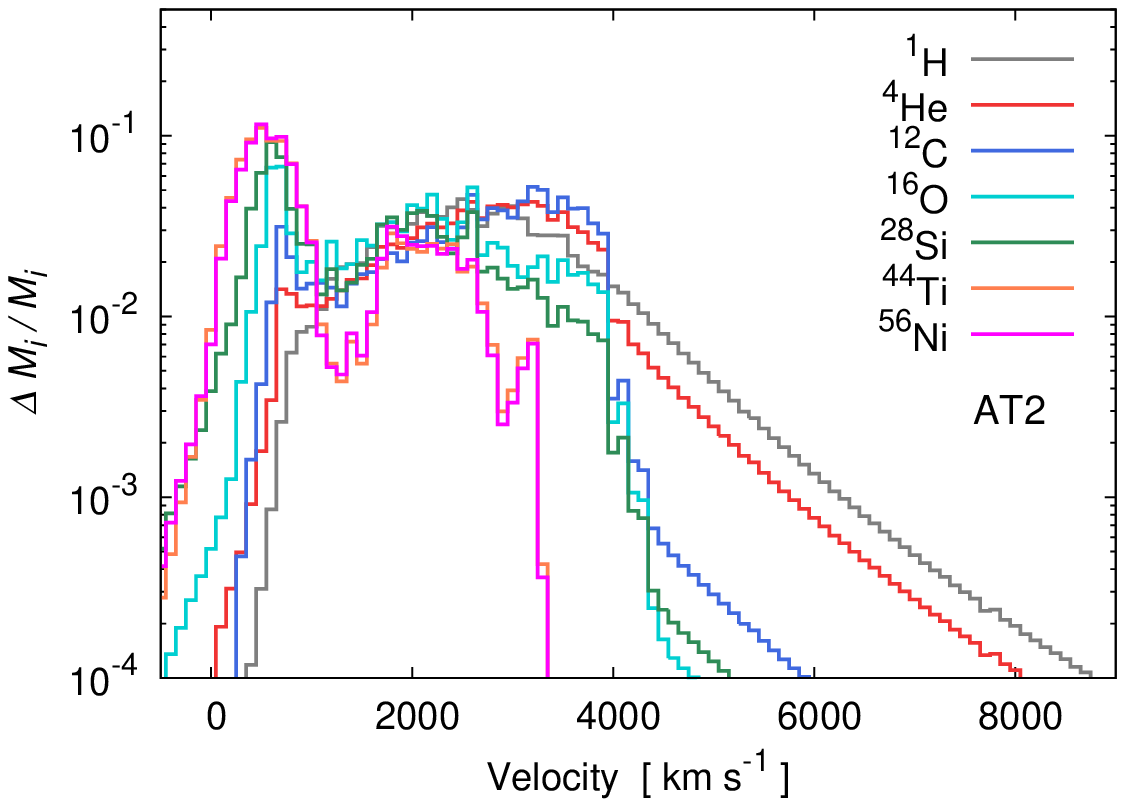}
\end{center}
\end{minipage}
\begin{minipage}{0.5\hsize}
\begin{center}
\includegraphics[width=6cm,keepaspectratio,clip]{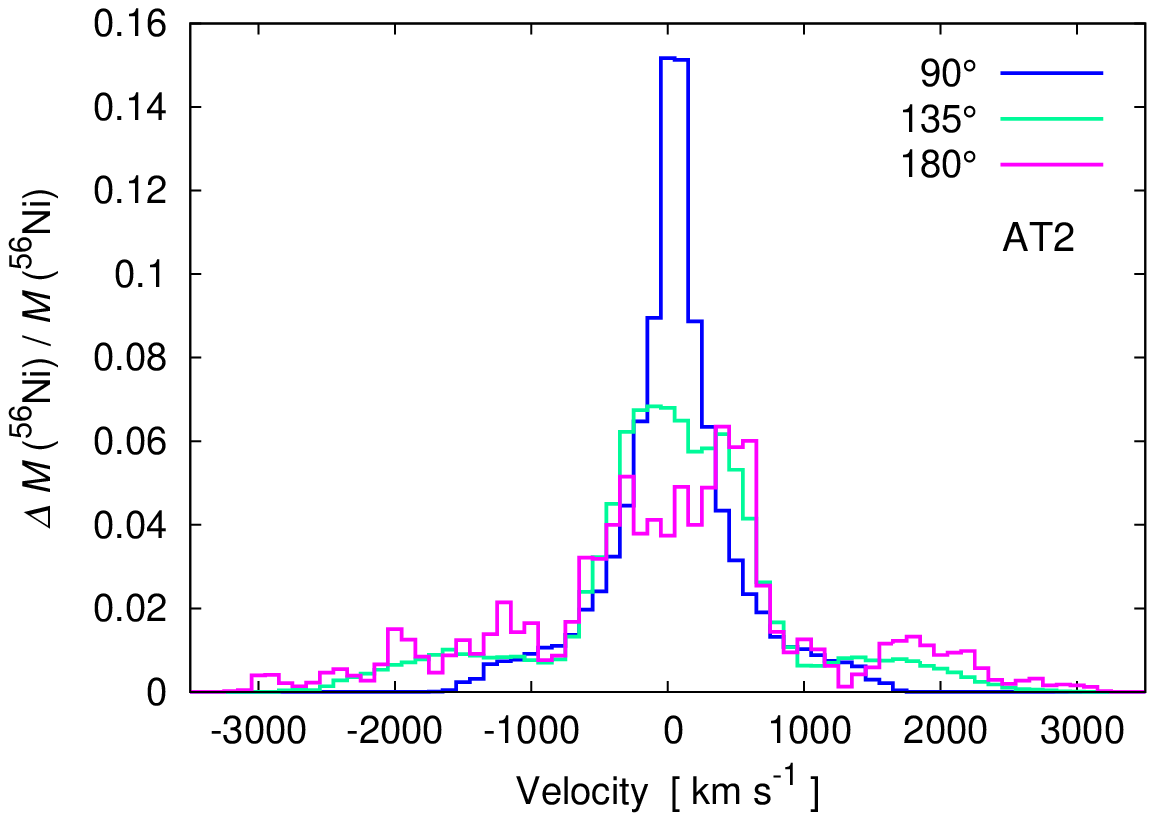}
\end{center}
\end{minipage}
\end{tabular}
\caption{Same as Figure~\ref{fig:AT1} but for model AT2 and the time of 5567 s.}
\label{fig:AT2}
\end{figure*}

The summary of the model AT2 results is shown in Figure~\ref{fig:AT2}. 
The `diode' boundary condition is employed for the inner radial boundary 
in later phases and gravity is turned on as noted in \S 3.3. The appearance 
of RT fingers is quite similar to that of model AT1 (the top left panel). 
However, the density distribution of inner regions is different from that 
of model AT1 due to the effects of fallback. The distribution of $^{56}$Ni is also 
different from that of model AT1. $^{56}$Ni is distributed only in regions apart from 
the equatorial plane. The mass distributions of elements as a function of radial velocity 
are shown in the bottom left panel of Figure~\ref{fig:AT2}. These mass distributions are 
similar to those of model AT1. The obtained maximum velocity of $^{56}$Ni 
is 3,100 km s$^{-1}$, which is somewhat reduced compared with that 
of model AT1 (see Table~\ref{table:results}). The minimum velocity of $^1$H (600 km s$^{-1}$) 
is similar to that of model AT1. From the bottom right panel of Figure~\ref{fig:AT2}, 
we see that the distributions of $^{56}$Ni as a function of line of sight velocity 
are clustered around the null velocity point compared with those in model AT1. 
From above results, even if the effects of fallback are included in the simulation, 
the high velocity of $^{56}$Ni can be reproduced by model AT2. 
However, as \citet{nag00} stated, such large perturbations (amplitude of 30\%) 
might not be introduced in the pre-collapse star. Hence, we calculate the model AS1 
that has same setups and model parameters as model AT2 but perturbations 
(amplitude of 30\%) are introduced in the initial radial velocities. 
Hereafter, we show the results of model AS1. 

\begin{figure*}[htbp]
\begin{tabular}{cc}
\hspace{-0.5cm}
\begin{minipage}{0.5\hsize}
\begin{center}
\includegraphics[width=6cm,keepaspectratio,clip]{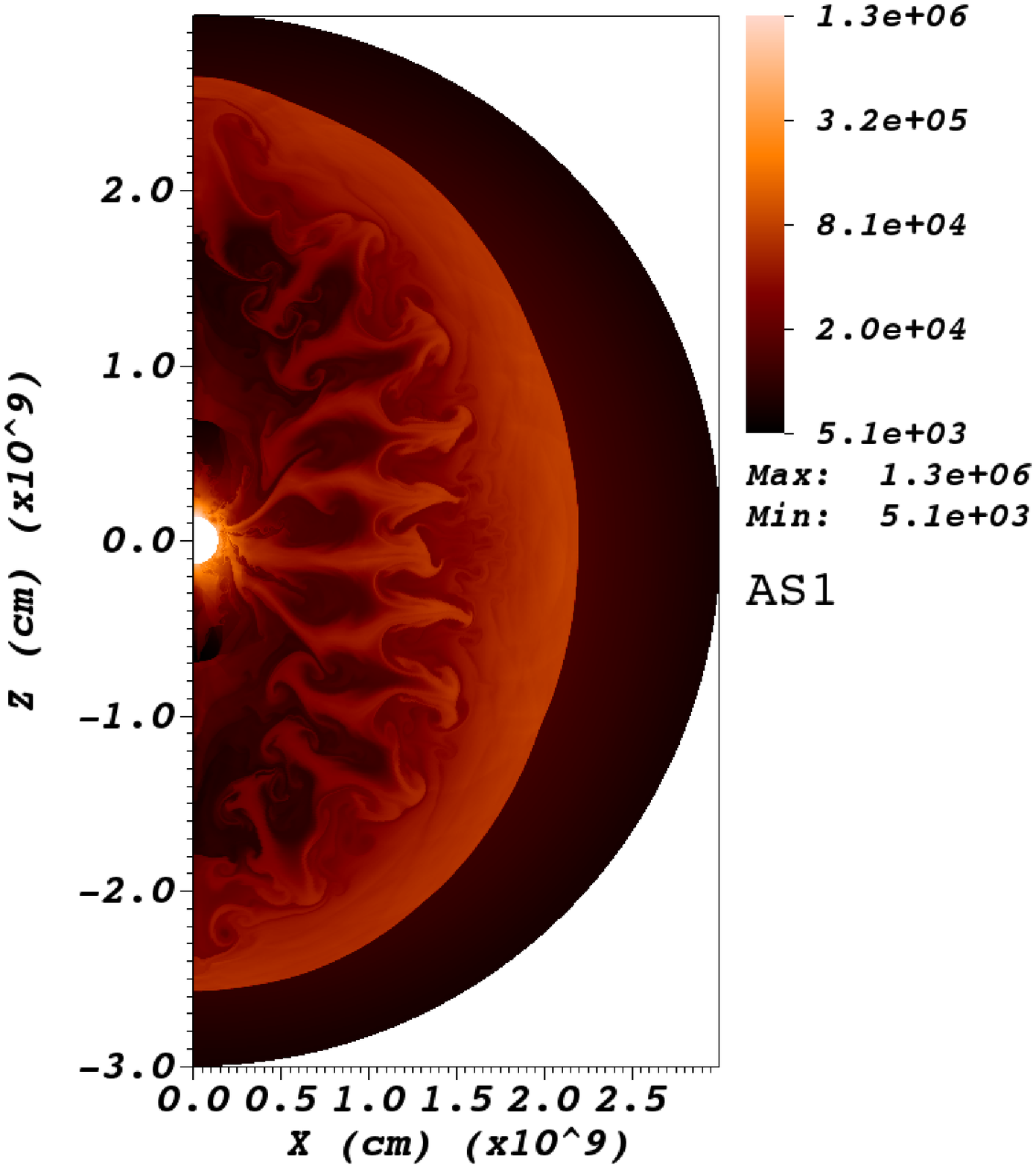}
\end{center}
\end{minipage}
\begin{minipage}{0.5\hsize}
\begin{center}
\includegraphics[width=6cm,keepaspectratio,clip]{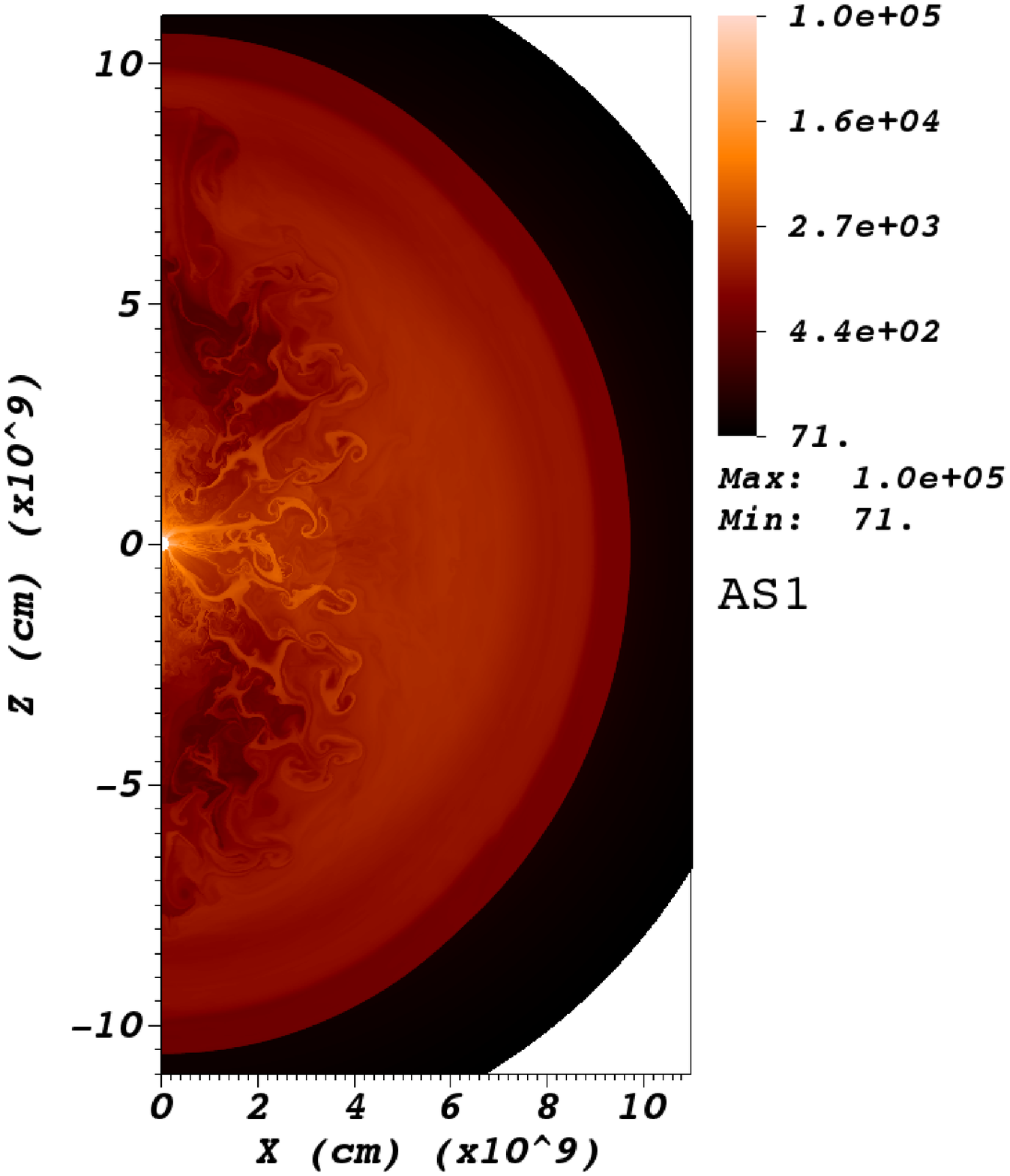}
\end{center}
\end{minipage}
\\
\hspace{-0.5cm}
\begin{minipage}{0.5\hsize}
\begin{center}
\includegraphics[width=6cm,keepaspectratio,clip]{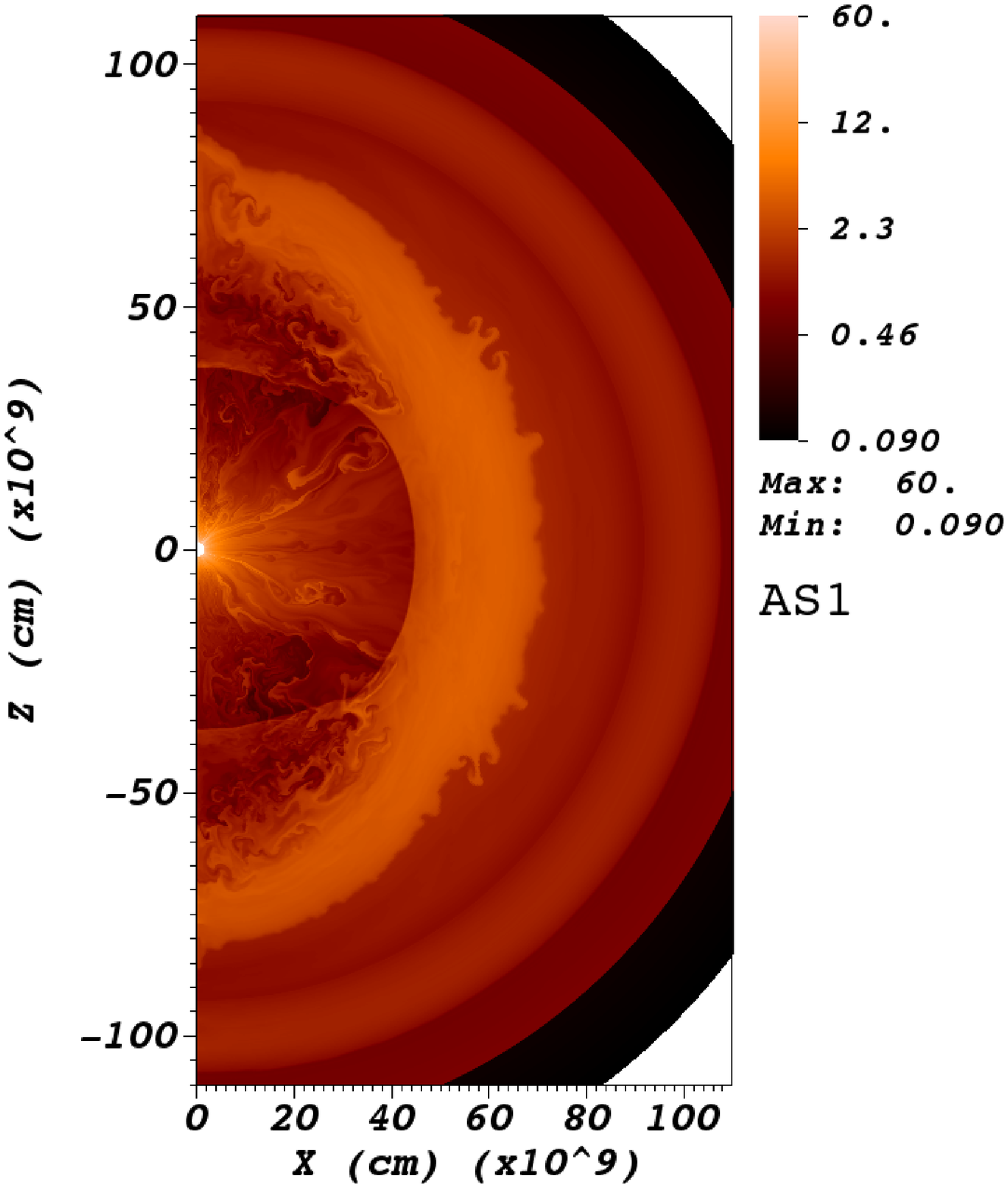}
\end{center}
\end{minipage}
\begin{minipage}{0.5\hsize}
\begin{center}
\includegraphics[width=6cm,keepaspectratio,clip]{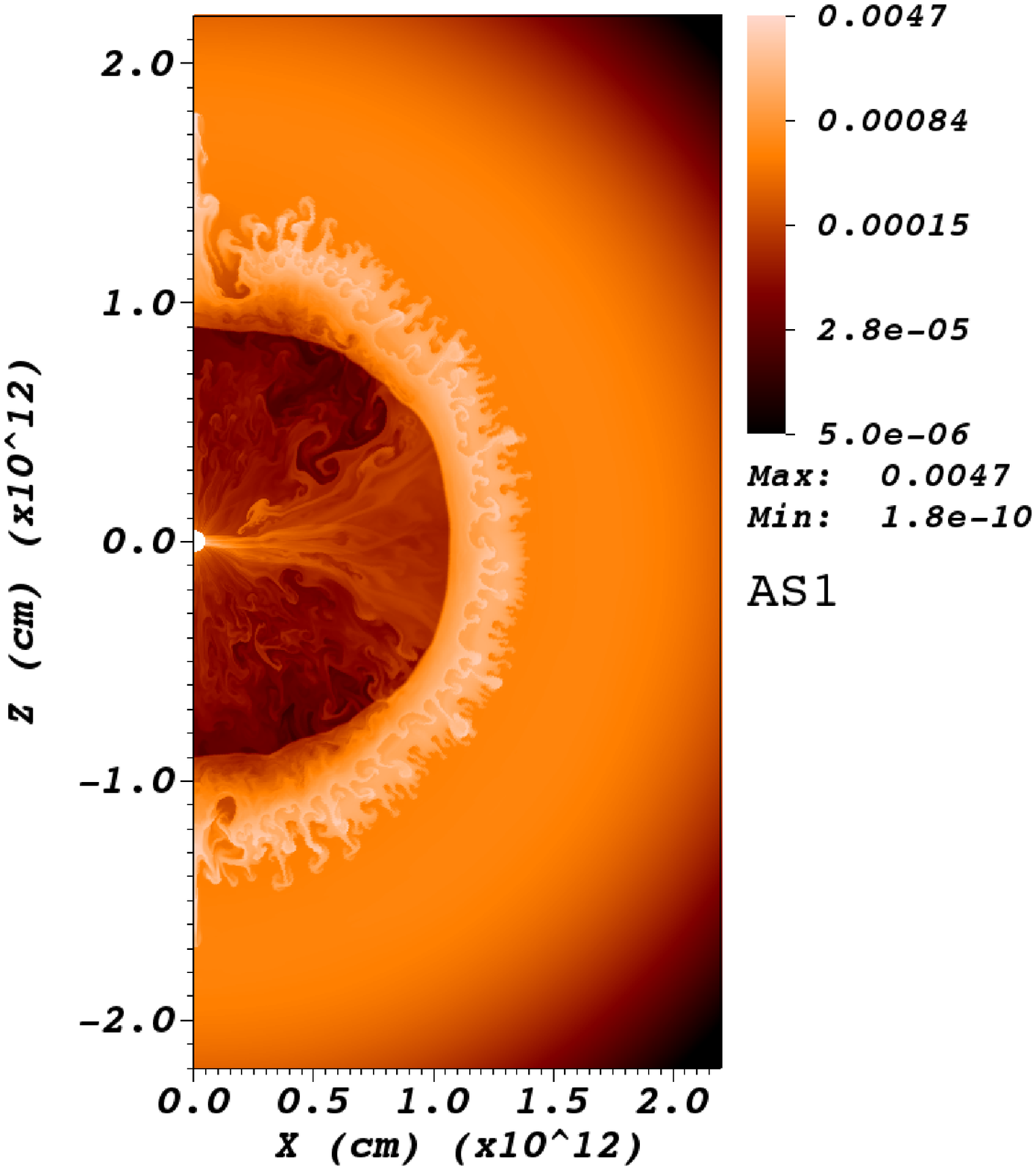}
\end{center}
\end{minipage}
\end{tabular}
\caption{Snap shots of distributions of density at the time of 
1.36 s (top left), 8.57 s (top right), 147.4 s (bottom left), and 5753 s (bottom right) for model AS1. 
The unit of values in color bars is g cm$^{-3}$. The values in color bars are logarithmically scaled.}
\label{fig:dens_AS1}
\end{figure*}

\begin{figure}
\hspace{-2cm}
\begin{center}
\includegraphics[width=6cm,keepaspectratio,clip]{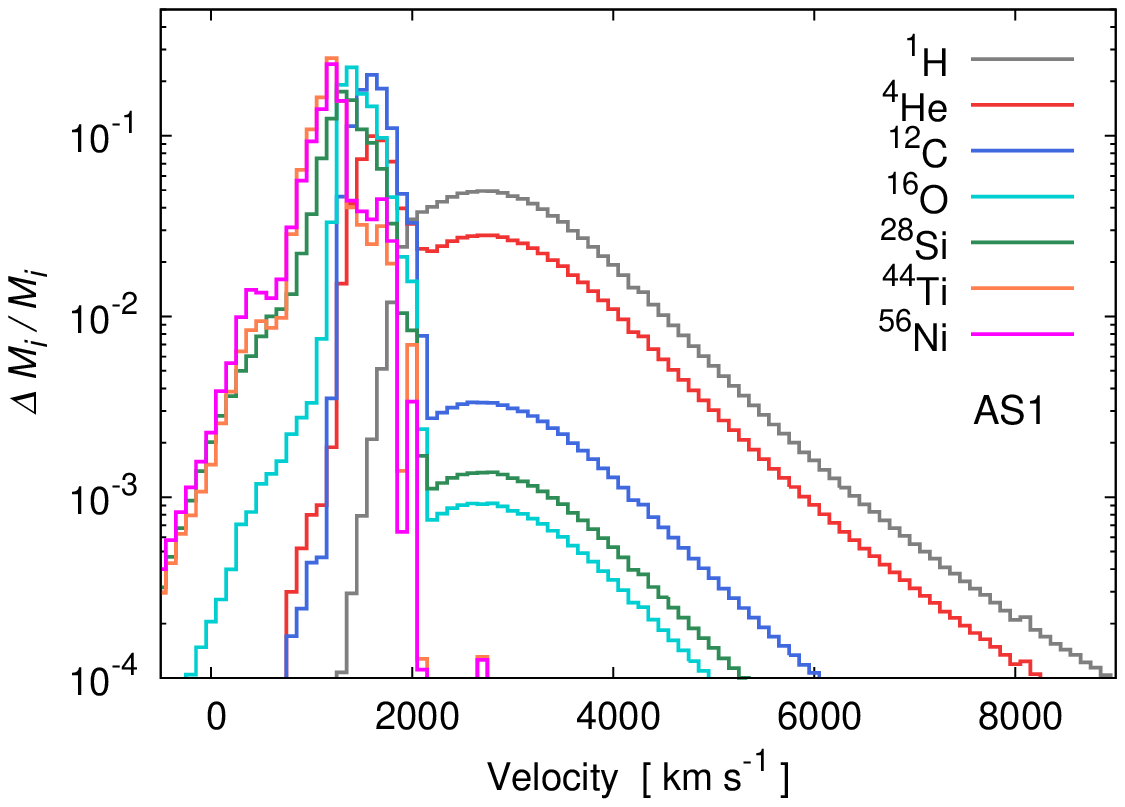}
\caption{Same as Figure~\ref{fig:vel_SPM} but for model AS1 at the time of 5753 s.}
\label{fig:vel_AS1}
\end{center}
\end{figure}

\begin{figure}
\hspace{-2cm}
\begin{center}
\includegraphics[width=6.5cm,keepaspectratio,clip]{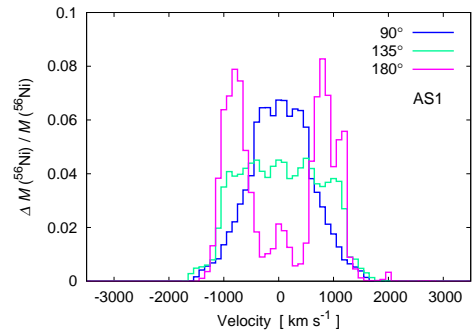}
\caption{Same as Figure~\ref{fig:line_AM1} but for model AS1 and the time of 5753 s}
\label{fig:line_AS1}
\end{center}
\end{figure}

In Figure~\ref{fig:dens_AS1}, we show the time evolution of density distribution of model AS1. 
Just after the initiation of the explosion (1.36 s after the explosion), outward finger structures, 
which is attributed to the imposed large perturbations, are clearly seen (the top left panel). 
We can also recognize inward finger structures adjacent to outward ones. 
The inward mixing may be caused by RT instabilities due to the inward gravitational force. 
However, after that, the finger structures are gradually broken up due to KH instability (the top right panel). 
Along the polar axis, relatively large-sale protrusions of inner matter are seen. 
This occurs physically because the explosion along the polar regions is the strongest, 
but this may be partly affected by numerical errors around the polar axis. 
After the formation of the dense helium shell, the fingers are almost destroyed due to 
the collision with the dense shell (the bottom left panel). Eventually, no protrusion of innermost 
metals is seen except for polar regions (the bottom right panel). 
The mass distributions of elements as a function of radial velocity are shown 
in Figure~\ref{fig:vel_AS1}. All metals $^{56}$Ni, $^{44}$Ti, $^{28}$Si, $^{12}$C, 
and $^{16}$O are limited at the velocity around 2,000 km s$^{-1}$, which 
corresponds to around the bottom of the dense helium shell. A part of innermost 
metals $^{56}$Ni and $^{44}$Ti can reach the dense shell but cannot penetrate the shell. 
The mass distributions of $^{56}$Ni as a function of line of sight velocity are shown 
in Figure~\ref{fig:line_AS1}. In all observer angles, a clear cut off of velocity around 
$\pm$ 1,500 km s$^{-1}$ is seen. The maximum radial velocity of $^{56}$Ni 
is 1,900 km s$^{-1}$ and the minimum radial velocity of $^{1}$H 
is 1,500 km s$^{-1}$. A strong inward mixing of $^1$H does not occur in this model. 

From the results in this section, we summarize as follows. The high velocity of $^{56}$Ni 
seen in models AT1 and AT2 cannot be reproduced if the same perturbations are imposed 
in the initial radial velocities. The initial perturbations cannot retain the structures 
in later phases in which RT instability around the composition interface of He/H grows. 
In other wards, if such structures remain and/or exist due to some unknown reasons, 
such high velocity of $^{56}$Ni might be reproduced. In the next section, we focus 
on the models in which large perturbations are introduced in initial radial velocities. 

\subsection{Aspherical explosions with clumpy structures}

\begin{figure*}[htbp]
\begin{tabular}{cc}
\hspace{-0.5cm}
\begin{minipage}{0.5\hsize}
\begin{center}
\includegraphics[width=6cm,keepaspectratio,clip]{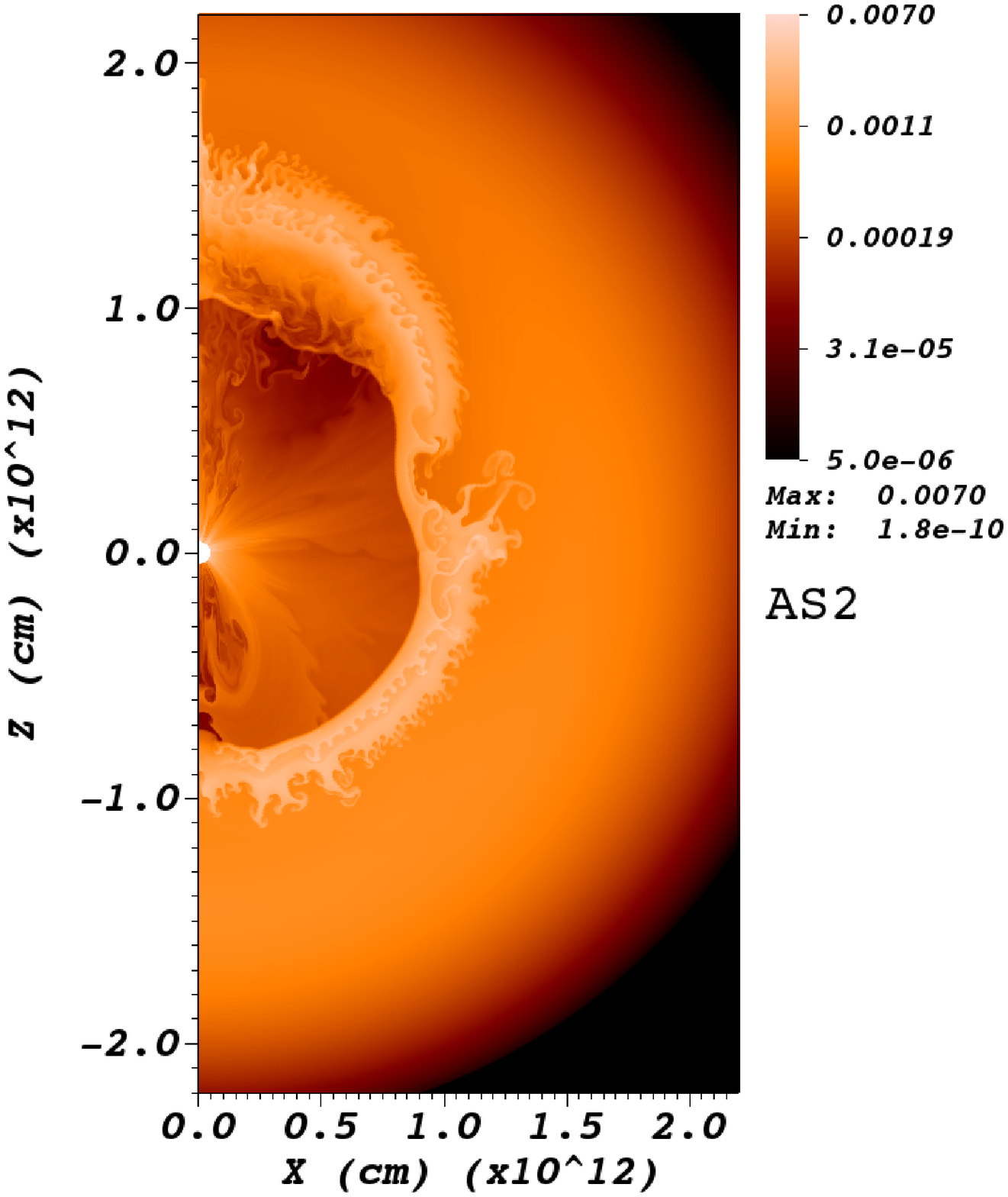}
\end{center}
\end{minipage}
\begin{minipage}{0.5\hsize}
\begin{center}
\includegraphics[width=6cm,keepaspectratio,clip]{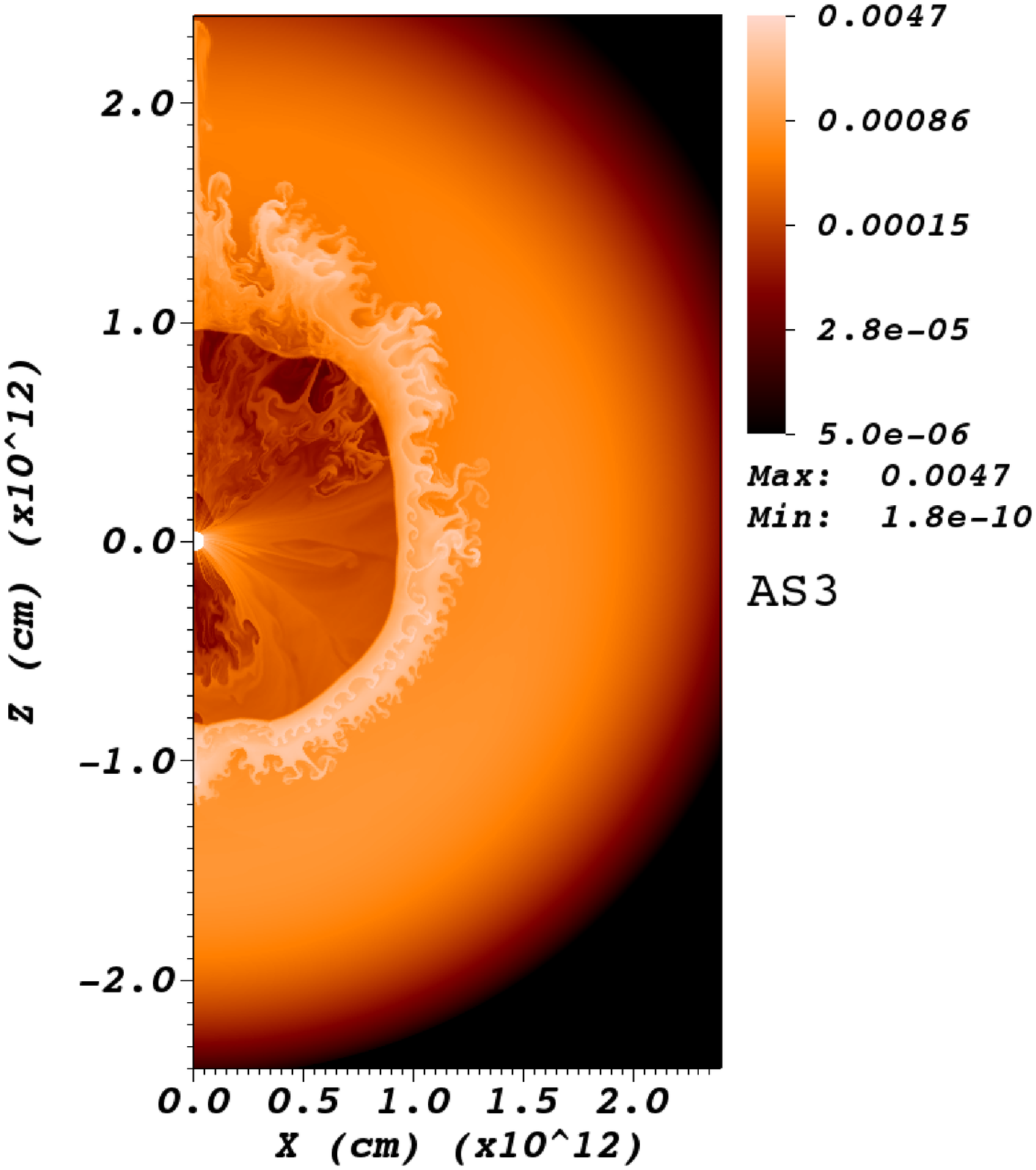}
\end{center}
\end{minipage}
\\
\hspace{-0.5cm}
\begin{minipage}{0.5\hsize}
\begin{center}
\includegraphics[width=6cm,keepaspectratio,clip]{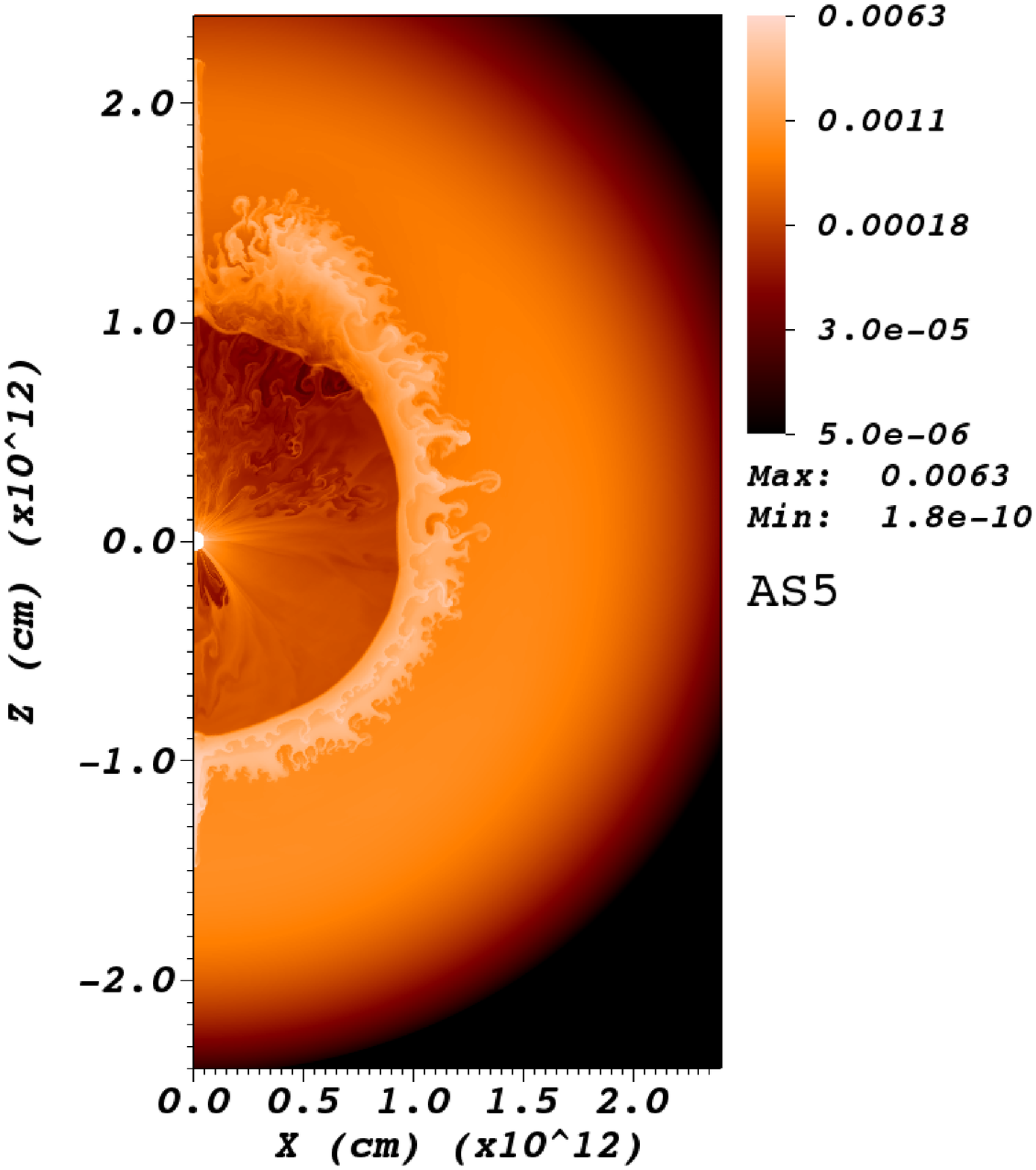}
\end{center}
\end{minipage}
\begin{minipage}{0.5\hsize}
\begin{center}
\includegraphics[width=6cm,keepaspectratio,clip]{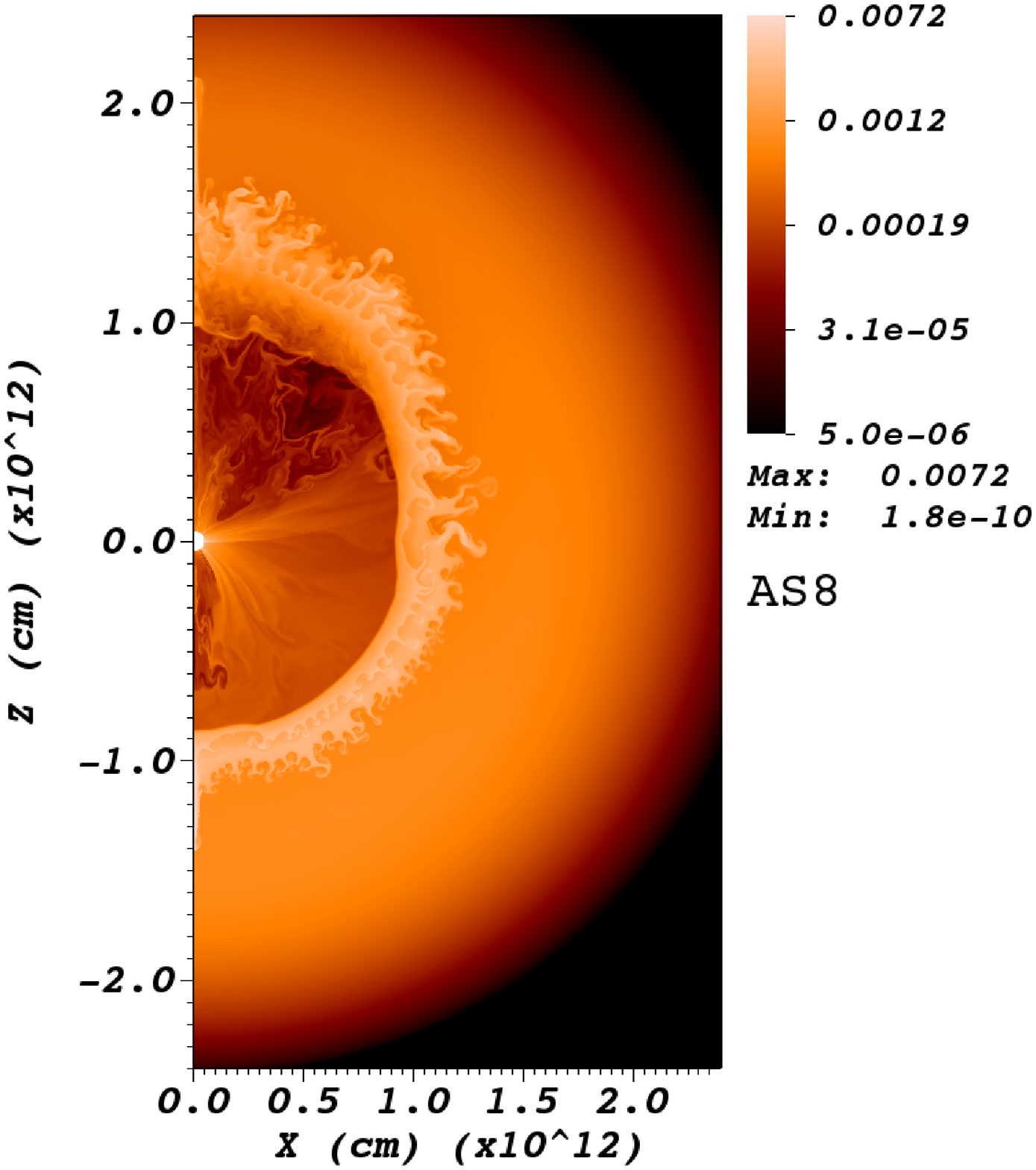}
\end{center}
\end{minipage}
\end{tabular}
\caption{Same as Figure~\ref{fig:dens_SPM} but for models 
AS2 (top left), AS3 (top right), AS5 (bottom left), and AS8 (bottom right)
and at the time of 5548 s, 5250 s, 5225 s, and 5231 s, respectively.}
\label{fig:dens_AS-1}
\end{figure*}

\begin{figure*}[htbp]
\begin{center}
\includegraphics[width=13cm,keepaspectratio,clip]{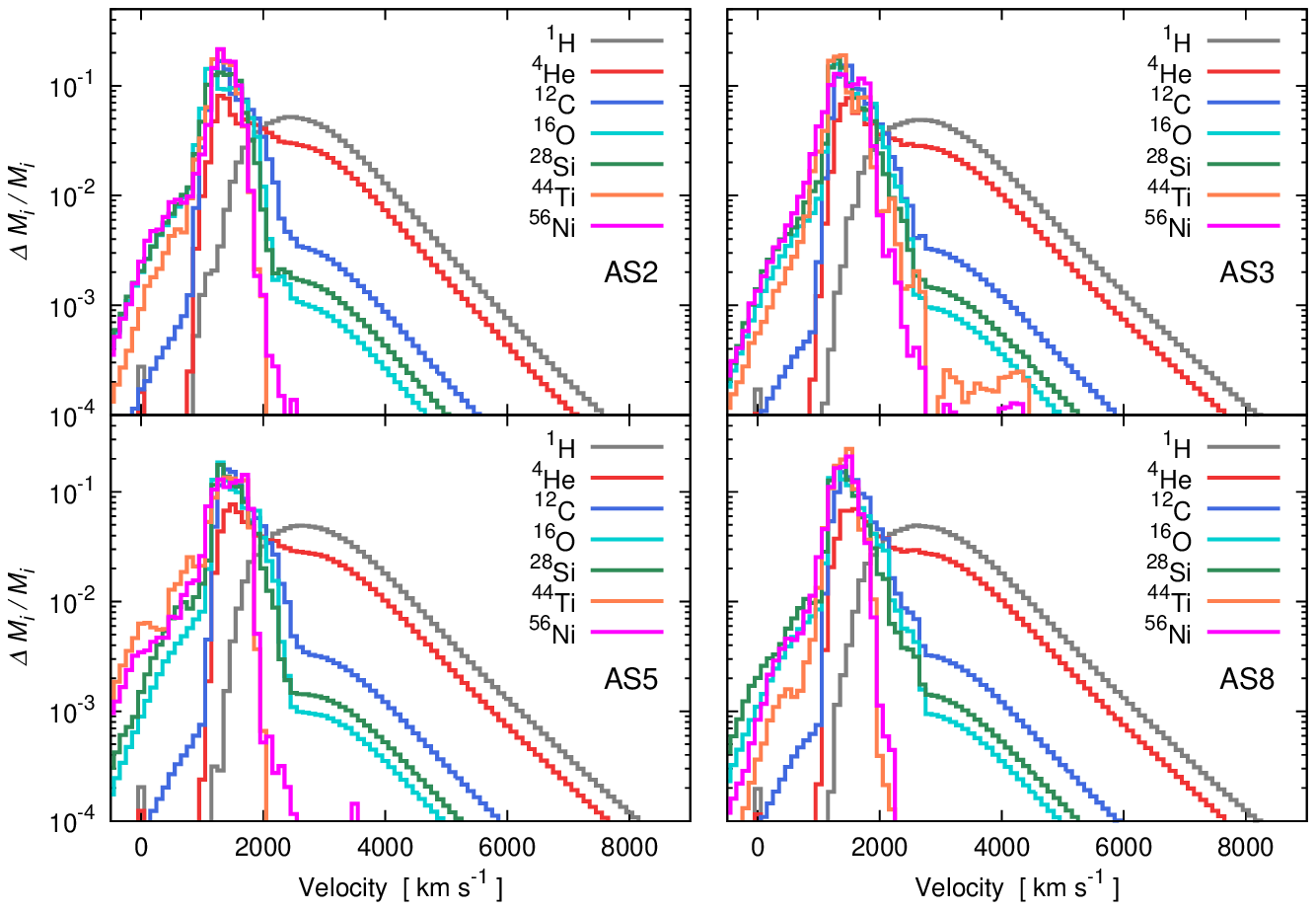}
\caption{Same as Figure~\ref{fig:vel_SPM} but for models AS2 (top left), AS3 (top right), 
AS5 (bottom left), AS8 (bottom right) and at the time of 5548 s, 5250 s, 5225 s, and 5231 s, 
respectively.}
\label{fig:vel_AS}
\end{center}
\end{figure*}

In this section, we show the results of aspherical explosion models with clumpy structures. 
In the previous sections, we consider bipolar explosions. In this section, explosions are 
also asymmetric across the equatorial plane, i.e, $v_{\rm up}/ v_{\rm down}$ = 2 
(see Table~\ref{table:models}). In models AS2 to AS8, we change the size of clumpy 
structure in the initial shock waves by setting different parameter $m$ in Equation 
(\ref{eq:clump}). The density distributions at the ends of simulation time for 
models AS2, AS3, AS5, and AS8 as representative models are shown 
in Figure~\ref{fig:dens_AS-1}. In all models AS2 to AS8, small-scale RT fingers 
are developed around the bottom of the dense helium shell and RT fingers 
in the upper hemisphere are slightly longer than those in the lower one. 
For models AS2 to AS5, the configurations of the fingers are different from each other 
in the upper hemisphere. While, for models that have smaller-scale clumps, i.e., AS6 to AS8, 
the differences of the configurations of fingers are not distinctive. 
In models AS3 and AS5, prominent extended fingers are seen very close to the polar axis. 
This is a common problem seen in a two-dimensional axisymmetric hydrodynamic simulation. 
This problem is partly attributed to the effects that flows cannot penetrate across the symmetry 
axis and discretization errors around the axis. However, it reflects the physical nature 
that the explosion is strongest in regions close to the polar axis. 
Unfortunately, we hardly speculate how the features are realistic in a two-dimensional 
axisymmetric calculation. Figure~\ref{fig:vel_AS} depicts the mass distributions of elements 
as a function of radial velocity at the ends of simulation time for models AS2, AS3, AS5, and AS8. 
For models of relatively larger-scale clumps, AS2 to AS5, the maximum velocity 
of innermost metals $^{56}$Ni and $^{44}$Ti are affected by the sizes of clumpy structures. 
In model AS3, the high velocity tails of $^{56}$Ni and $^{44}$Ti are smoothly extended 
around 3,000 km s$^{-1}$ and a small amount of high velocity clumps 
(up to 4,000 km s$^{-1}$) is recognized. Model AS5 has also a slightly extended high 
velocity wing and a small amount of high velocity $^{56}$Ni clump. On the other hand, 
in models AS6 to AS8, the mass distributions are similar to each other and the maximum 
velocity of innermost metals are limited to around 2,000 km s$^{-1}$. 
From above results, we know that the size of clump may affect the protrusion 
of innermost metals and the clump with a relatively larger size tend to penetrate the 
dense helium shell more easily. However, it is difficult to find a monotonic behavior 
with respect to the penetration of innermost metals. The results are somewhat sensitive 
to the clump size. Additionally, we find that the high velocity clumps of $^{56}$Ni 
is clustered only in regions very close to the polar axis. Therefore, the high velocity 
clumps of $^{56}$Ni seen in models AS3 and AS5 are doubtful. 
It is noted that strong RM instabilities around the composition interface of He/H 
obtained by \citet{kif06} (see \S 1 and \S 3.4) are not confirmed in models AS2 to AS8. 
In fact, as summarized in Table~\ref{table:results}, the minimum radial velocities 
of $^1$H range between 1,200 to 1,300 km s$^{-1}$ except for that for model AS2 
(that is about 900 km s$^{-1}$). Therefore, strong inward mixing of $^1$H due to 
RM instabilities is not realized in models AS2 to AS8. The differences may be due to 
the following facts: the progenitor model, a 15 $M_{\odot}$ blue supergiant star 
(see Figure~8 in \citet{kif03}), is different from ours and our models do not duplicate 
some features of a neutrino-driven explosion model, such as initial angular velocities 
and their gradients, thermal and density structures, and so on. 

\begin{figure*}[htbp]
\begin{tabular}{cc}
\hspace{-0.5cm}
\begin{minipage}{0.5\hsize}
\begin{center}
\includegraphics[width=6cm,keepaspectratio,clip]{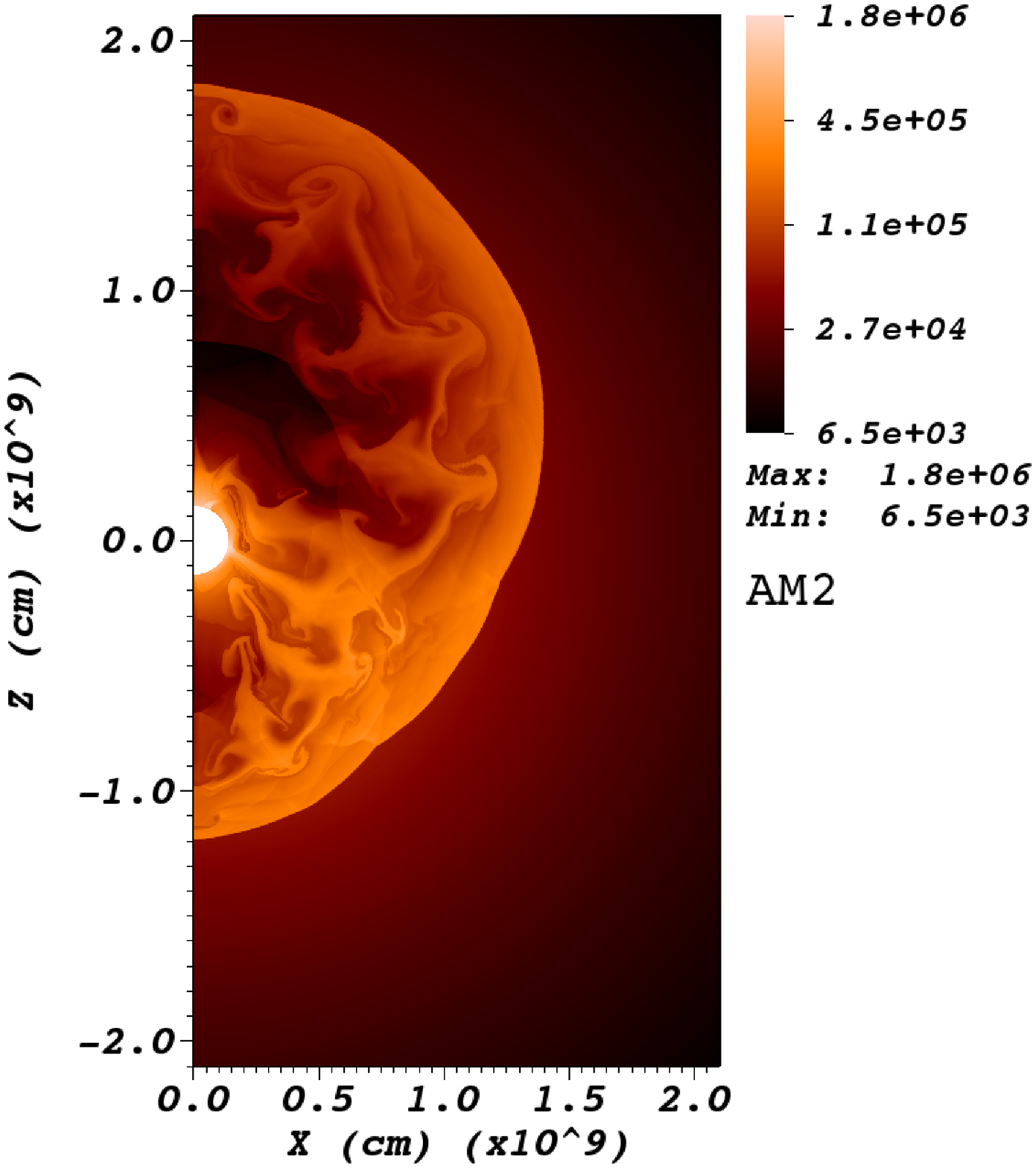}
\end{center}
\end{minipage}
\begin{minipage}{0.5\hsize}
\begin{center}
\includegraphics[width=6cm,keepaspectratio,clip]{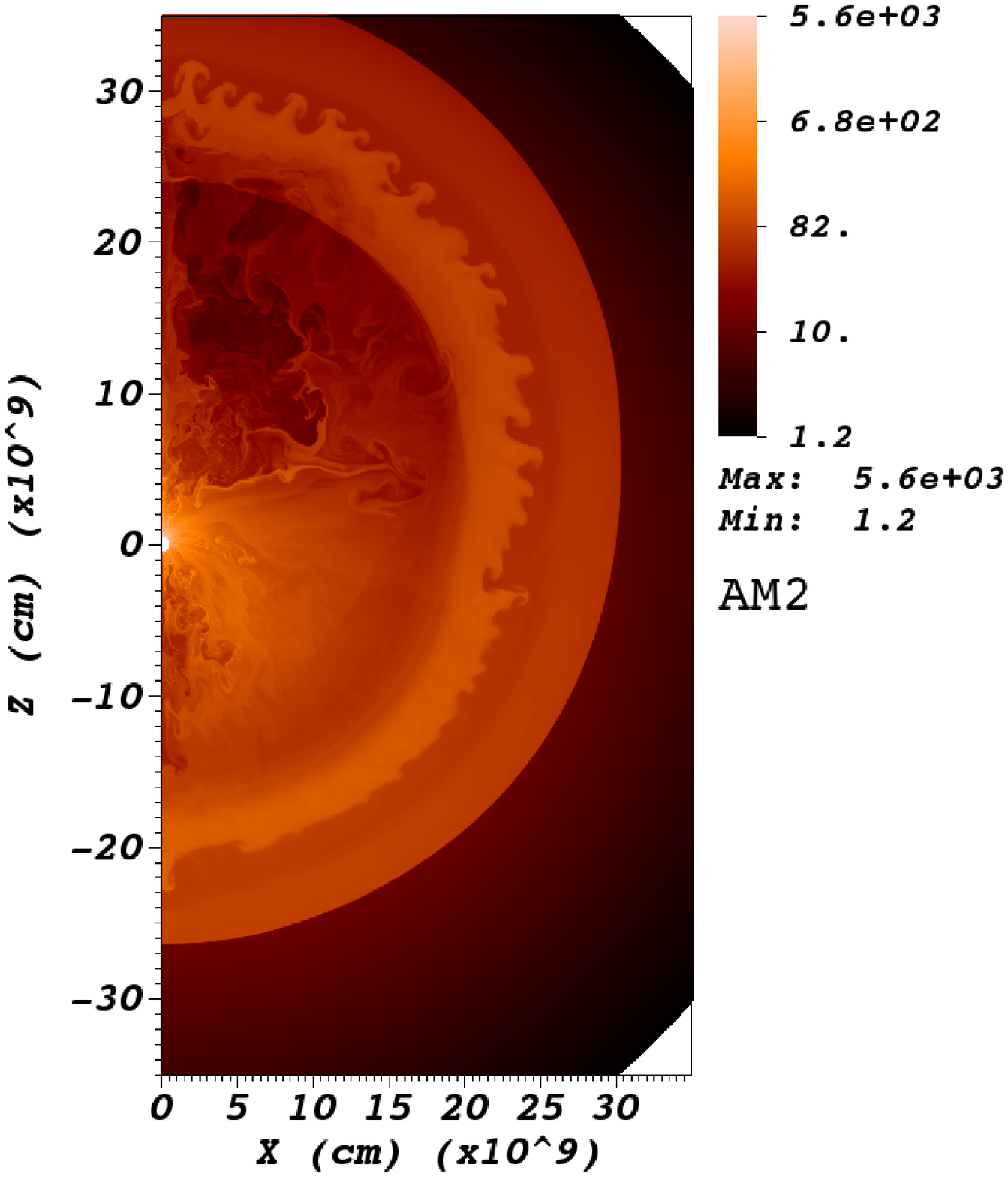}
\end{center}
\end{minipage}
\\
\hspace{-0.5cm}
\begin{minipage}{0.5\hsize}
\begin{center}
\includegraphics[width=6cm,keepaspectratio,clip]{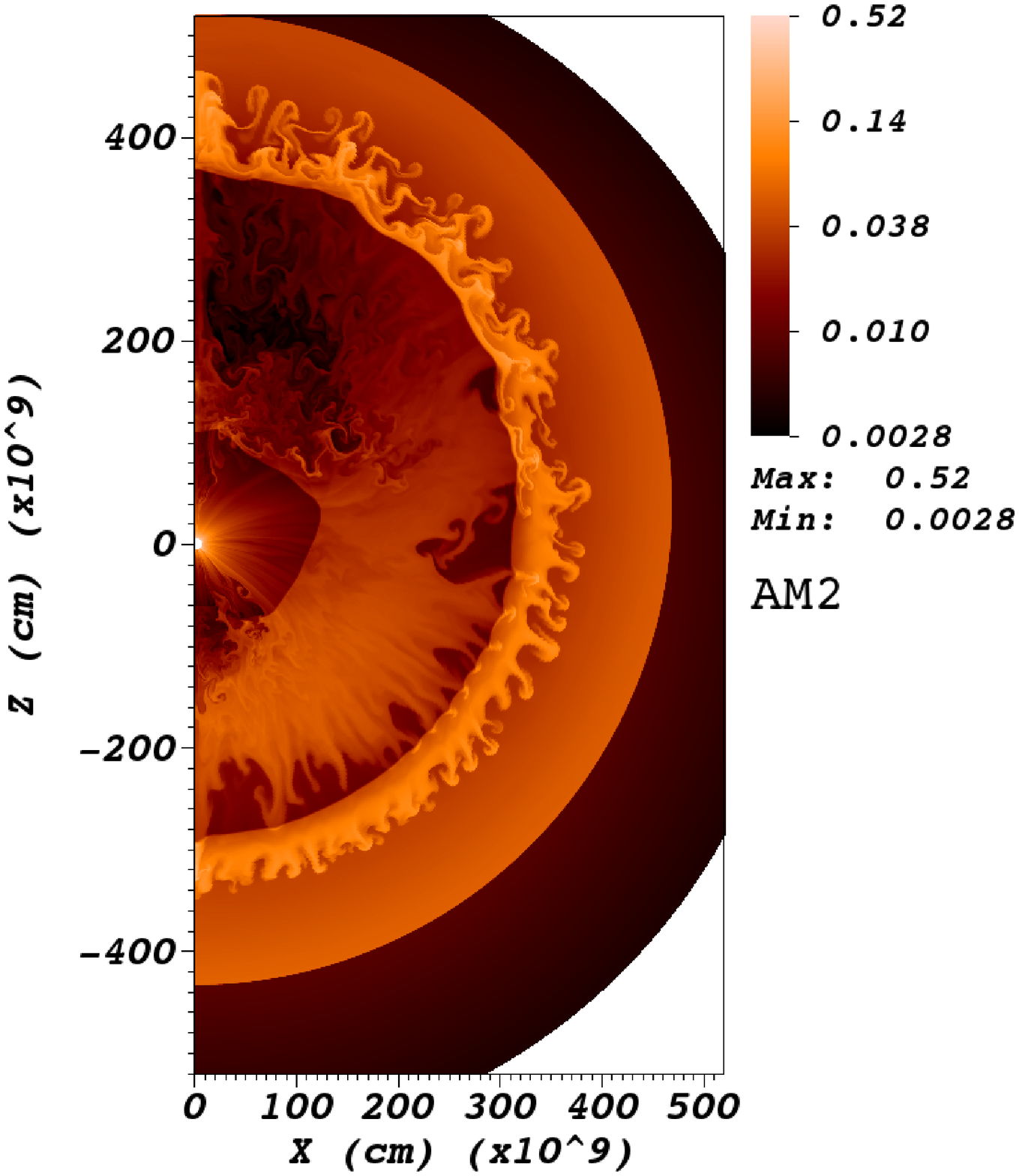}
\end{center}
\end{minipage}
\begin{minipage}{0.5\hsize}
\begin{center}
\includegraphics[width=6cm,keepaspectratio,clip]{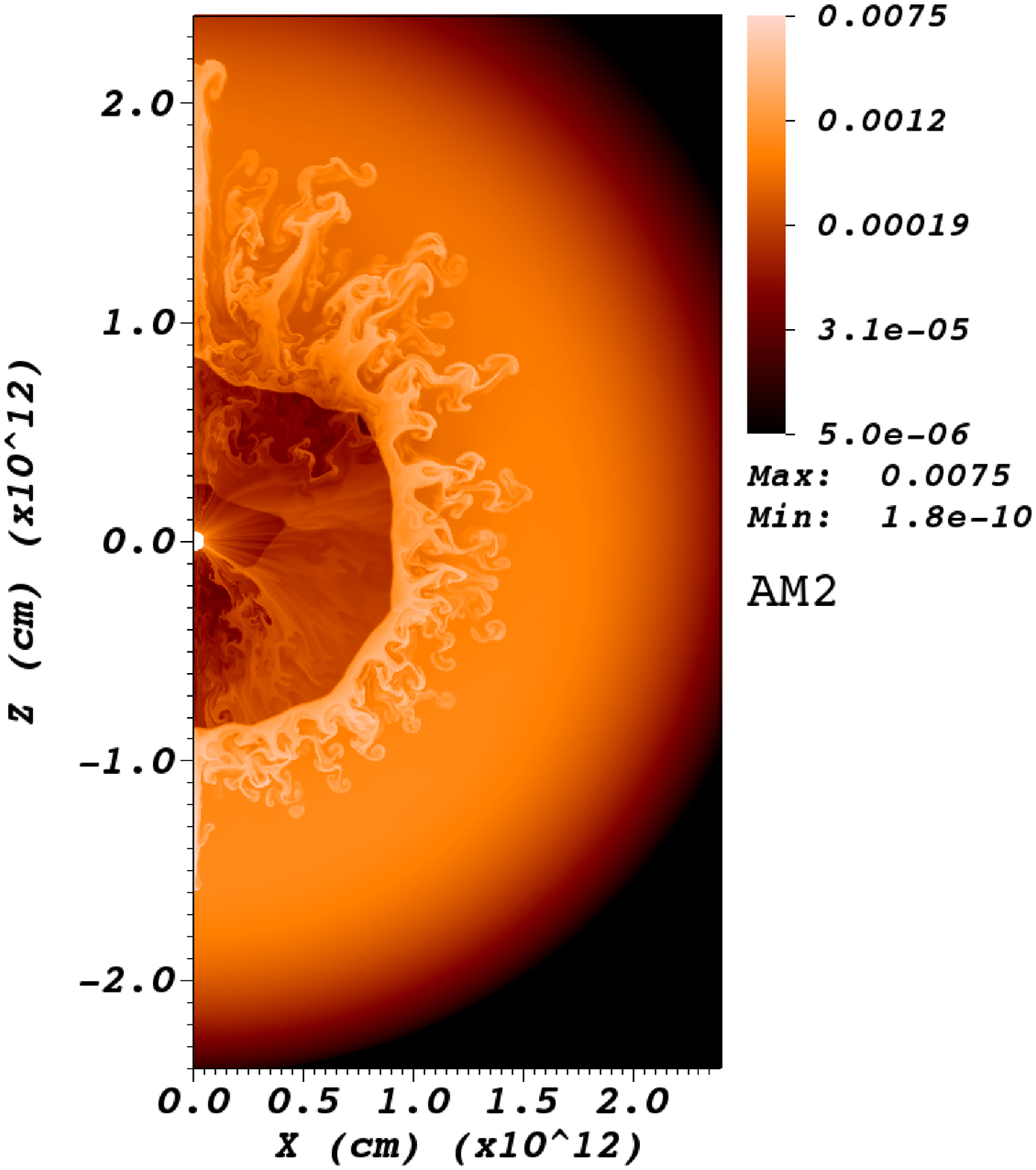}
\end{center}
\end{minipage}
\end{tabular}
\caption{Snap shots of distributions of density at the time of 
0.572 s (top left), 30.0 s (top right), 711.5 s (bottom left), and 5753 s (bottom right) for model AM2. 
The unit of values in color bars is g cm$^{-3}$. The values in color bars are logarithmically scaled.}
\label{fig:dens_AM2}
\end{figure*}

In the previous models in the paper, no high velocity of $^{56}$Ni ($\gtrsim$ 3,000 km s$^{-1}$) 
is obtained except the cases in the test models AT1 and AT2. Therefore, we finally consider 
the perturbations of both initial shock waves and pre-supernova origins, i.e., model AM2. 
The time evolution of the density distribution for model AM2 is shown 
in Figure~\ref{fig:dens_AM2}. After the initiation of the explosion, a globally anisotropic 
shock wave asymmetric across the equatorial plane propagates outward (the top left panel). 
Inside the shock wave, smaller-sale clumpy structures, i.e., outward and inward fingers, 
are also seen. After the shock wave passes through the composition interface of C+O/He, 
perturbations grow due to RT instabilities around the composition interface (the top right panel). 
At this phase, first moving clumps of $^{56}$Ni reach the interface and are conveyed outward 
with the aid of RT instabilities. Then, RT instabilities around the composition interface of He/H 
are developed (the bottom left panel). We find that the multiply introduced perturbations make 
some fractions of innermost metals including $^{56}$Ni reach around the bottom of 
the dense helium shell and penetrate it. Eventually, prominent RT fingers are developed 
in particular in the upper hemisphere (the bottom right panel). 

\begin{figure*}[htbp]
\begin{tabular}{cc}
\hspace{-0.5cm}
\begin{minipage}{0.5\hsize}
\begin{center}
\includegraphics[width=6cm,keepaspectratio,clip]{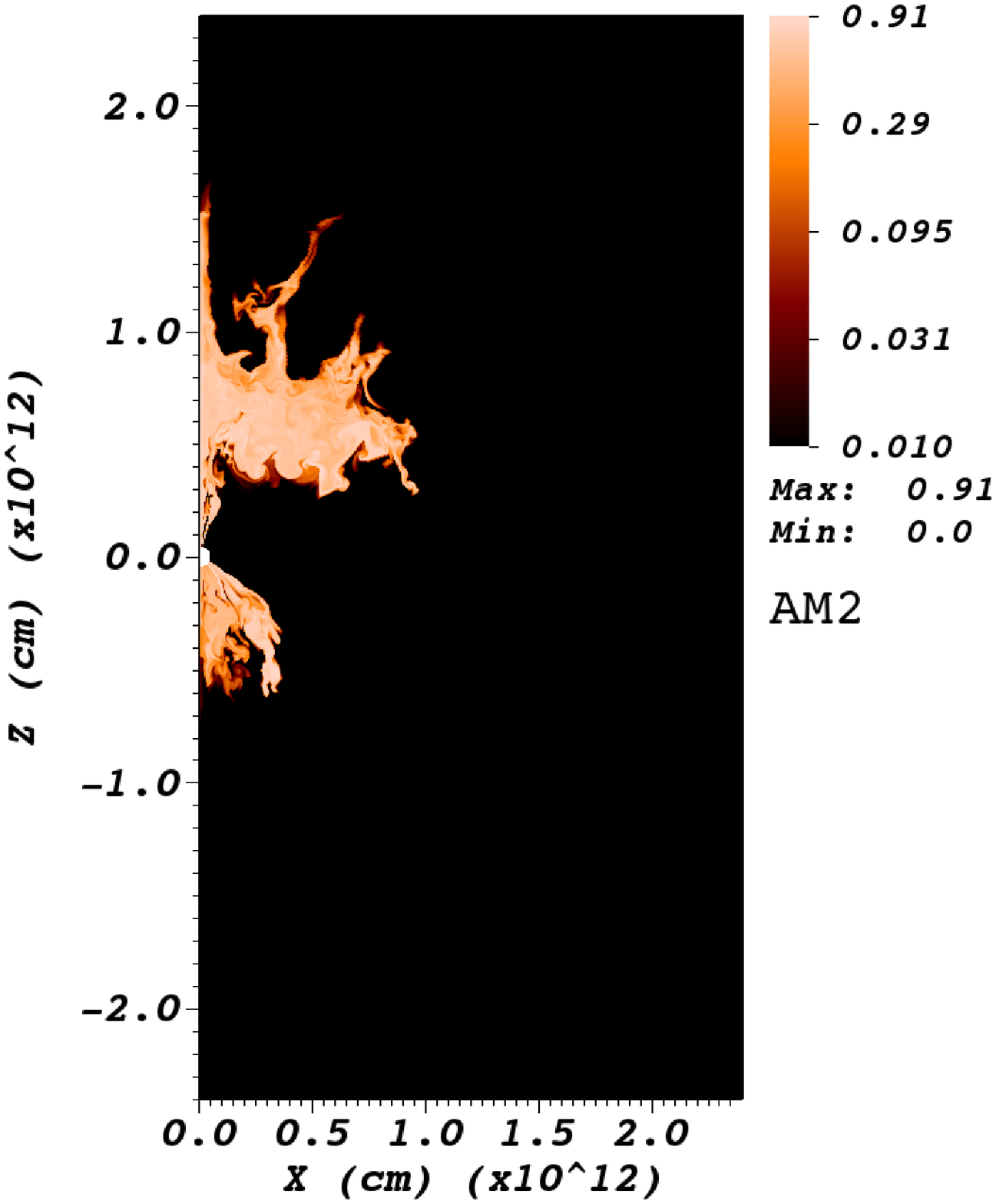}
\end{center}
\end{minipage}
\begin{minipage}{0.5\hsize}
\begin{center}
\includegraphics[width=6cm,keepaspectratio,clip]{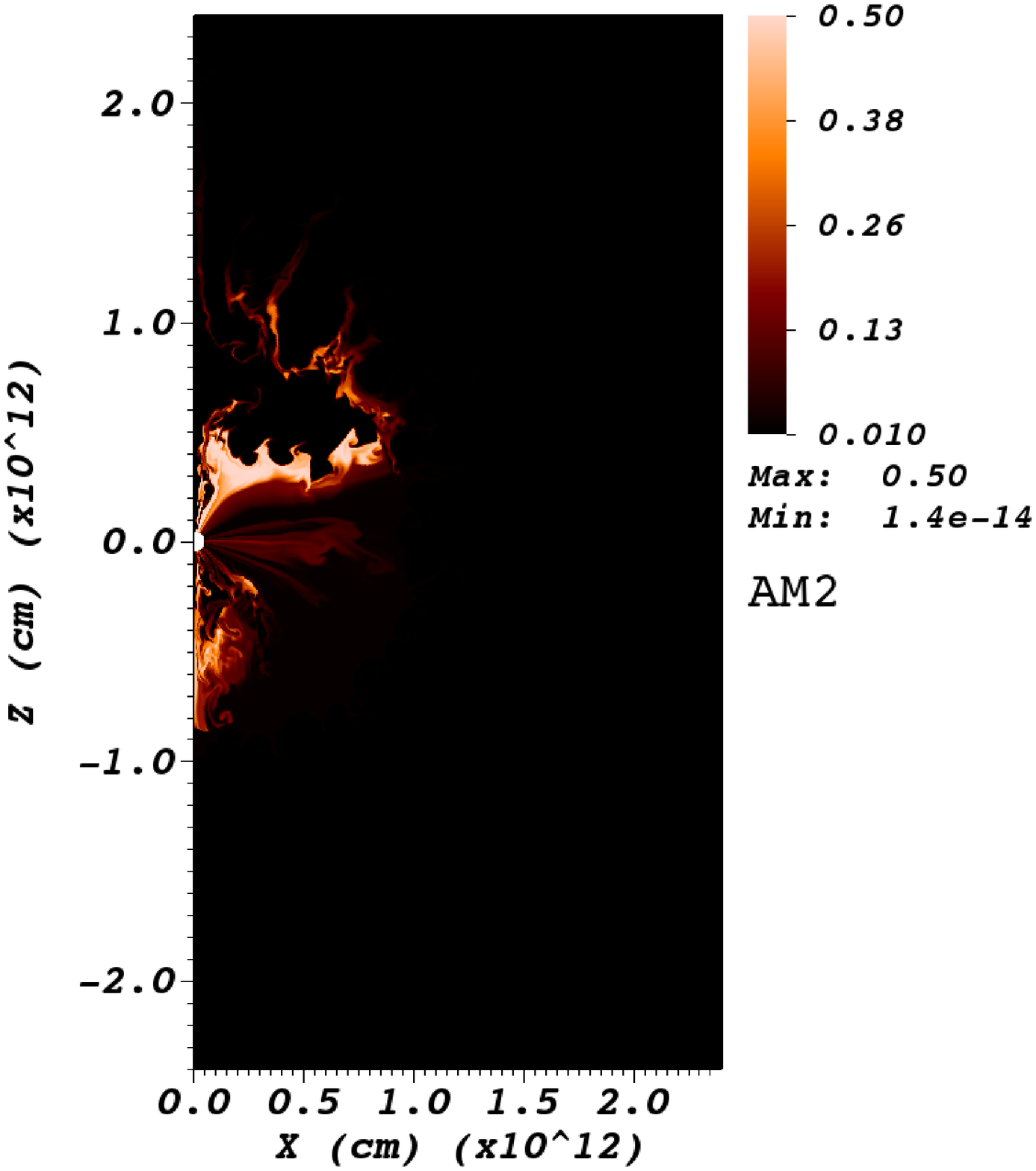}
\end{center}
\end{minipage}
\\
\hspace{-0.5cm}
\begin{minipage}{0.5\hsize}
\begin{center}
\includegraphics[width=6cm,keepaspectratio,clip]{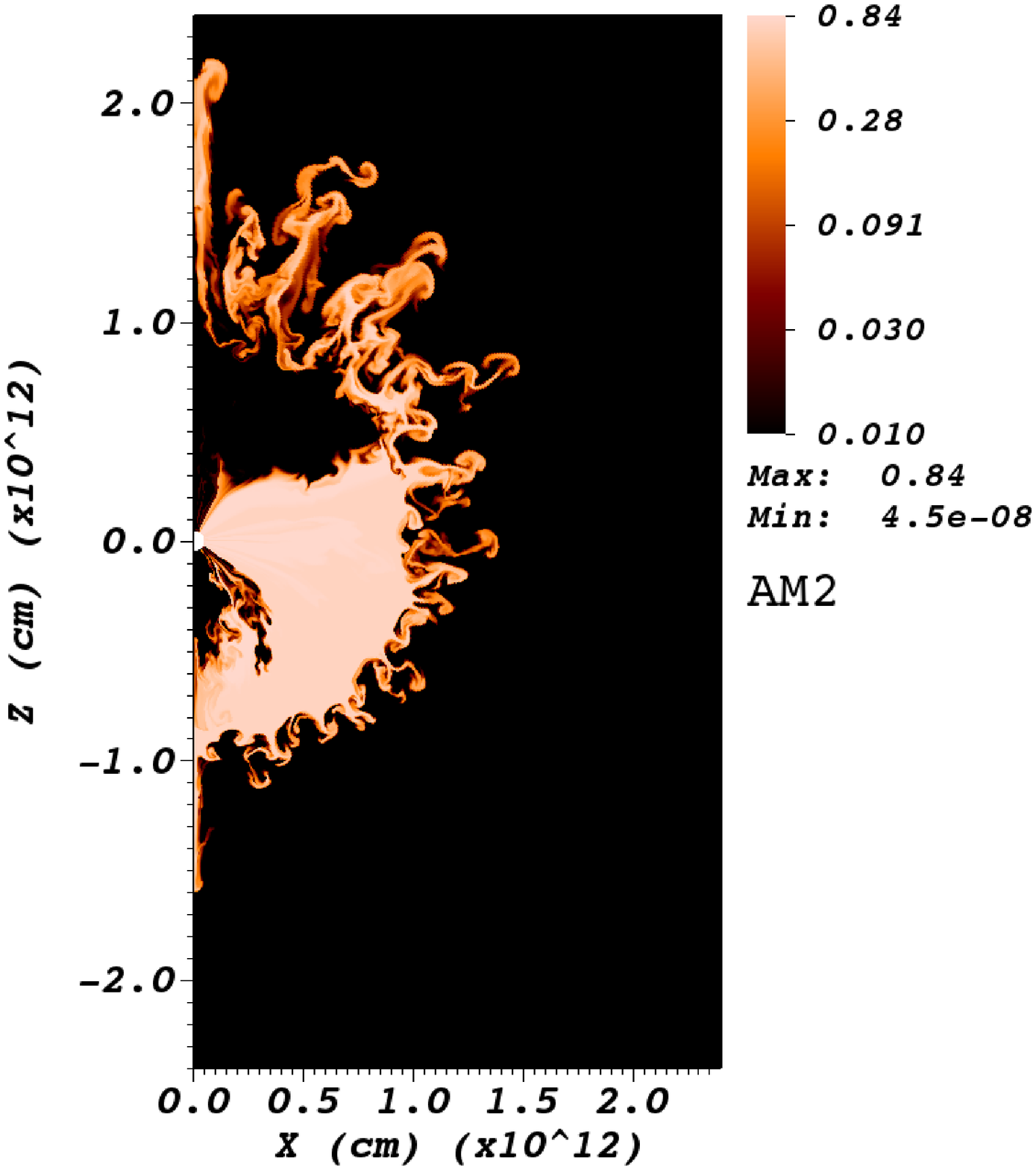}
\end{center}
\end{minipage}
\begin{minipage}{0.5\hsize}
\begin{center}
\includegraphics[width=6cm,keepaspectratio,clip]{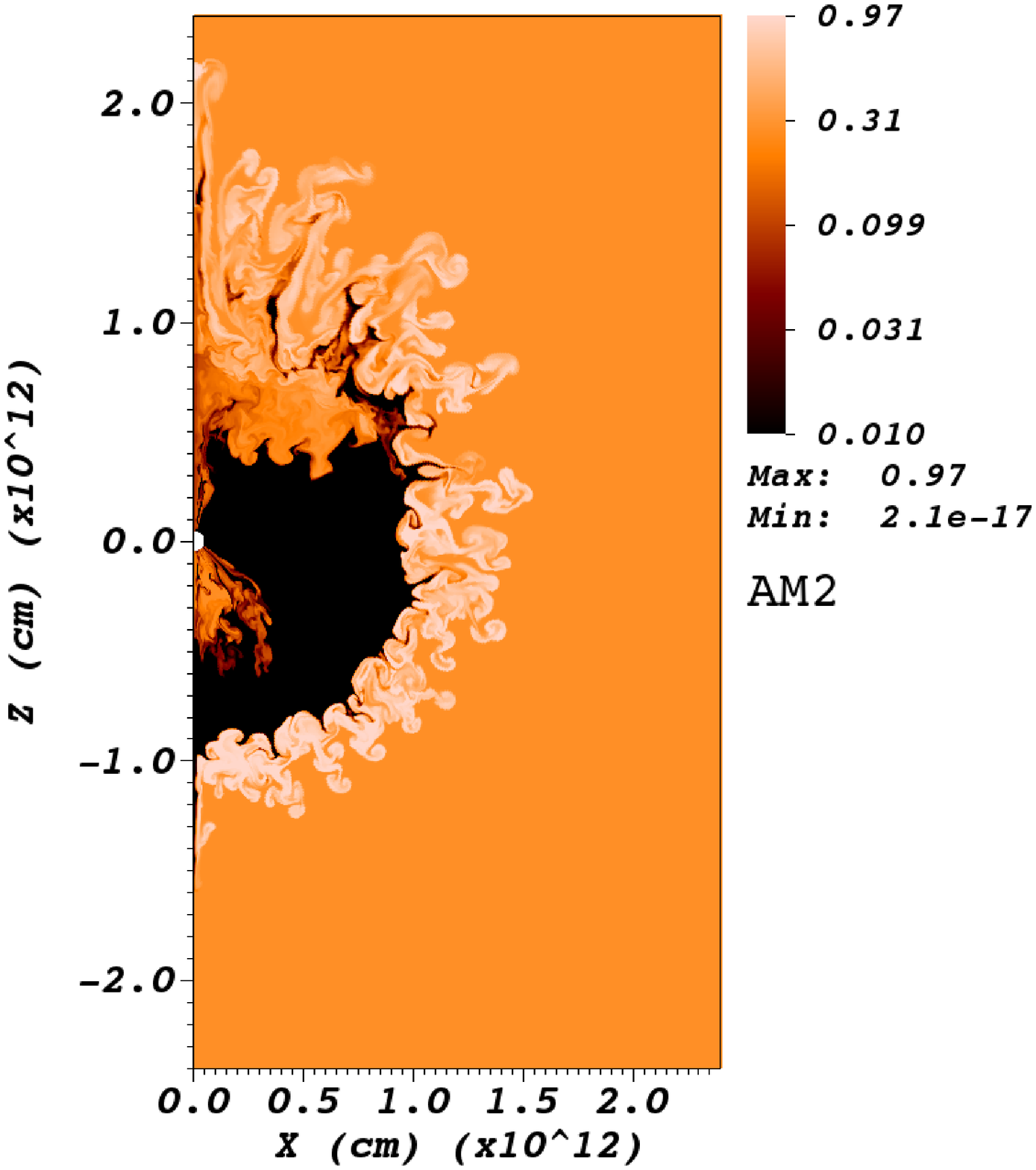}
\end{center}
\end{minipage}
\end{tabular}
\caption{Same as Figure~\ref{fig:element_SM} but for elements, $^{56}$Ni, $^{28}$Si, $^{16}$O, 
and $^{4}$He and the time of 4578 s.}
\label{fig:element_AM2}
\end{figure*}

The distributions of mass fractions of elements $^{56}$Ni, $^{28}$Si, $^{16}$O, and $^4$He 
are shown in Figure~\ref{fig:element_AM2}. We can see that most of $^{56}$Ni is confined 
inside the dense helium shell. However, some fraction of $^{56}$Ni penetrates the shell along 
RT fingers (the top left panel). We emphasize that the penetrations of $^{56}$Ni are seen not 
only in regions close to the polar axis but also in regions away from the polar axis. 
$^{28}$Si is prominent around the $^{56}$Ni (the top left panel). 
$^{16}$O is outstanding in regions inside the dense helium shell in the 
lower hemisphere and in RT fingers (the bottom left panel). $^{4}$He is mixed inward 
due to RT instabilities (the bottom right panel). It should be noted that the obtained 
morphology of inner ejecta such as $^{56}$Ni, $^{28}$Si, and $^{16}$O, is roughly elliptical 
and the ratio of the major to minor axes is approximately 2. These are roughly consistent 
with the recent observation of supernova remnant SN~1987A \citep[1.8 $\pm$ 0.17:][]{kja10}.    

\begin{figure}[htbp]
\hspace{-2cm}
\begin{center}
\includegraphics[width=6cm,keepaspectratio,clip]{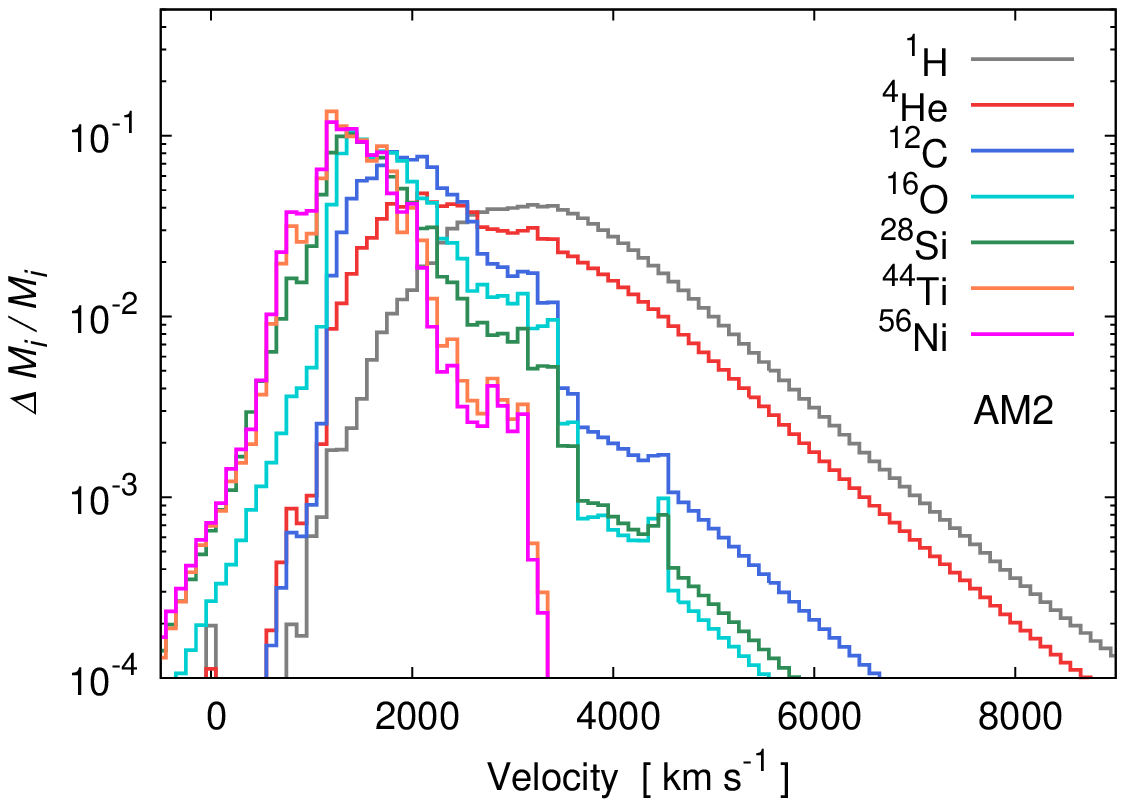}
\caption{Same as Figure~\ref{fig:vel_SPM} but for model AM2 at the time of 4578 s.}
\label{fig:vel_AM2}
\end{center}
\end{figure}

The mass distributions of elements at the end of simulation time as a function 
of radial velocity for model AM2 are shown in Figure \ref{fig:vel_AM2}. 
The largest value of the velocity of $^{56}$Ni clump achieves around 3,000 km s$^{-1}$. 
We find that the amount of $^{56}$Ni with velocity over 2,700 km s$^{-1}$ 
is approximately 1.4 $\times$ 10$^{-3}$ $M_{\odot}$. The high velocity tails 
of other metals, $^{28}$Si, $^{12}$C, and $^{16}$O, are also enhanced compared 
with those in models AS2 to AS8. Note that models AS2 to AS8 have no perturbation 
of pre-supernova origins (see Figure~\ref{fig:vel_AS}). The minimum velocity 
of $^{1}$H (1,100 km s$^{-1}$, see Table~\ref{table:results}) is slightly smaller 
than that of models AS2 to AS8 (1,200 -- 1,300 km s$^{-1}$). This indicates inward 
mixing of $^{1}$H is slightly enhanced compared with models AS2 to AS8. 
But strong inward mixing of $^{1}$H seen in e.g., models AM1 and AT2 
is not realized in model AM2. The possible reason are that the higher 
explosion energy ($\sim$ 2 $\times$ 10$^{51}$ erg) minifies the time 
for RT instabilities to grow and a globally aspherical explosion makes 
the amount of inward $^1$H to be small. 

\begin{figure}[htbp]
\hspace{-2cm}
\begin{center}
\includegraphics[width=6cm,keepaspectratio,clip]{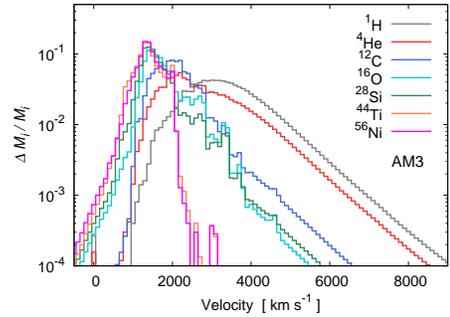}
\caption{Same as Figure~\ref{fig:vel_SPM} but for model AM3 at the time of 4562 s.}
\label{fig:vel_AM3}
\end{center}
\end{figure}

To see the effects of clumpy structures, we perform the model AM3 which has 
no clumpy structure given by Equation (\ref{eq:clump}) as a counterpart of model AM2. 
The mass distributions of elements as a function of radial velocity are shown 
in Figure~\ref{fig:vel_AM3}. We can see very small fraction of $^{56}$Ni is clustered 
around 3,000 km s$^{-1}$.  But the fraction is rather smaller than that of model AM2. 
This means that initial clumpy structures are important for $^{56}$Ni to be conveyed 
into high velocity regions. 

\begin{figure}[htbp]
\hspace{-2cm}
\begin{center}
\includegraphics[width=6.5cm,keepaspectratio,clip]{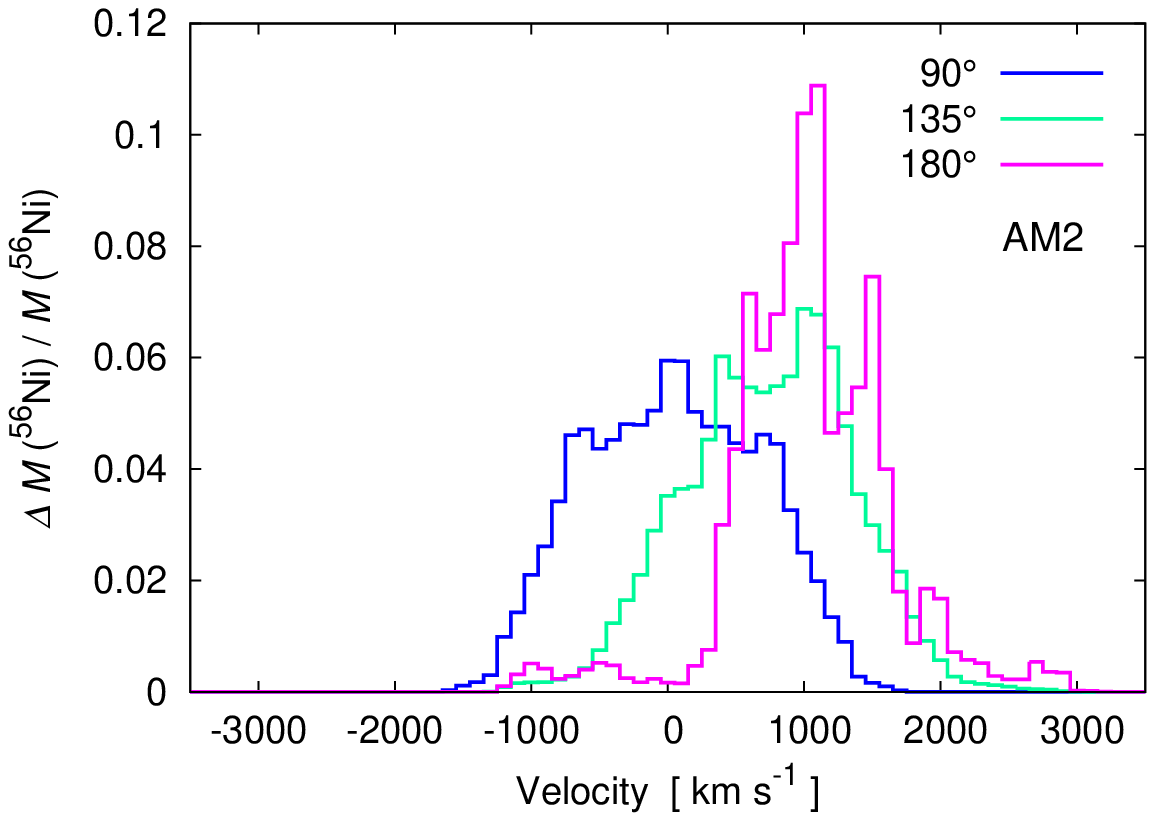}
\caption{Same as Figure~\ref{fig:line_AM1} but for model AM2 and the time of 4578 s.}
\label{fig:line_AM2}
\end{center}
\end{figure}

The mass distributions of $^{56}$Ni as a function of line of sight velocity are shown 
in Figure~\ref{fig:line_AM2}. If $\theta_{\rm ob} =$ 90$^{\circ}$, the distribution 
is clustered around the null velocity point and the high velocity tails reach 
around $\pm$ 1,500 km s$^{-1}$. If the observer see the explosion from the 
opposite direction that the explosion is the strongest, i.e., 
$\theta_{\rm ob} =$ 180$^{\circ}$, the tail of the red-sifted side reaches 
the velocity of around 3,000 km s$^{-1}$ while the fraction of the blue-shifted side 
is significantly small. This is because the amount of $^{56}$Ni moving 
in the direction where the explosion is stronger is larger than that in other directions. 
This reflects that $^{56}$Ni tend to be concentrated in the regions where 
the explosion is stronger. In the case of $\theta_{\rm ob} =$ 135$^{\circ}$, 
the distribution seems to be the combination from the case of 
$\theta_{\rm ob} =$ 90$^{\circ}$ and the case of $\theta_{\rm ob} =$180$^{\circ}$, 
and the tail of the red-shifted side is slightly reduced compared 
with the case of $\theta_{\rm ob} =$ 180$^{\circ}$. In both cases 
of $\theta_{\rm ob} =$ 135$^{\circ}$ and $\theta_{\rm ob} =$ 180$^{\circ}$, 
the peaks of the distributions are located around 1,000 km s$^{-1}$. 
Hence, the observed sifts of peaks in the line profiles of [Fe II] in SN~1987A are 
reproduced in the case of $\theta_{\rm ob} =$ 135$^{\circ}$ and 180$^{\circ}$ in this model. 

We finally find that the aspherical explosion asymmetric across the equatorial plane 
with clumpy structures with the aid of perturbations of pre-supernova origins can 
convey $^{56}$Ni into high velocity regions of $\sim 3,000$ km s$^{-1}$. 
In the next section,  we discuss about the possible ingredients to obtain higher 
velocity of $^{56}$Ni ($\sim$ 4,000 km s$^{-1}$) and some  implications from several aspects. 

\begin{figure}[htbp]
\hspace{-2cm}
\begin{center}
\includegraphics[width=6.5cm,keepaspectratio,clip]{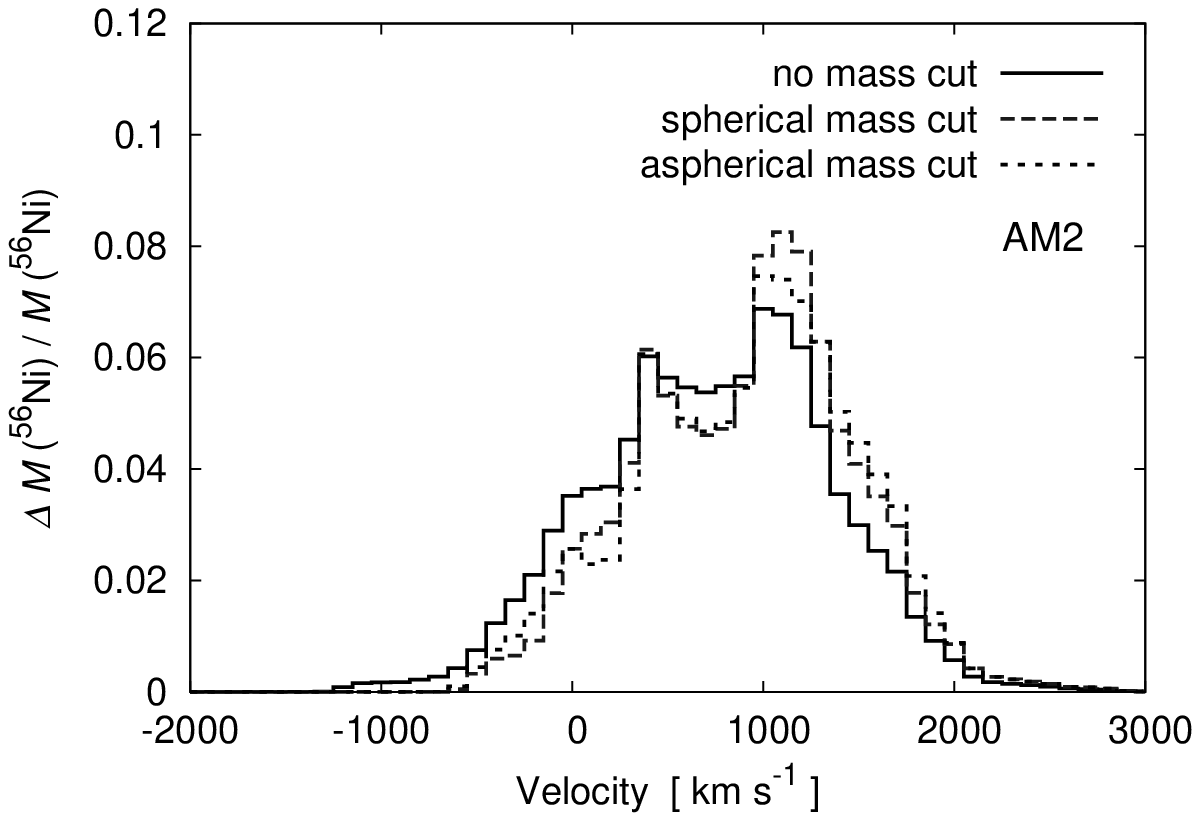}
\caption{
Mass distributions of $^{56}$Ni as a function of line of sight velocity 
($\theta_{\rm ob}$ = 135$^{\circ}$) at the end of simulation time (4578 s) for model AM2. 
Three cases without mass cut (solid), with the spherical mass cut (dashed), 
and with the aspherical mass cut (dotted) are  shown. See the text in \S 5.1 
for the explanation of the mass cut. 
}
\label{fig:line_AM2_mass_cut}
\end{center}
\end{figure}

\section{Discussion}

\subsection{Effects of the mass cut}

In our calculations, the masses of $^{56}$Ni in the computational domain at the end 
of simulation time are slightly overestimated compared with the value 
for SN~1987A \citep[0.07 $M_{\odot}$: e.g.,][]{shi88}. The obtained masses of $^{56}$Ni 
range over 1.91 -- 1.97 $\times$ 10$^{-1}$  $M_{\odot}$ for spherical models and model AT1. 
Those for the other models range over 
7.23 $\times$ 10$^{-2}$ -- 1.09 $\times$ 10$^{-1}$ $M_{\odot}$. 
The effects of the fallback of matter in our calculations may have some uncertainties 
because those could depend on the location of the inner boundary, the adopted 
inner boundary condition and the treatments of gravity. Therefore, in this section, 
we consider so-called the `mass cut' which determines arbitrarily the location that 
divides the ejecta from the compact remnant. In the previous sections, we do not 
take into account the mass cut in deriving the mass of $^{56}$Ni. Generally, 
the mass cut is determined so that the mass of $^{56}$Ni in the ejecta 
is consistent with the observed one. We consider two types of the mass cut. 
One is the spherical mass cut which is determined to be the location that the sum 
of the mass of $^{56}$Ni in from regions with larger radii to regions with 
smaller radii reaches 0.07 $M_{\odot}$. In this case the shape of the mass cut 
is almost spherical symmetric. The other is the aspherical mass cut which 
is determined to be the location that the sum of the mass of $^{56}$Ni in from 
regions that have larger explosion energy 
(the sum of the specific kinetic, internal, and gravitational energies in the cell) 
to regions that have smaller explosion energy reaches 0.07 $M_{\odot}$ 
in a similar way in \citet{nag00}. The shape of the aspherical mass cut could 
be aspherical due to an aspherical explosion.  
Figure \ref{fig:line_AM2_mass_cut} shows the mass distributions of $^{56}$Ni 
as a function of line of sight velocity ($\theta_{\rm ob} =$ 135$^{\circ}$) 
at the end of simulation time for model AM2 with the effects of the mass cut. 
The case without mass cut is also depicted for reference. 
Both cases with the spherical mass cut (the dashed line) and aspherical 
mass cut (the dotted line), high velocity tails ($\gtrsim$ 1,000 km s$^{-1}$) are 
slightly enhanced compared with that without mass cut (the solid line). 
While the low velocity tails are slightly reduced compared with that without mass cut. 
The differences between the cases with the spherical mass cut and the aspherical 
mass cut are not much distinctive except for around the peak ($\sim$ 1,000 km s$^{-1}$). 
From the estimations in this section, we conclude that the effects of the mass cut 
on the results are not much large unless the total mass of $^{56}$Ni is too overestimated.

\subsection{Neutron star kick and its observational implication}

As mentioned in \S 3.4, an aspherical explosion has been thought to be the one of promising 
triggers of neutron star (NS) kicks \citep[e.g.,][]{sch06,won10}. The recoil velocity 
of a compact remnant (neutron star) can be estimated simply considering the momentum 
conservation. Initially, the total momentum of the progenitor is zero in the frame 
of the center of gravity (i.e., the center of the progenitor). 
Then, the neutron star recoil velocity is given by 
\begin{equation}
\vec{v}_{\rm NS}\,(t) = - \vec{P}_{\rm gas}\,(t) /M_{\rm NS}\,(t), 
\label{eq:kick}
\end{equation}
where $\vec{P}_{\rm gas} = \int_{r_{\rm in}}^{r_{\rm out}} \rho \,\vec{v} \, dV$ 
is the total momentum outside the neutron star and $M_{\rm NS}$ is the mass 
of the neutron star. $r_{\rm in}$ ($r_{\rm out}$) is the radius of the inner (outer) boundary. 
We regard simply the mass of the neutron star $M_{\rm NS}$ as the mass inside 
the inner boundary. We estimate the recoil velocity of the compact remnant for model AM2. 
Note that since our simulations are axisymmetric, only $Z$-component has the non-zero value. 
The obtained neutron star velocity at the end of simulation time is $-\,734$ km s$^{-1}$. 
Since the sign is negative, the nascent neutron star is kicked in the opposite to stronger 
explosion direction. The averaged observed values of young pulsars are several 
hundred km s$^{-1}$ \citep[e.g.,][]{fau06} and some of them have over 
500 km s$^{-1}$, even 1,000 km s$^{-1}$ \citep{cha05}. 
Therefore, the estimated value is within the observational values but it is somewhat larger 
than that of a typical one. In our estimation the values of $r_{\rm in}$ is rather larger than 
the surface radius of the compact remnant. Therefore, the estimated value may have large 
uncertainty. Hence, it is safe to say that our estimation includes uncertainty of 
several tens of \% or more. Nonetheless, the estimated value may be within 
the observed range. From the analysis, the asymmetry roughly represented by 
$v_{\rm pol}/ v_{\rm eq}$, $v_{\rm up}/ v_{\rm down} =$ 2 may be sufficient to explain 
the observed velocities of young pulsars. For a typical velocity of young pulsars, 
$v_{\rm up}/ v_{\rm down} <$ 2 may be preferable.

For SN~1987A, the compact remnant has not been found so far. However, if the explosion 
is stronger in some direction, the nascent neutron star will be kicked in the opposite 
to the strongest explosion direction as discussed above. Actually, the observed 
line profiles of [Fe II] as a function of Doppler velocity for SN~1987A are asymmetric 
across the null velocity point. The peak is located in the red-shifted side \citep{haa90}. 
It is commonly known that the images of SN~1987A have three rings. The inner ring 
is inclined at about 45$^{\circ}$ to the sky, and the north (south) part of the ring 
is closer to (away from) us and blue-sifted (red-shifted) \citep[e.g.,][]{tzi11}. 
Moreover, recent near infrared spectroscopic observations have revealed that 
the inner ejecta of SN~1987A is elongated and it is roughly confined to the same plane 
as the inner ring \citep{kja10}. From above considerations, we can speculate the 
direction of the velocity of the compact remnant of SN~1987A. The south part 
of the explosion of SN~1987A is stronger than the north part of it because 
the south part of inner ejecta is red-shifted and the line profile of [Fe II] in SN~1987A 
implies that the explosion is stronger in the red-shifted side. 
Therefore, the compact remnant of SN~1987A may be kicked 
in a northern direction. It should be noted that a similar discussion 
have been done by \citet{nag00} but the conclusion is opposite to ours. 
This may be because at that time, it had been thought that the explosion 
of SN~1987A had been a jetlike or bipolar and the north part of the inner ejecta 
of SN~1987A had been red-shifted \citep{wan02}. However, as noted above, 
recent observations contradict it. 

The observed features of line profiles of [Fe II] in SN~1987A have not been excellently 
reproduced by not only our models but also recent previous studies with 
a neutrino-driven model (e.g., \citet{kif06,gaw10}). This is one of remained enigmas 
to be explained in the future. If the line profiles are reproduced by a more sophisticated model, 
it will be a good diagnostic to speculate the direction and magnitude 
of the recoil velocity of the compact remnant of SN~1987A. 

\subsection{Limitations of simulations and possible ingredients to obtain a higher velocity of $^{56}$Ni}

In this section, we consider the limitation of our simulations in the present paper and some possible 
ingredients to convey $^{56}$Ni into higher velocity regions. 

Self-gravity is implemented in our code assuming the spherical symmetry to save CPU time. 
However, as we have seen in previous sections, the distributions of density and mass fractions 
of metals are rather anisotropic and self-gravity due to anisotropic matter distributions 
could potentially affect the fallback and protrusions of innermost metals. Hence, it is desirable 
that the Poisson equation for self-gravity is solved in more sophisticated manner including 
multi-dimensional effects. Additionally, if matter distributions are changed from the results 
in the paper, it could affect the estimation of the recoil velocity of the nascent neutron star. 

Energy depositions due to decays of radioactive nuclei $^{56}$Ni and $^{56}$Co are 
the one of possible mechanisms to accelerate innermost metals including $^{56}$Ni 
in later phases after the shock breakout. If we assume the ejected mass of $^{56}$Ni 
is 0.07 $M_{\odot}$, the total released energy estimated by Equations (\ref{eq:ni}) 
and (\ref{eq:co}) reaches $\sim\,$1.3 $\times$ 10$^{49}$ erg 400 day after the explosion. 
In our models, the radial velocity of $^{56}$Ni is clustered around 1,000 km s$^{-1}$ 
at the ends of simulation time. Hence, the kinetic energy of bulk $^{56}$Ni 
is roughly estimated as 
\begin{equation}
E_{\rm kin} \sim 1.6 \times 10^{48} \,{\rm erg} \left( \frac{M}{0.07 \, M_{\odot}} \right) \, 
\left( \frac{v}{1,000 \,{\rm km \,s^{-1}}}\right)^2.
\label{eq:ekin}
\end{equation}
Therefore, if we assume that all the released energy from decaying $^{56}$Ni and $^{56}$Co 
is converted to the kinetic energy of itself, it becomes eight times larger than 
that before the heating, which corresponds to a threefold increase in velocity. 
Of course, part of the gamma-rays from the metals may escape without heating and 
the estimation of Equation (\ref{eq:ekin}) may have a large uncertainty. 
Then, the above estimation is kind of the upper bound. Nonetheless, the peak velocity 
of the $^{56}$Ni could increase by about 30\% due to decays of $^{56}$Ni 
and $^{56}$Co \citep{her91}. As mentioned in \S1, the heating due to the decay 
of $^{56}$Ni and $^{56}$Co could be the seed of perturbations in a later phase, i.e., 
`nickel bubble'. 
For SN 1987A, if 10\% of $^{56}$Ni ($\sim$ 0.007$M_{\odot}$) had $\sim$ 1,800 km s$^{-1}$ 
at the stage of $\sim$ 10$^4$ s after the explosion, the velocity of $^{56}$Ni of $\sim$ 3,000 km s$^{-1}$ 
could be explained \citep{bas94}. 
Our simulations are stopped just after the shock breakout due to 
the limitation of time. However, in order to determine the final velocity of $^{56}$Ni, 
more long term simulations are required. 

Three-dimensional effects could be the most important to convey innermost metals 
into high velocity regions. The differences of the growth of a single-mode perturbation 
between two- and three-dimension was investigated by \citet{kan00}. The authors found 
that the growth of the perturbations in three-dimension is 30\% -- 35\% faster than 
that in two-dimension. \citet{ham10} demonstrated that the drag force to clumps 
of innermost metals in three-dimension is less than that in two-dimension. 
In their three-dimensional simulation, the clump can penetrate the dense helium shell 
(`wall') at the bottom of the hydrogen envelope even in the absence of RM instabilities. 
The authors insisted that in two-dimensional simulations, the motion of a clump 
is severely restricted to keep the `torus'-like structure due to the axisymmetric assumption, 
and the drag force to the clump becomes larger as the distance of the clump from 
the polar axis becomes larger. Therefore, our two-dimensional axisymmetric simulations 
may overestimate the drag force to clumps moving away from the polar axis, and 
protrusions of clumps of metals in a direction away form the polar axis could 
be changed in three-dimension. 
\citet{jog10b} investigated RT mixing in supernovae in three-dimension. 
Their finding is as follows. RT instabilities grow faster in three-dimension than 
in two-dimension at first, but in later phases, small-scale perturbations cause 
so-called `inverse cascading', mergers of smaller-scale structures into larger-scale ones, 
in three-dimension, which reduce the local 
Atwood number
\footnote{Atwood number $A$ is defined at the interface of fluids that have different densities 
$\rho_1$ and $\rho_2$ as $A \equiv (\rho_2 - \rho_1)/(\rho_1 + \rho_2)$, where $\rho_2 > \rho_1$.}, 
and eventually the resultant mixing lengths are not changed by the difference of the dimension. 
From the results of \citet{ham10} and \citet{jog10b}, the dimensional effects on matter mixing 
may be important only if the scale of perturbations is large. 

In some models in the paper, outstanding protrusions of matter including $^{56}$Ni along 
with the polar axis are seen. As mentioned before, those may reflect the combination 
of several effects as follows. 
1. No penetration of matter across the polar axis due to the `reflection' boundary condition. 
2. Discretization errors close to the polar axis. 
3. The physical nature that the explosions are the strongest in regions around the axis in our modes. 
However, as far as in two-dimensional axisymmetric simulation, we hardly know which point 
is the dominant effect. Moreover, \citet{nor10} pointed out that strong bipolar asymmetry 
seen in two-dimensional neutrino-driven explosions aided by convection and/or SASI 
may not survive in three-dimension. Therefore, high-resolution, three-dimensional 
simulations are ultimately required to conclude the mixing of innermost metals 
and dimensional effects. We plan to extend our simulations for a prolonged time 
including multi-dimensional effects of self-gravity in the near future. 

\subsection{$^{44}$Ti as an indicator of asphericity?}

$^{44}$Ti is a relatively long-lived radioactive nucleus 
(the half life is 58.9$\pm$0.3 yr: \citet{ahm06}) and accounts for the energy source 
of the light curve of a core-collapse supernova after the heating due to decays of 
$^{56}$Ni and $^{56}$Co cease. Inner ejecta of the remnant of SN~1987A may currently 
be heated due to decays of $^{44}$Ti \citep{kja10}. Recently, direct-escape lines from 
the decay of $^{44}$Ti were detected in the remnant of SN~1987A \citep{gre12}. 
So far, direct-escape lines from the decay of $^{44}$Ti have been clearly detected 
\citep[e.g.,][]{ren06} only in Cassiopeia A except for the remnant of SN~1987A. 
The obtained mass of $^{44}$Ti at the ends of simulation time in our models 
range over 1.71 $\times$ 10$^{-4}$ -- 5.16 $\times$ 10$^{-4}$ $M_{\odot}$. 
This is roughly consistent with the value (3.1$\pm$0.8) $\times$ 10$^{-4}$ $M_{\odot}$ 
derived from the detected direct-escape lines of $^{44}$Ti in SN~1987A \citep{gre12}. 
$^{44}$Ti is synthesized by incomplete silicon burning in the explosive nucleosynthesis. 
As mentioned in \S 1 and \S 4.2, $^{44}$Ti is enhanced by an aspherical explosion 
due to the strong alpha-rich freeze-out \citep{nag98a,nag00}. We estimate the ratio of the 
masses of $^{44}$Ti to $^{56}$Ni. For spherical explosion models, the values are 
1.24 $\times$ 10$^{-3}$. For bipolar explosion case, the values of models 
of $v_{\rm pol}/ v_{\rm eq}$ = 2, are approximately 2.1 $\times$ 10$^{-3}$. 
While  the values of models of $v_{\rm pol}/ v_{\rm eq}$ = 4 are 2.57 $\times$ 10$^{-3}$. 
Therefore, models that have clear aspherical feature enhance the mass of $^{44}$Ti 
relative to that of $^{56}$Ni. For aspherical models of $v_{\rm pol}/ v_{\rm eq}$ = 2 
and $v_{\rm up}/ v_{\rm down}$ = 2, the values range over 
2.36 $\times$ 10$^{-3}$ -- 5.22 $\times$ 10$^{-3}$. 
The value of representative model AM2 is 4.97 $\times$ 10$^{-3}$. Therefore, 
the values of models of $v_{\rm pol}/ v_{\rm eq}$ = 2 and $v_{\rm up}/ v_{\rm down}$ = 2 
tend to be enhanced compared with those of bipolar explosion models. 
Note that the obtained mass of $^{44}$Ti in our models may be overestimated 
due to the small nuclear reaction network. The mass of synthesized $^{44}$Ti is roughly 
three orders of magnitude smaller than that of $^{56}$Ni. 
Therefore, neglecting other elements in the network may more affect 
the mass fraction of $^{44}$Ti than that of $^{56}$Ni relatively. Therefore, 
the values of the ratio should be regarded as a guide. 
Nonetheless, the qualitative tendencies may be correct. Thus, the value of mass ratio 
of $^{44}$Ti and $^{56}$Ni could be a good indicator of the asphericity of the explosion. 

\section{Summary}

We investigate matter mixing in a series of aspherical core-collapse supernova explosions 
of a 16.3 $M_{\odot}$ with a compact hydrogen envelope using a two-dimensional 
axisymmetric AMR hydrodynamic code, FLASH. We revisit RT mixing in spherical and/or 
mildly aspherical (bipolar jetlike) explosions with perturbations of pre-supernova origins. 
The effects of initial clumpy structures and multiply introduced perturbations are also studied.   
Our main findings are as follows. 

In spherical explosion models, the obtained maximum velocities of $^{56}$Ni range 
over 1,500 -- 1,600 km s$^{-1}$ and the minimum velocities of $^{1}$H range 
over 800 -- 1,400 km s$^{-1}$. The growth of RT mixing depends on the timing 
that perturbations are introduced. If perturbations are introduced when the shock 
wave reaches the composition interface of C+O/He, RT instabilities grow around 
the interface of C+O/He. On the other hand, if perturbations are introduced just before 
the shock wave reaches the composition interface of He/H, RT instabilities grow around 
the interface of He/H. RT instabilities around the interface of C+O/He account for 
conveying innermost metals including $^{56}$Ni into high velocity regions. 
While RT instabilities around the interface of He/H results in strong inward mixing 
of $^{1}$H and which can explain observed minimum velocity of $^{1}$H. 
Multiply introduced perturbations of pre-supernova origins, i.e., RT instabilities 
around the composition interfaces of both C+O/He and He/H, 
do not affect the maximum velocity of $^{56}$Ni. 

In the case of mildly aspherical bipolar explosions the qualitative features are 
the same as in the spherical models, and the maximum velocities of $^{56}$Ni and 
the minimum velocities of $^1$H range over 1,200 -- 1,700 km s$^{-1}$, 
and 700  -- 1,300 km s$^{-1}$, respectively. Both the maximum velocity of $^{56}$Ni 
and the minimum velocity of $^{1}$H are obtained in the model which has the most 
aspherical explosion ($v_{\rm pol}/v_{\rm eq} = $ 4) and perturbations are 
introduced two times. The growth of RT instabilities is enhanced slightly in the direction 
of stronger explosion. The distributions of elements, e.g., $^{56}$Ni, $^{28}$Si, $^{16}$O, 
and $^{4}$He, are rather different from those of spherical models. $^{56}$Ni is concentrated 
on regions closer to the polar axis. On the other hand, bulk of $^{16}$O is concentrated in equatorial regions. 

As a revisiting model, we consider the model that has a mild explosion 
$v_{\rm pol}/v_{\rm eq} = $ 2 with large perturbations (30\% amplitude) around 
the composition interface of He/H. In this model rather high velocity of $^{56}$Ni 
($\sim$ 3,000 km s$^{-1}$) is obtained and strong inward mixing 
of $^1$H can be explained simultaneously. However, if the same perturbations 
are introduced in initial radial velocities, the shape of perturbations can not 
survive up until the shock wave reaches at the interface of He/H, and eventually, 
no high velocity $^{56}$Ni is obtained. This implies that if such large and clear 
perturbations survive or exist due to some unknown mechanisms, 
the high velocity of $^{56}$Ni around 3,000 km s$^{-1}$ can be reproduced. 

To mimic a neutrino-driven explosion aided by convection and/or SASI, aspherical 
explosions asymmetric across the equatorial plane ($v_{\rm up}/v_{\rm down} = $ 2) 
with clumpy structures in initial shock waves (30\% amplitude) are investigated. 
The obtained maximum velocities of $^{56}$Ni and minimum velocity of $^1$H range 
over 1,800 -- 2,200 km s$^{-1}$ and 900 -- 1,300 km s$^{-1}$, respectively. 
Overall the maximum velocities of $^{56}$Ni are larger than those of bipolar explosion 
models but strong inward mixing of $^1$H are not obtained. The protrusions of $^{56}$Ni 
into the dense helium shell are sensitive to sizes of initial clumps. However, without RT 
instabilities due to perturbations of pre-supernova origins, the obtained maximum velocity 
of $^{56}$Ni does not reach the observed level. 

Finally, the combination of multiply introduced perturbations of pre-supernova origins 
and the aspherical explosion asymmetric across the equatorial plane with clumpy structures 
can cause the high velocity of $^{56}$Ni (3,000 km s$^{-1}$) 
without strong RM instabilities around the composition interface of H/He. 

To obtain a higher velocity of $^{56}$Ni ($\sim$ 4,000 km s$^{-1}$), some additional 
ingredients may be required. As mentioned before, in two-dimensional 
axisymmetric simulations, it is difficult to assess whether features seen along 
the polar axis are realistic or not. Therefore, to conclude the mixing of innermost 
metals and final velocity of $^{56}$Ni, more robust long-term, 
ultimately three-dimensional, simulations are required. 

\acknowledgments

We thank Shin-ichiro Fujimoto, Nobuya Nishimura, 
Yudai Suwa, and Yasuhide Matsuo for useful discussions. 
The software used in this work was in part developed by the DOE NNSA-ASC OASCR Flash Center 
at the University of Chicago. 
The numerical calculations were carried out on SR16000 at YITP in Kyoto University. 
This work is supported by the Ministry 
of Education, Culture, Sports, Science and Technology 
(No. 23105709 and No. 24540278), the Japan Society for the Promotion 
of Science (No. 19104006, No. 23340069, and No. 25610056), and 
the Grant-in-Aid for the Global COE Program 
"The Next Generation of Physics, Spun from Universality and Emergence" 
from the Ministry of Education, Culture, Sports, Science and Technology (MEXT) of Japan. 
S.-H. L. and J. M. acknowledge support from Grants-in-Aid 
for Foreign JSPS Fellow (No. 2503018 and No. 24.02022). 
We thank RIKEN for providing the facilities and financial support.

\end{document}